\documentclass[a4paper,fleqn]{cas-dc}

\usepackage[authoryear]{natbib}

\usepackage{tabularx}
\usepackage{algorithm}
\usepackage{booktabs}
\usepackage{algpseudocode}
\usepackage{subcaption}

\usepackage{xcolor}
\usepackage{booktabs}
\usepackage{multirow}
\usepackage{siunitx}
\usepackage{graphicx} 

\def\tsc#1{\csdef{#1}{\textsc{\lowercase{#1}}\xspace}}
\tsc{WGM}
\tsc{QE}

\begin{document}
\let\WriteBookmarks\relax
\def\floatpagepagefraction{1}
\def\textpagefraction{.001}

\shorttitle{}    

\shortauthors{}  

\title [mode = title]{GRPO-GCC: Enhancing Cooperation in Spatial Public Goods Games via Group Relative Policy Optimization with Global Cooperation Constraint}  



%

\author[1,2]{Zhaoqilin Yang}[style=chinese,orcid=0000-0002-3676-4761]


\ead{zqlyang@gzu.edu.cn}

\tnotetext[1]{https://github.com/geek12138/GRPO-GCC}

\affiliation[1]{organization={State Key Laboratory of Public Big Data, College of Computer Science and Technology},
	addressline={Guizhou University}, 
	city={Guiyang},
	postcode={550025}, 
	state={Guizhou},
	country={China}}

\affiliation[2]{organization={Institute of Cryptography and Data Security},
	addressline={Guizhou University}, 
	city={Guiyang},
	postcode={550025}, 
	state={Guizhou},
	country={China}}

\author[3]{Chanchan Li}[style=chinese,orcid=0000-0002-2395-0928]

\ead{ccli@gzu.edu.cn}

\affiliation[3]{organization={State Key Laboratory of Public Big Data, College of Mathematics and Statistics},
	addressline={Guizhou University}, 
	city={Guiyang},
	postcode={550025}, 
	state={Guizhou},
	country={China}}

\author[4]{Tianqi Liu}[style=chinese,orcid=0009-0000-2491-5190]

\ead{23115073@bjtu.edu.cn}

\affiliation[4]{organization={School of Computer Science and Technology},
	addressline={Beijing Jiaotong University}, 
	city={Beijing},
	postcode={100044}, 
	state={Beijing},
	country={China}
	}

\author[5]{Hongxin Zhao}[style=chinese,orcid=0000-0002-9168-6145]

\ead{hxzhao@amss.ac.cn}

\affiliation[5]{organization={Institute of Applied Mathematics, Academy of Mathematics and Systems Science},
	addressline={Chinese Academy of Sciences}, 
	city={Beijing},
	postcode={100044}, 
	state={Beijing},
	country={China}}

\author[6,2]{Youliang Tian\corref{cor1}}[style=chinese,orcid=0000-0002-5974-1570]
\ead{yltian@gzu.edu.cn}
\cortext[cor1]{Corresponding author}

\affiliation[6]{organization={State Key Laboratory of Public Big Data, College of Big Data and Information Engineering},
	addressline={Guizhou University}, 
	city={Guiyang},
	postcode={550025}, 
	state={Guizhou},
	country={China}}


\begin{abstract}
Inspired by the principle of self-regulating cooperation in collective institutions, we propose the Group Relative Policy Optimization with Global Cooperation Constraint (GRPO-GCC) framework. This work is the first to introduce GRPO into spatial public goods games, establishing a new deep reinforcement learning baseline for structured populations. GRPO-GCC integrates group relative policy optimization with a global cooperation constraint that strengthens incentives at intermediate cooperation levels while weakening them at extremes. This mechanism aligns local decision making with sustainable collective outcomes and prevents collapse into either universal defection or unconditional cooperation. The framework advances beyond existing approaches by combining group-normalized advantage estimation, a reference-anchored KL penalty, and a global incentive term that dynamically adjusts cooperative payoffs. As a result, it achieves accelerated cooperation onset, stabilized policy adaptation, and long-term sustainability. GRPO-GCC demonstrates how a simple yet global signal can reshape incentives toward resilient cooperation, and provides a new paradigm for multi-agent reinforcement learning in socio-technical systems.
\end{abstract}





\maketitle

\section{Introduction}
\label{sec1}

The Paris Agreement demonstrates the pivotal role of cooperation in addressing global challenges. Nations coordinate emission reductions to limit climate change while balancing individual interests with collective objectives \citep{dawes_1988_anomalies, perc_2016_phase, perc_2017_statistical}. Such large-scale coordination shows that certain outcomes cannot be achieved by any single actor alone. It highlights the necessity of structured collective strategies. This tension between self-interest and collective welfare illustrates how coordinated mechanisms can generate long-term benefits. It also provides a conceptual foundation for studying and designing systems where multiple agents must align their actions to achieve socially optimal outcomes.

Cooperation in complex adaptive systems poses a persistent challenge, as individual incentives often diverge from collective welfare \citep{pennisi_2005_did, kennedy_2005_don}. Evolutionary game theory offers a rigorous framework to examine these conflicts, representing the balance between private gains and group benefits through structured payoff models \citep{nowak_1993_spatial, macy_2002_learning, wang_2015_universal}. Real-world cases such as transboundary fisheries management illustrate this dilemma. States can either cooperate to limit catches or pursue short-term profits independently. This situation mirrors public goods problems that are prone to resource depletion. These analyses show how population structures and strategic interactions influence the emergence and stability of cooperation \citep{nowak_1992_evolutionary, hauert_2005_game, szabo_2007_evolutionary}. They provide a theoretical foundation for designing institutions that promote sustained collective action.

Institutional mechanisms sustain cooperation by translating theoretical principles into enforceable governance structures \citep{chen_2015_first, dos_2015_evolution, HUA_2024_eswa}. Market-based instruments, exemplified by the European Union Emissions Trading System, provide economic incentives for verified emission reductions. Reputation-based mechanisms \citep{tang_2024_cooperative, quan_2020_information,Li_2021_JSM} reinforce compliance by linking organizational performance to investor and stakeholder evaluations. Legal sanctions \citep{chen_2014_probabilistic, chen_2015_competition, liu_2018_synergy} constrain noncompliant behavior, as in the enforcement of international fishing quotas. Social exclusion strategies \citep{liu_2017_competitions, szolnoki_2017_alliance} operate through certification bodies such as the Marine Stewardship Council, which withdraw accreditation for unsustainable practices. Policy interventions \citep{griffin_2017_cyclic, wang_2021_tax, lee_2024_supporting}, including energy efficiency subsidies in Germany, establish structured economic pressures that promote low-carbon technologies. Collectively, these institutional designs operationalize evolutionary principles into systematic tools that reinforce cooperation in complex collective action settings.

In this paper, we propose the Group Relative Policy Optimization with Global Cooperation Constraint (GRPO-GCC) framework. The Group Relative Policy Optimization (GRPO) \citep{Shao_2024_GRPO} was originally introduced in the DeepSeekMath project to enhance the reasoning ability of large language models through reinforcement learning (RL). It extends Proximal Policy Optimization (PPO) by incorporating relative group-based comparisons for advantage estimation, which enables more efficient and stable policy updates. As one of the most advanced deep reinforcement learning (DRL) algorithms, GRPO enables agents to make more intelligent and human-like decisions. It outperforms conventional evolutionary or learning-based approaches by capturing more adaptive and context-aware behaviors. Building on this foundation, GRPO-GCC enhances agent intelligence and human-like behavior in spatial public goods games (SPGG). It is inspired by cooperative dynamics in the Paris Agreement, aligning self-interest with collective welfare. The framework integrates a global cooperation constraint into GRPO, guiding agents toward coordinated group behavior. Strategic knowledge is adjusted via relative policy comparisons across populations. This promotes collective-aligned behaviors while preserving individual adaptability. Analogous mechanisms exist in social insects, where foraging strategies optimize colony efficiency. Systematic evaluations show GRPO-GCC improves cooperation rates and adaptive behavior compared to standard GRPO, Q-learning \citep{watkins_1992_q}, and the Fermi update rule. To our knowledge, this is the first application of GRPO to SPGG. GRPO-GCC establishes a foundation for agents reflecting realistic human cooperative behavior.

Our research makes three main contributions:
\begin{itemize}
	\item We propose the GRPO-GCC framework for SPGG. The framework embeds a global cooperation signal into the DRL process, enabling agents to dynamically balance local incentives with collective welfare. By combining structured policy optimization with adaptive coordination, GRPO-GCC improves cooperative stability and exhibits human-like adaptability in complex social dilemmas.
	
	\item We are the first to introduce GRPO into SPGG. GRPO employs group-wise normalization and KL-constrained updates to support collective reasoning while maintaining individual adaptability. Compared with traditional evolutionary and learning paradigms, it enables agents to make more intelligent and context-aware cooperative decisions. These decisions are also more human-like, offering a modeling approach that is grounded in cognitive principles.
	
	\item We design a Global Cooperation Constraint (GCC) tailored for GRPO. The mechanism adaptively regulates cooperative incentives based on the overall cooperation level, forming a self-regulating feedback loop that sustains system balance. This integration stabilizes cooperation dynamics, enhances interpretability, and extends GRPO’s adaptability to broader coordination tasks, establishing a unified framework for sustainable collective behavior.
\end{itemize}

\section{Related Work}
\label{sec:related}

Conventional evolutionary game theory analyses employing Fermi update rules or replicator dynamics effectively characterize localized competitive interactions and near-term payoff consequences \citep{szabo_1998_evolutionary, schuster_1983_replicator}. Recent research has further examined multi-level public goods games. The findings show that global cooperation often falters when local decisions dominate. In these studies, the Fermi update rule serves as the modeling framework \citep{ZHAO_2026_eswa}. Other evolutionary approaches have explored alternative sanctioning mechanisms. One example is the pool exclusion strategy, where prosocial and antisocial exclusion can significantly influence cooperation dynamics through replicator equations \citep{Liu_2019_ND}. Another example is sampling punishment, which selectively penalizes defectors when their fraction in a sampled subgroup exceeds a threshold. This mechanism improves cooperation cost-effectively in both public goods and collective-risk dilemmas \citep{Xiao_2023_PLA}. Nevertheless, such paradigms largely omit the intricate cognitive mechanisms underpinning adaptive strategy development.

RL reconceptualizes strategic choice as an ongoing process optimized through sequential state evaluations, action selections, and reward feedback \citep{sutton_1998_reinforcement, izquierdo_2007_transient, lipowski_2009_statistical}. As a foundational RL method, Q-learning maintains cooperative equilibria even when strong individual temptations threaten group stability across diverse environments \citep{watkins_1992_q, han_2021_evolutionary, shi_2022_analysis}. Similarly, hypergraph-based Q-learning models have been developed to capture higher-order interaction structures in multi-agent systems, offering richer dynamics for cooperation in public goods settings \citep{Shi_2024_TNSE}. Recent refinements in Q-learning increasingly focus on enhancing algorithmic adaptability within dynamic environments \citep{yan_2024_periodic} and fortifying collaborative incentives among agents \citep{shen_2024_learning}. However, Q-learning still struggles in high-dimensional state-action spaces due to limited representation power.

Early DRL methods, such as deep Q-learning \citep{Mnih_2015_nature}, demonstrated that neural networks could approximate value functions and reproduce human-like cooperative tendencies in large-scale public goods games \citep{Tamura_2024_RSOC}. Building on this foundation, the A3C framework \citep{Mnih_2016_ICML} introduced asynchronous actor–critic optimization, improving stability and scalability in multi-agent cooperation tasks. Multi-agent extensions of A3C have further explored inequity aversion in intertemporal social dilemmas \citep{hughes_2018_NeurIPS}, highlighting the role of social preference modeling in sustaining cooperation. Subsequent approaches, particularly PPO \citep{John_2017_arxiv}, further enhanced training stability through clipped policy objectives. PPO and its variants have been successfully applied to multi-agent systems, including SPGG \citep{YANG_2025_116762, YANG_2025_116928}, where curriculum learning and team-level objectives foster cooperation. Other DRL studies have examined pursuit strategies based on mean DDPG \citep{Wang_2025_ESWA} and complex network models for knowledge dissemination and cooperative dynamics \citep{Chen_2024_ESWA}.

These prior efforts provide the foundation for our work. GRPO represents one of the most advanced DRL algorithms, originally introduced to enhance reasoning capabilities in large language models \citep{Shao_2024_GRPO}. It is capable of generating more intelligent and human-like decisions than conventional evolutionary or learning-based methods. However, GRPO has not yet been applied to SPGG, and the sustainability of cooperation in such settings remains insufficiently examined through advanced RL frameworks. Our GRPO-GCC framework fills this gap by extending GRPO with a global cooperation constraint. It systematically balances local incentives with collective welfare, achieving more realistic and enduring cooperative behavior.

\section{Model}
\label{sec:model}

\subsection{SPGG}

We consider a population of agents arranged on an $L \times L$ toroidal lattice, where $L$ denotes the side length of the lattice, resulting in $L^2$ agents in total. Each agent is identified by an index $i$ and adopts a binary strategy $s_i \in \{0\ (\text{defect}),\ 1\ (\text{cooperate})\}$, where $s_i = 1$ indicates a cooperator and $s_i = 0$ denotes a defector.  

Each agent participates in five overlapping groups, one centered on itself and four centered on its von Neumann neighbors. Let $g$ denote one such group and $N_C^g$ the number of cooperators within that group. The payoff of agent $i$ within a single group $g$ is defined as
\begin{equation}
	\Pi_i^g(s_i, N_C^g) = 
	\begin{cases} 
		\frac{r}{5} N_C^g - 1, & s_i = 1, \\[6pt]
		\frac{r}{5} N_C^g, & s_i = 0,
	\end{cases}
\end{equation}
where $\Pi_i^g(s_i, N_C^g)$ represents the payoff earned by agent $i$ in group $g$, and $r>1$ is the enhancement factor that amplifies the total contributions of cooperators. The subtraction of $1$ for $s_i=1$ reflects the unit cost paid by each cooperator to the group it belongs to, while defectors pay no cost.  

Since every agent participates in five overlapping groups, the total payoff of agent $i$ across all groups is expressed as
\begin{equation}
	\Pi_i(\mathbf{S}) = 
	\begin{cases} 
		\frac{r}{5} \sum_{g \in \mathcal{G}_i} N_C^g - 5, & s_i = 1, \\[6pt]
		\frac{r}{5} \sum_{g \in \mathcal{G}_i} N_C^g, & s_i = 0,
	\end{cases}
\end{equation}
where $\Pi_i(\mathbf{S})$ denotes the total payoff received by agent $i$ under the global strategy configuration $\mathbf{S} = (s_1, s_2, \dots, s_{L^2}) \in \{0,1\}^{L \times L}$, and $\mathcal{G}_i$ represents the set of five groups that include agent $i$. The summation term $\sum_{g \in \mathcal{G}_i} N_C^g$ therefore accumulates the number of cooperators across all local neighborhoods that $i$ participates in.  

A cooperator thus contributes a total cost of $5$ units across the five groups it belongs to, while a defector contributes nothing. The resulting payoff $\Pi_i(\mathbf{S})$ captures the spatial structure of the interaction network. It also reflects the coupling between local and global cooperation levels. This payoff serves as the fundamental environment on which policy optimization and learning mechanisms are applied in subsequent sections.

This payoff landscape defines a dynamic and spatially coupled decision environment where agents balance individual and collective incentives through repeated interactions. Such a setting naturally aligns with DRL, allowing agents to refine strategies using both local and global information. These iterative adjustments foster stable cooperation patterns over time. In the following subsection, we introduce the GRPO framework as the core decision mechanism governing adaptive strategy evolution in the SPGG context.

\subsection{GRPO}

Building upon the SPGG described above, we introduce GRPO as the reinforcement learning framework for policy updates. GRPO is adopted because it enables agents to exhibit intelligent, adaptive, and human-like decision-making patterns through deep reinforcement learning, thereby allowing them to emulate structured social interactions more faithfully. To the best of our knowledge, this study constitutes the first application of GRPO to the SPGG, establishing a conceptual bridge between advanced DRL mechanisms and evolutionary cooperation modeling.

GRPO extends the standard PPO framework through two key modifications designed to enhance both stability and fairness in multi-agent learning. First, it performs group-wise normalization of advantages, ensuring that the comparison among candidate actions sampled within the same evaluation group is fair and scale-invariant. Second, it incorporates a frozen reference policy combined with a Kullback–Leibler (KL) divergence penalty, which constrains excessive policy updates and mitigates instability in structured environments.

Formally, for each agent $i$ in the global state $\mathbf{S} = (s_1, s_2, \dots, s_{L^2})$, a set of $G$ candidate actions $\{a^g\}_{g=1}^G$ is sampled from the previous policy $\pi{\theta_{\mathrm{old}}}$. Each candidate action $a^g$ yields a cumulative reward $R^g$, representing the total payoff obtained by the agent through its interactions with neighboring groups in the spatial public goods game. The normalized advantage $\hat{A}^g $ is computed as
\begin{equation}
	\hat{A}^g = \frac{R^g - \mu}{\sigma},
\end{equation}
where $\mu = \frac{1}{G}\sum_{g=1}^{G}R^g$ and $\sigma$ denote the mean and standard deviation of the sampled rewards.
This normalization ensures that variations in reward magnitude do not bias gradient estimation, thereby maintaining both numerical stability and reliable policy improvement.

The clipped surrogate objective $\mathcal{L}_{\text{clip}}(\theta)$ is defined as
\begin{equation}
	\mathcal{L}_{\text{clip}}(\theta) =
	\mathbb{E}\left[
	\min \big( r^g(\theta)\hat{A}^g,;
	\operatorname{clip}(r^g(\theta), 1-\epsilon, 1+\epsilon)\hat{A}^g \big)
	\right],
\end{equation}
where $r^g(\theta) = \frac{\pi\theta(a^g \mid \mathbf{S})}{\pi_{\theta_{\mathrm{old}}}(a^g \mid \mathbf{S})}$ denotes the policy ratio between the new policy $\pi_\theta$ and the old policy $\pi_{\theta_{\mathrm{old}}}$, and $\epsilon$ controls the clipping threshold.

To further stabilize training, a KL divergence penalty $\mathcal{L}_{\text{KL}}(\theta)$ is introduced to regularize deviation from a frozen reference policy $\pi_{\theta_{\mathrm{ref}}}$:
\begin{equation}
	\mathcal{L}_{\text{KL}}(\theta) =
	- \beta D_{\mathrm{KL}}\left(\pi_\theta \parallel \pi_{\theta_{\mathrm{ref}}}\right),
\end{equation}
where $\beta$ determines the regularization strength and $D_{\mathrm{KL}}$ represents the Kullback–Leibler divergence.
The negative sign indicates that greater divergence decreases the overall objective, discouraging excessive policy shifts.

The final GRPO objective $\mathcal{L}_{\text{GRPO}}(\theta)$ is expressed as
\begin{equation}
	\mathcal{L}^{\text{GRPO}}(\theta) =
	\mathcal{L}_{\text{clip}}(\theta) + \mathcal{L}_{\text{KL}}(\theta),
\end{equation}
which jointly optimizes the clipped PPO objective and the regularization term.

This formulation allows agents to refine their strategies through distributed evaluation, maintaining local adaptability while enhancing group-level coordination.
Compared with conventional evolutionary and reinforcement-based approaches, GRPO provides a more flexible and cognitively interpretable framework for modeling cooperative decision-making within structured populations.

\subsection{GRPO-GCC}
Although GRPO enhances stability and cognitive capability in multi-agent environments, it does not explicitly promote or sustain cooperative tendencies among agents. In collective-action systems, the emergence of cooperation requires mechanisms that encourage agents to act beyond immediate self-interest. Without such regulation, agents may converge to individually rational but collectively inefficient strategies. 
To address this limitation, we integrate the GCC into GRPO, enabling adaptive modulation of cooperative payoffs according to the global state of the population.  

Let
\begin{equation}
	g = \frac{1}{L^2}\sum_{j=1}^{L^2}s_j
\end{equation}
denote the global cooperation rate, where $s_j \in \{0,1\}$ represents the strategy of agent $j$. 
The adjusted payoff for agent $i$ is defined as
\begin{equation}
	R_i(\mathbf{S}) =
	\begin{cases}
		\Pi_i(\mathbf{S}) \big(1 + \rho g(1-g)\big), & s_i = 1, \\[6pt]
		\Pi_i(\mathbf{S}), & s_i = 0,
	\end{cases}
\end{equation}
where $\Pi_i(\mathbf{S})$ denotes the total payoff obtained from overlapping groups, and $\rho \ge 0$ controls the strength of the cooperation constraint. The self-limiting factor $g(1-g)$ ensures that the cooperative incentive reaches its peak at intermediate levels of $g$. This mechanism adaptively promotes cooperation while preventing extreme behaviors such as unconditional cooperation or total defection.

Replacing the original reward in GRPO with the GCC-adjusted payoff yields the final optimization objective:
\begin{equation}
	\mathcal{L}^{\text{GRPO-GCC}}(\theta) =
	\mathcal{L}_{\text{clip}}\!\left(\theta; R_i(\mathbf{S})\right) +
	\mathcal{L}_{\text{KL}}(\theta).
\end{equation}

This formulation empowers GRPO-GCC to balance exploration, stability, and cooperative motivation. It provides a unified framework for studying how globally constrained incentives drive the emergence of intelligent collective behavior in spatially structured populations.

\subsection{Policy Network}
\label{sec:policy_net}

The policy network is designed to model the decision-making process of agents within the SPGG under the GRPO-GCC framework. It takes as input the local spatial configuration surrounding each agent and outputs a stochastic policy representing the probability distribution over possible actions. The architecture enables agents to capture both local spatial interactions and global behavioral trends, supporting context-aware cooperation aligned with population-level coordination.

\begin{figure}[htbp]
	\centering
	\includegraphics[width=0.95\linewidth]{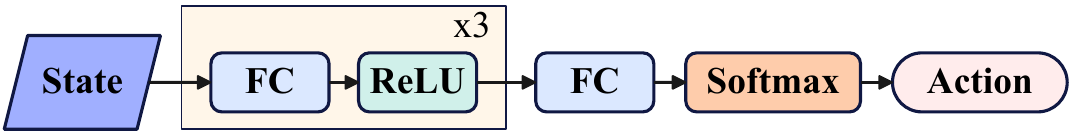}
	\caption{The policy network outputs a stochastic probability distribution over cooperation and defection within the GRPO-GCC framework.}
	\label{fig:policy_net}
\end{figure}

The network consists of four fully connected (FC) layers with ReLU \citep{Glorot_2011_relu} activation functions for nonlinear feature extraction. The first three layers are responsible for hierarchical spatial representation learning, while the fourth FC layer transforms the latent representation into output logits.  
A subsequent softmax layer then converts these logits into a normalized probability distribution over cooperative and defective strategies, as illustrated in Fig.~\ref{fig:policy_net}.

Formally, given the input state vector $s_t^i$ of agent $i$ at iteration $t$, the forward computation is represented as
\begin{align}
	h_1^i &= \sigma(\text{FC}_1(s_t^i)), \\
	h_2^i &= \sigma(\text{FC}_2(h_1^i)), \\
	h_3^i &= \sigma(\text{FC}_3(h_2^i)), \\
	z^i &= \text{FC}_4(h_3^i), \\
	\pi_\theta(a_t^i|s_t^i) &= \text{softmax}(z^i),
\end{align}
where $\text{FC}_k(\cdot)$ denotes the $k$-th fully connected layer parameterized by its trainable weights and biases, $\sigma(\cdot)$ represents the ReLU activation function, and $\text{softmax}(\cdot)$ converts the network output logits $z^i$ into a normalized probability distribution over the available actions $\{0,1\}$ corresponding to defection and cooperation.  

The multilayer nonlinear mapping allows the network to learn hierarchical spatial representations and encode subtle contextual differences between cooperative and defective behaviors. By embedding this four-layer structure into the GRPO-GCC framework, agents gain both local situational awareness and globally consistent coordination. This integration yields decision patterns that closely approximate human cooperative reasoning in structured environments.

Algorithm~\ref{alg:grpo_gcc_extreme} presents the GRPO-GCC training process, illustrating the main loop of sampling, reward adjustment, advantage computation, and policy updating.

\begin{algorithm}[H]
	\caption{GRPO-GCC for SPGG}
	\label{alg:grpo_gcc_extreme}
	\begin{algorithmic}[1]
		\State Initialize policy $\pi_\theta$, reference $\pi_\theta^{\text{ref}}$, state $\mathbf{S}_0$
		\For{epoch $t=1$ to $T$}
		\For{each agent $i$}
		\State Sample $G$ candidates $\{a^{g,i}_t\} \sim \pi_{\theta_{\text{old}}}$, compute GCC rewards $R_i^g(\mathbf{S}_t)$
		\State Compute advantages $\hat{A}_i^g = (R_i^g-\mu)/\sigma$, policy ratios $r_i^g(\theta)$
		\State Update $\theta \gets \theta + \nabla_\theta \big(\mathcal{L}_{\text{clip}} + \mathcal{L}_{\text{KL}}\big)$
		\EndFor
		\State Update global state $\mathbf{S}_{t+1}$, reference policy $\pi_\theta^{\text{ref}} \gets \pi_\theta$
		\EndFor
	\end{algorithmic}
\end{algorithm}

\section{Experimental results}
\label{sec:exp}
\subsection{Experimental setup}

Experiments were conducted on a $200\times200$ lattice with von Neumann neighbors. Each agent updated its policy via GRPO-GCC with learning rate $\alpha=1\times10^{-4}$, KL penalty weight $\beta=0.04$, and clip $\epsilon=0.2$. Global cooperation coefficient $\rho=1.0$ controlled cooperation sustainability. GRPO-GCC sampled $\eta=8$ candidates per agent and performed $\zeta=3$ inner updates. Policy optimization used Adam \citep{Diederik_2015_ICLR} with StepLR, halving $\alpha$ every 1,000 iterations. Advantages were normalized across sampled candidates, and reference policies were updated periodically. In all experimental result figures, C refers to cooperators and D refers to defectors. Training ran for 1,000 iterations. Computations were implemented in PyTorch 2.2.1 with CUDA 12.8 on a 32-core CPU and Titan RTX GPU.

\subsection{GRPO Hyperparameter Sensitivity Analysis}
\label{sec:hyperparam}

This section investigates the sensitivity of the model to key hyperparameters $\beta$, $\eta$, and $\zeta$. Experimental results show that when $r<5$, the cooperation rate remains at $0\%$, while when $r>5$, the cooperation rate stabilizes at 100\%. This indicates that hyperparameter choices primarily influence performance near the critical threshold $r=5.0$ without altering the overall cooperation boundary. The experiments are initialized on a $200 \times 200$ lattice, where the upper half of the grid is filled with defectors and the lower half with cooperators. The following provides a detailed analysis of each hyperparameter.

\begin{figure}
	\centering
	\includegraphics[width=0.95\linewidth]{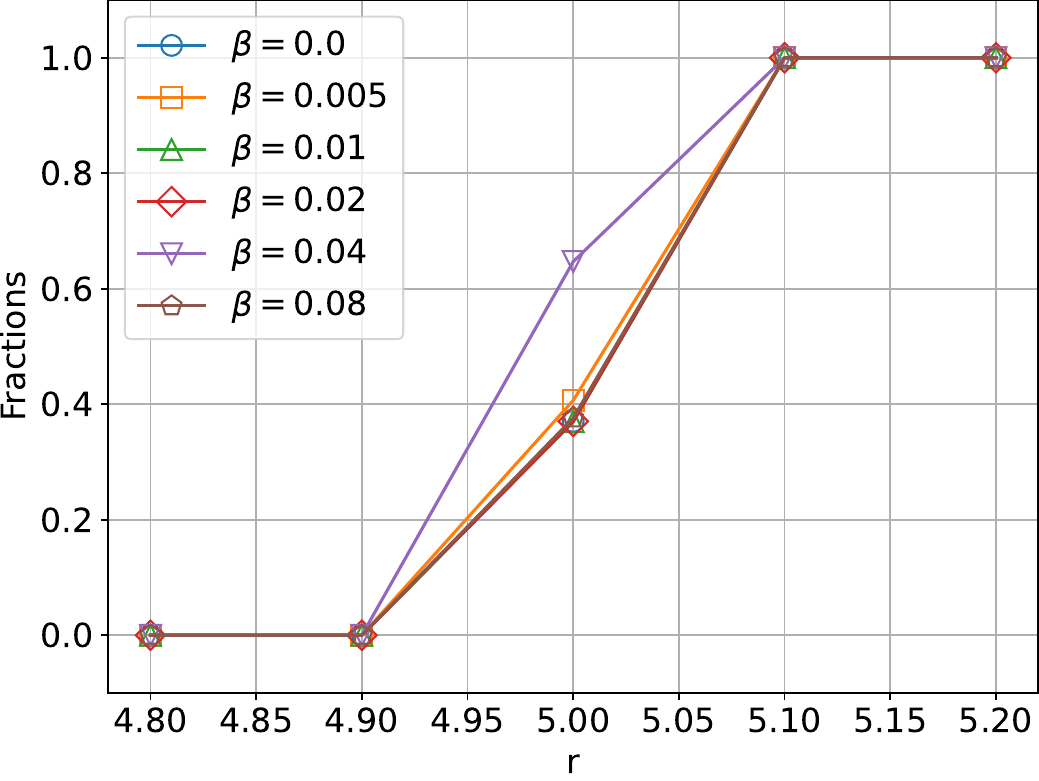}
	\caption{Cooperation rates under different $\beta$ values. $\beta$ controls KL penalty strength, with $\beta=0.04$ yielding the highest cooperation rate at $r=5.0$.}
	\label{fig:beta}
\end{figure}

The parameter $\beta$ controls the strength of the KL penalty term. As shown in Fig.~\ref{fig:beta}, when $\beta$ is too small, the reference policy imposes insufficient constraint, leading to reduced cooperation rates near the threshold. Conversely, an excessively large $\beta$ restricts policy updates, hindering exploration of cooperative strategies. The results show that $\beta=0.04$ achieves the best balance between constraint and exploration, yielding the highest cooperation rate at $r=5.0$.

\begin{figure}
	\centering
	\includegraphics[width=0.95\linewidth]{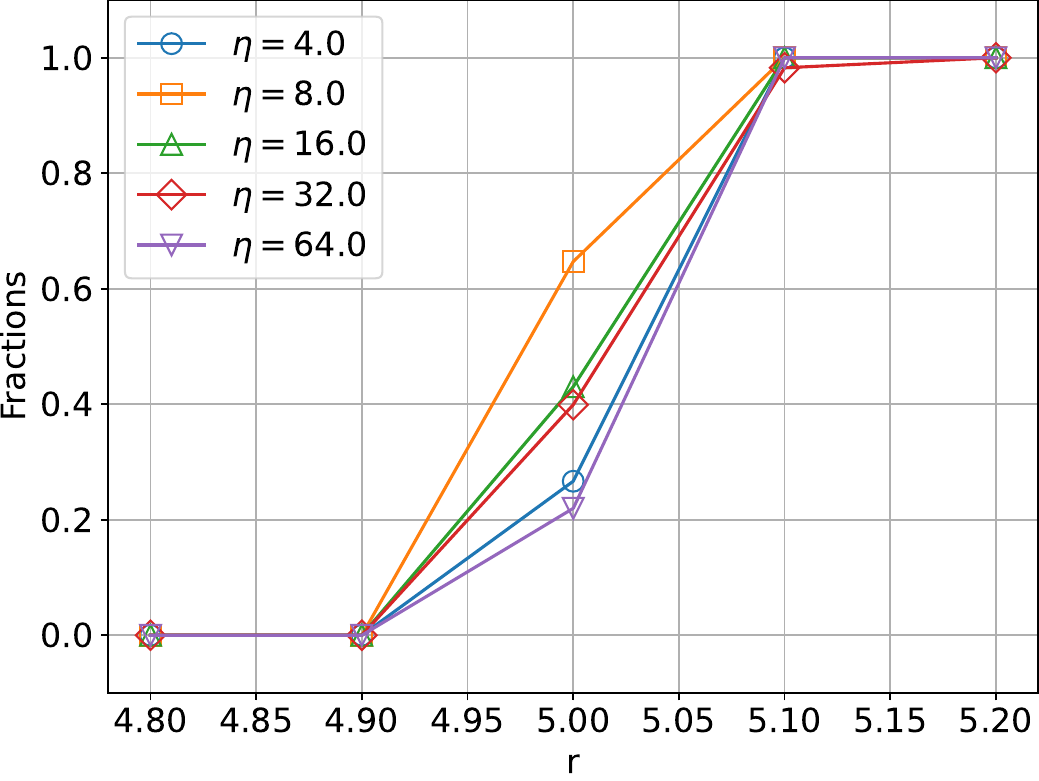}
	\caption{Cooperation rates under different $\eta$ values. $\eta$ controls the number of sampled candidates, with $\eta=8$ achieving the best performance at $r=5.0$.}
	\label{fig:eta}
\end{figure}

The parameter $\eta$ determines the number of sampled candidates per agent. Fig.~\ref{fig:eta} shows the impact of $\eta$ on cooperation. Too small a value leads to insufficient sampling, limiting the available information for policy updates and resulting in weaker cooperation performance. In contrast, very large values increase computational overhead without improving cooperation. The results demonstrate that $\eta=8$ provides adequate diversity of candidates to foster cooperation, achieving the highest cooperation rate at $r=5.0$.

\begin{figure}
	\centering
	\includegraphics[width=0.95\linewidth]{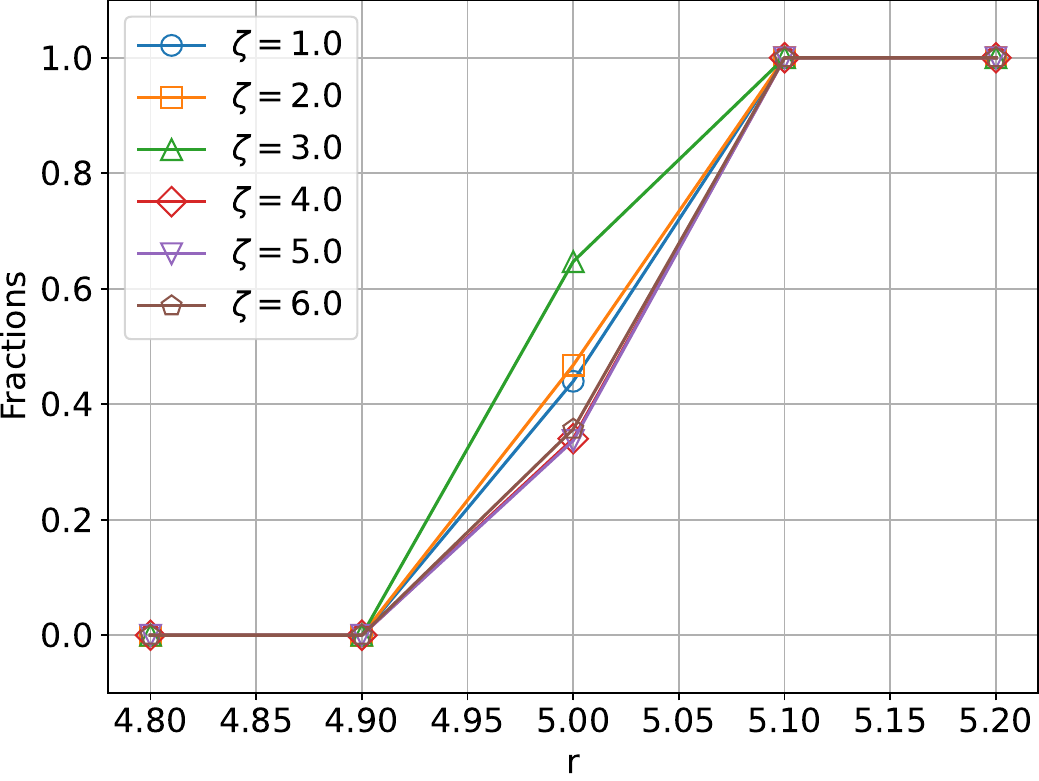}
	\caption{Cooperation rates under different $\zeta$ values. $\zeta$ controls the number of inner updates, with $\zeta=3$ achieving the best performance at $r=5.0$.}
	\label{fig:zeta}
\end{figure}

The parameter $\zeta$ reflects the number of inner updates. As shown in Fig.~\ref{fig:zeta}, with $\zeta$ set too small, policy optimization remains insufficient, resulting in lower cooperation rates. Excessively large values, however, yield diminishing returns while incurring additional computational cost. The experiments indicate that $\zeta=3$ strikes the optimal balance between optimization effectiveness and efficiency, producing the highest cooperation rate at $r=5.0$.

In summary, the optimal hyperparameter configuration ($\beta=0.04$, $\eta=8$, $\zeta=3$) ensures complete defection when $r<5$, full cooperation when $r>5$, and the strongest cooperative performance at the threshold $r=5.0$. All subsequent experiments are conducted under this configuration.

\subsection{GRPO-GCC Hyperparameter Sensitivity Analysis}

\begin{figure}
	\centering
	\includegraphics[width=0.95\linewidth]{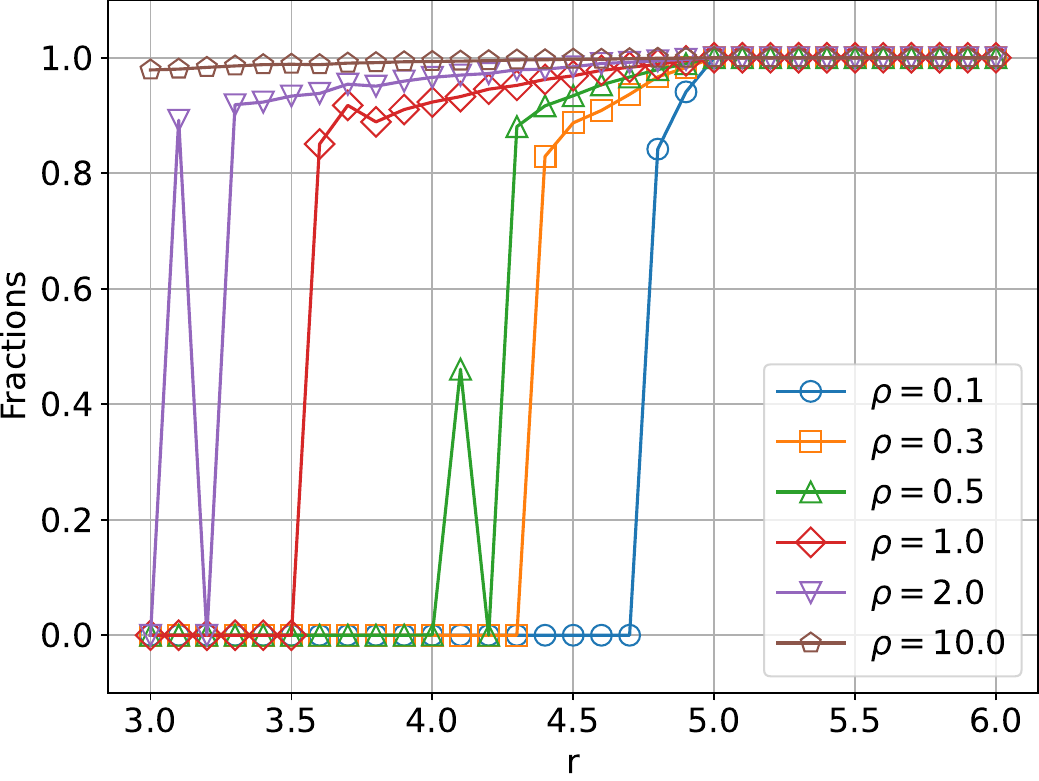}
	\caption{Cooperation rate with varying global cooperation coefficient $\rho$ in GRPO-GCC. Higher $\rho$ values promote cooperation even at smaller $r$.}
	\label{fig:rho}
\end{figure}

We investigate the effect of the global cooperation coefficient $\rho$ in GRPO-GCC. The initial state is set on a $200 \times 200$ lattice, We investigate the effect of the global cooperation coefficient $\rho$ in GRPO-GCC. The initial state is set on a $200 \times 200$ lattice, with defectors in the upper half and cooperators in the lower half. The enhancement factor $r$ ranges from 3.0 to 6.0, and $\rho$ takes values 0.1, 0.3, 0.5, 1.0, 2.0, and 10.0. Results indicate that under the GCC mechanism, GRPO agents choose to cooperate even when $r<5$. Moreover, larger values of $\rho$ encourage agents to adopt cooperative strategies at smaller $r$. For example, when $\rho=1.0$, more than $80\%$ of agents converge to cooperation once $r>3.5$. As $\rho$ increases, cooperation is reinforced, reaching over $95\%$ across nearly all $r$ values at $\rho=10.0$. This demonstrates that the GCC mechanism effectively promotes cooperation under GRPO. For fair comparisons with Q-learning and the Fermi update rule, we set $\rho=1.0$ in subsequent experiments, ensuring stable cooperative behavior without excessive bias. Fig.~\ref{fig:rho} illustrates the cooperation fractions for different $\rho$ values across the range of $r$.

\subsection{Algorithm performance evaluation under varying enhancement factors $r$}

\begin{figure*}[htbp!]
	\begin{minipage}{0.48\linewidth}
		\centering
		\includegraphics[width=0.95\linewidth]{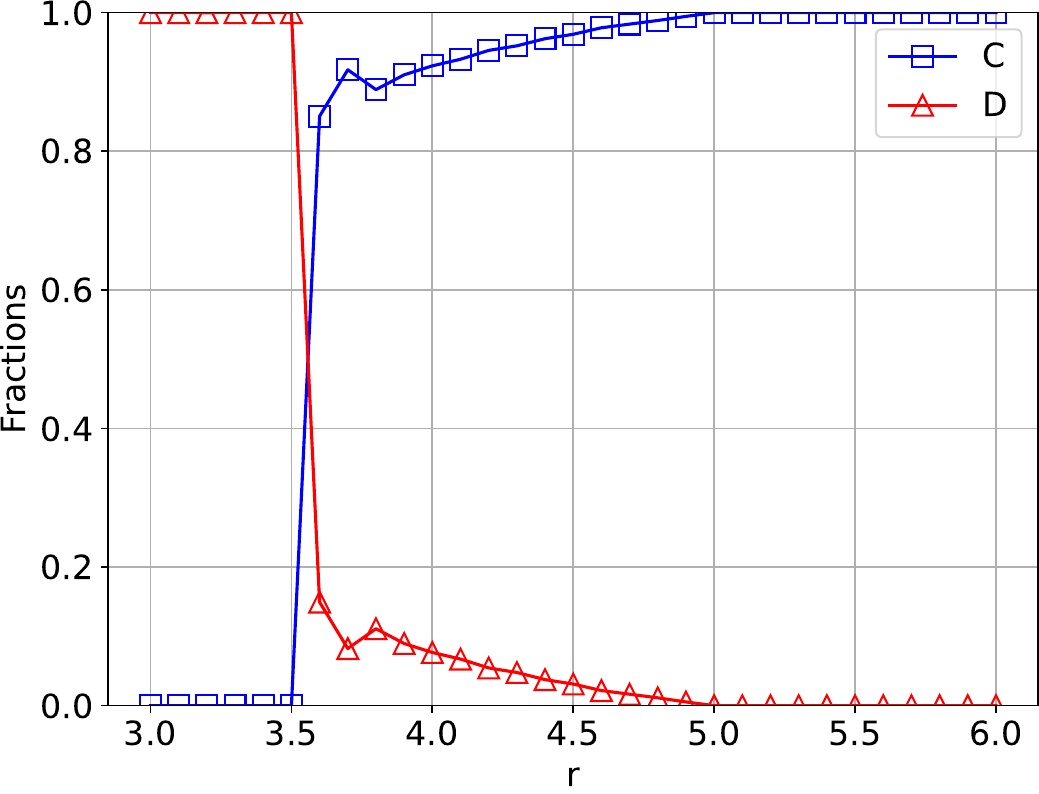}\\
		{\footnotesize (a) GRPO-GCC}
	\end{minipage}
	\hfill
	\begin{minipage}{0.48\linewidth}
		\centering
		\includegraphics[width=0.95\linewidth]{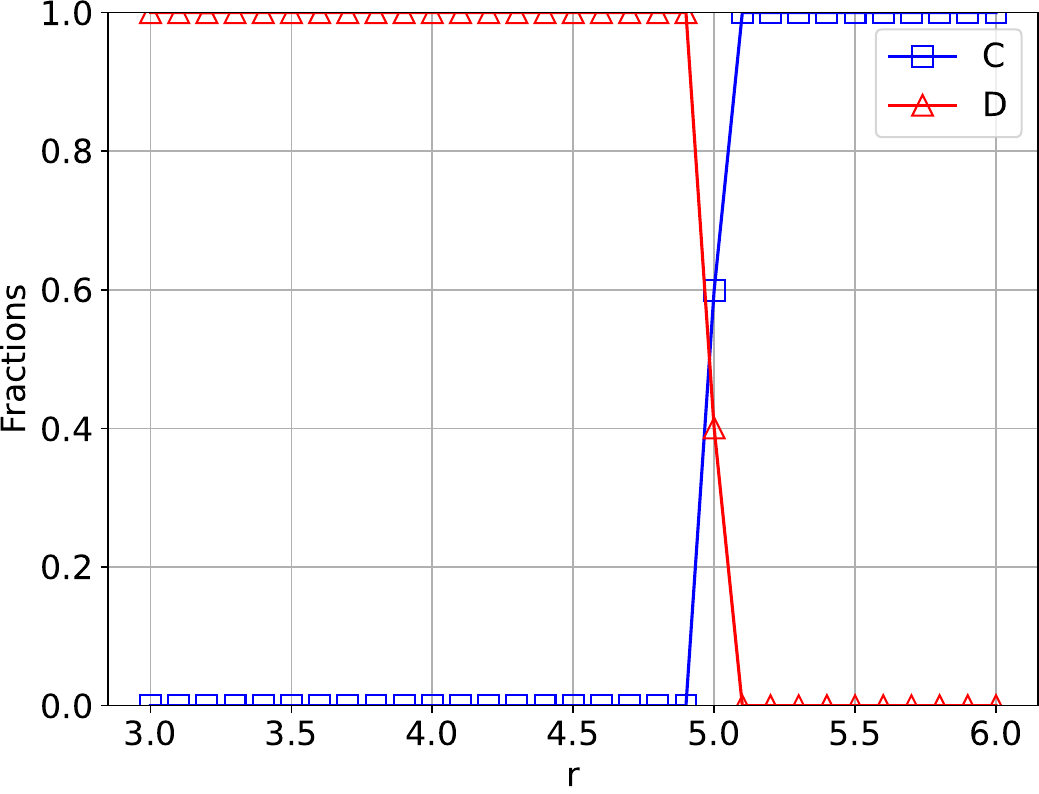}\\
		{\footnotesize (b) GRPO}
	\end{minipage}
	\\
	[3mm]
	\begin{minipage}{0.48\linewidth}
		\centering
		\includegraphics[width=0.95\linewidth]{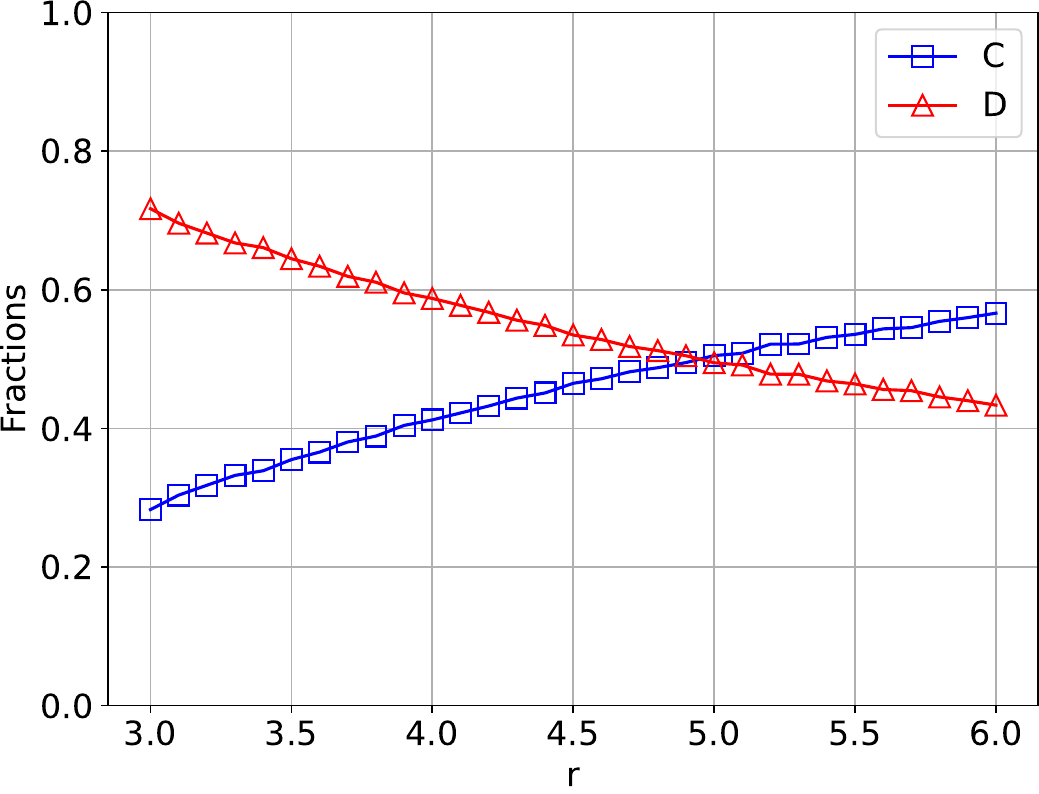}\\
		{\footnotesize (c) Q-learning}
	\end{minipage}	
	\hfill
	\begin{minipage}{0.48\linewidth}
		\centering
		\includegraphics[width=0.95\linewidth]{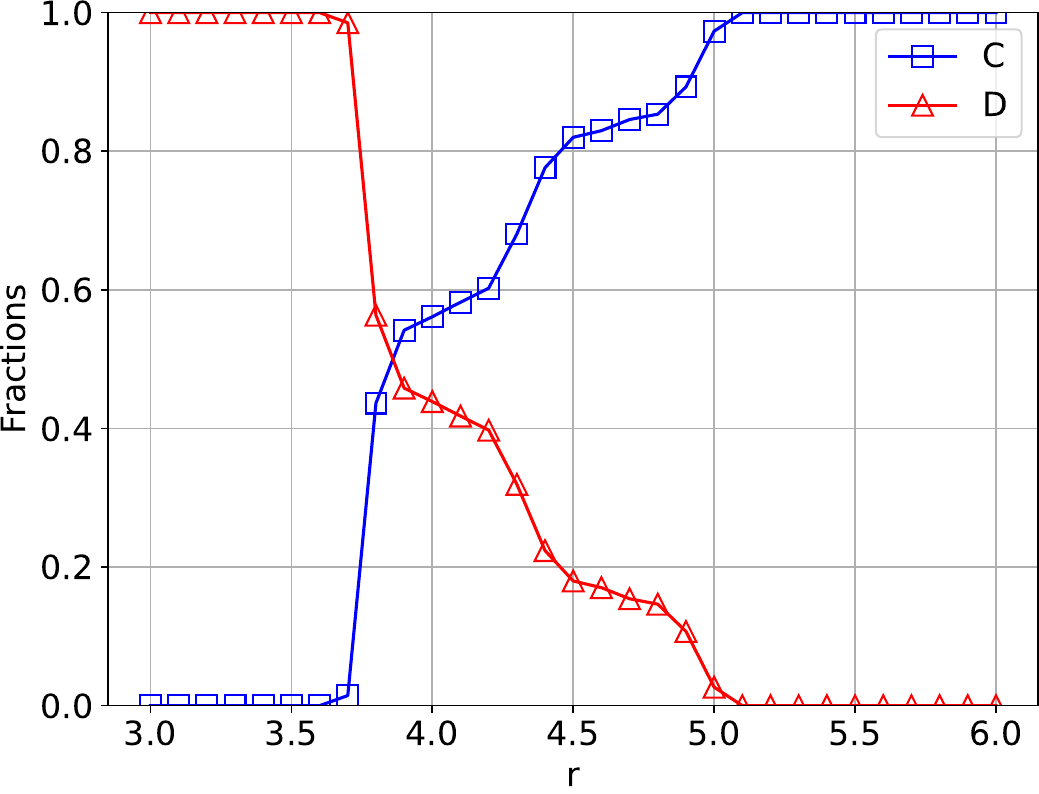}\\
		{\footnotesize (d) Fermi update rule}
	\end{minipage}	
	\caption{Performance comparison under varying enhancement factors $r$ for GRPO-GCC, GRPO, Q-learning, and Fermi update rule. GRPO-GCC achieves stable cooperation above 80\% when $r \geq 3.6$ and reaches 100\% at $r \geq 5.0$, outperforming baseline methods. Cooperators (C) are represented by blue squares and defectors (D) by red triangles.}
	\label{fig:algo_r_comparison}
\end{figure*}

This section evaluates the performance of different algorithms under varying enhancement factors $r$. The compared models include GRPO-GCC, GRPO, Q-learning, and the Fermi update rule. The enhancement factor $r$ ranges from 3.0 to 6.0, and the global cooperation coefficient in GRPO-GCC is fixed at $\rho=1.0$. The horizontal axis of the figure represents $r$, while the vertical axis indicates the fraction of cooperators. Agents are initialized such that defectors occupy the upper half of the lattice and cooperators the lower half. Cooperators (C) are represented by blue squares and defectors (D) by red triangles, as illustrated in Fig.~\ref{fig:algo_r_comparison}.

Experimental results show that GRPO-GCC achieves cooperation as early as $r \geq 3.6$, with the cooperation rate consistently exceeding 80\%. When $r \geq 5.0$, the cooperation rate reaches 100\%. This demonstrates that the global cooperation coefficient effectively encourages cooperative behavior even under relatively weak enhancement. It highlights the robustness of the GRPO-GCC framework in maintaining stable cooperation. In contrast, GRPO without the GCC mechanism fails to sustain cooperation when $r < 5.0$, with the final cooperation rate remaining at 0\%. At the threshold $r=5.0$, cooperation emerges but only reaches about 60\%. For $r > 5.0$, cooperation eventually stabilizes at 100\%. This result indicates that GRPO can eventually achieve full cooperation. However, it requires stronger external incentives compared with GRPO-GCC. This highlights the necessity of the cooperation coefficient in sustaining efficient cooperative behavior. Q-learning exhibits a different pattern, with approximately 30\% cooperation even at $r=3.0$. The cooperation rate increases with larger $r$, but never exceeds 60\% even when $r=6.0$. This suggests that Q-learning enables the early emergence of cooperation through direct reward-based adaptation. However, it struggles to achieve stable large-scale cooperation because policy coordination among agents is limited. Finally, the Fermi update rule shows the onset of cooperation around $r > 3.6$, similar to GRPO-GCC. However, the increase in cooperation is slower, reaching only about 60\% at $r=4.0$, and stabilizing at 100\% only when $r > 5.0$. This behavior reflects the probabilistic nature of the Fermi update rule. Strategy adoption occurs gradually and depends strongly on payoff differences. As a result, the system exhibits delayed but eventual convergence to full cooperation. In summary, GRPO-GCC outperforms all baseline methods by enabling high levels of cooperation even at smaller enhancement factors. This confirms the effectiveness of the global cooperation coefficient in accelerating and stabilizing cooperative dynamics.

\subsection{Comparative analysis of algorithms}
\label{exp:Compare}

\begin{figure*}[htbp!]
	\begin{minipage}{\linewidth}
		\begin{minipage}{0.27\linewidth}
			\centering
			\includegraphics[width=\linewidth]{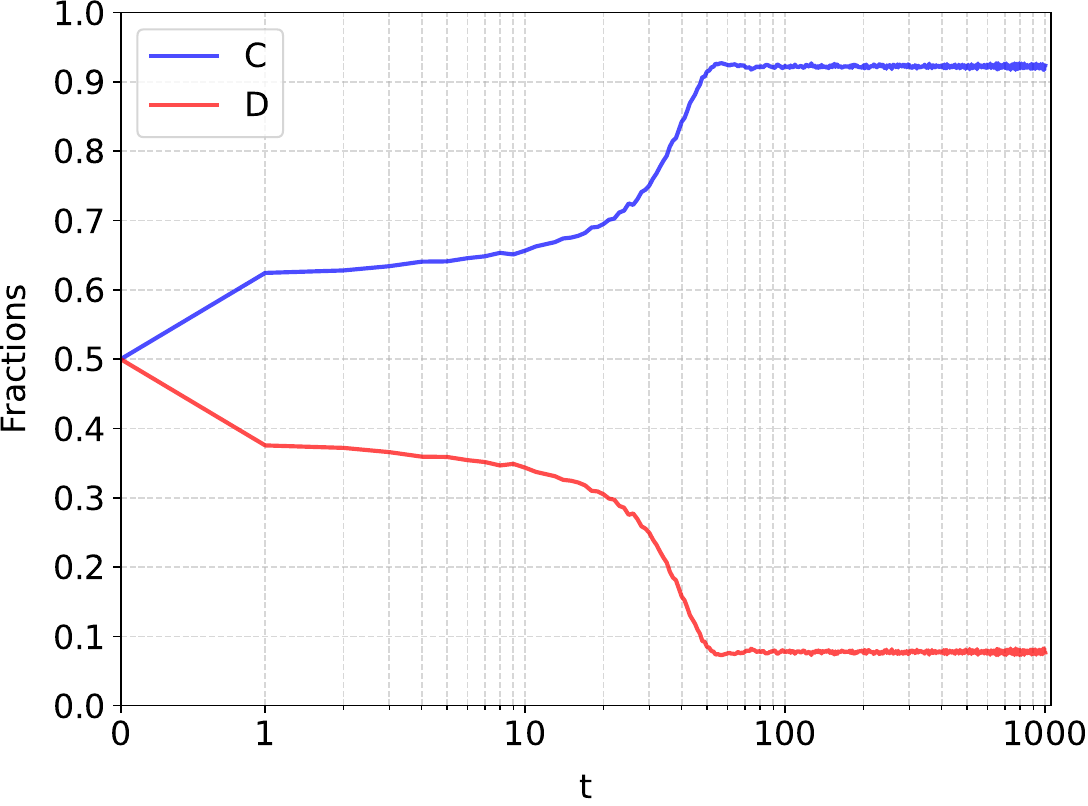}\\
		\end{minipage}
		\begin{minipage}{0.14\linewidth}
			\centering
			\fbox{\includegraphics[width=\linewidth]{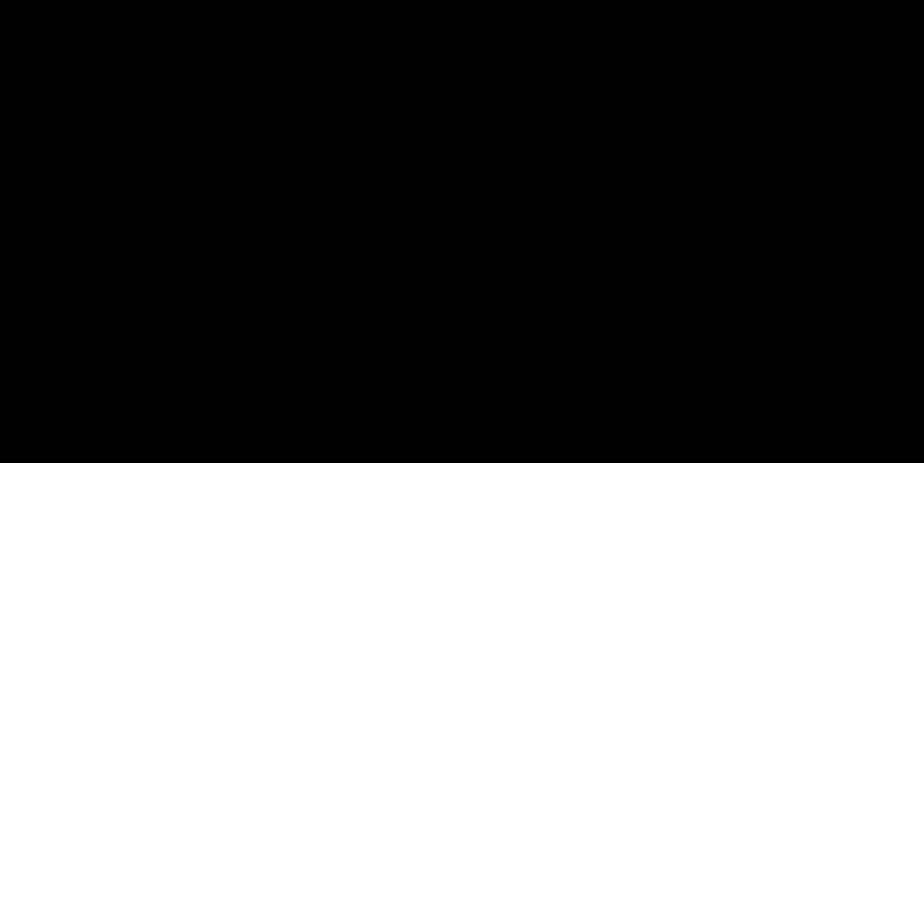}}\\
			{\footnotesize t=0}
		\end{minipage}
		\begin{minipage}{0.14\linewidth}
			\centering
			\fbox{\includegraphics[width=\linewidth]{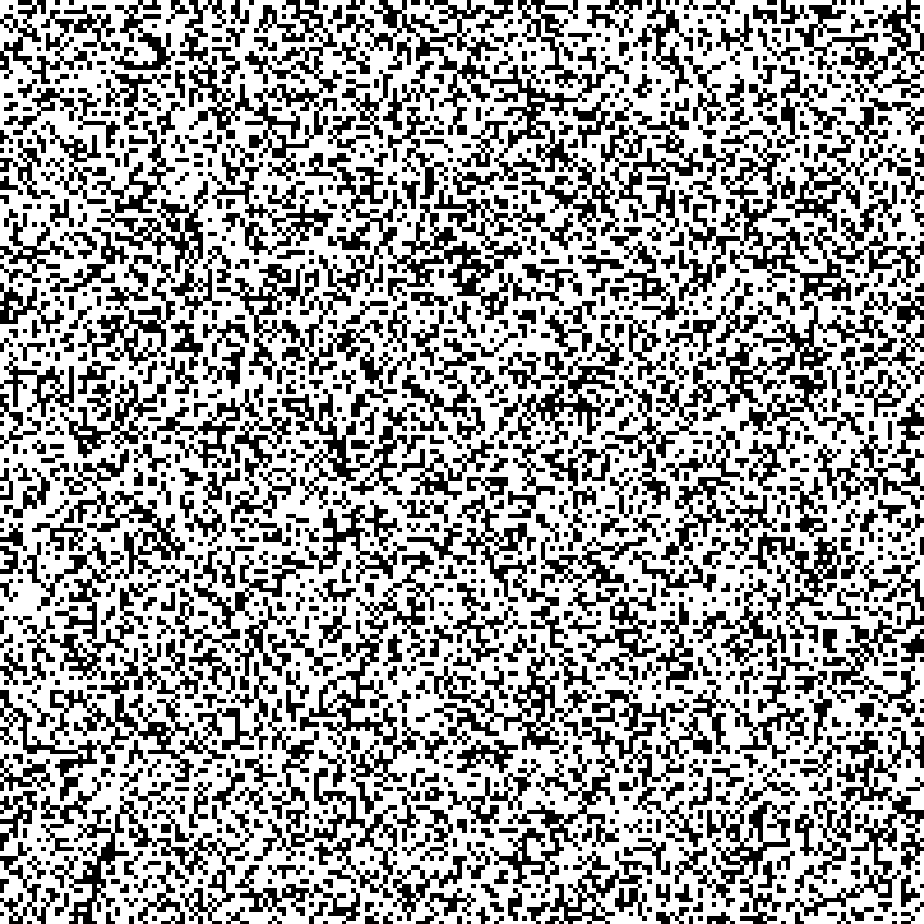}}\\
			{\footnotesize t=1}
		\end{minipage}
		\begin{minipage}{0.14\linewidth}
			\centering
			\fbox{\includegraphics[width=\linewidth]{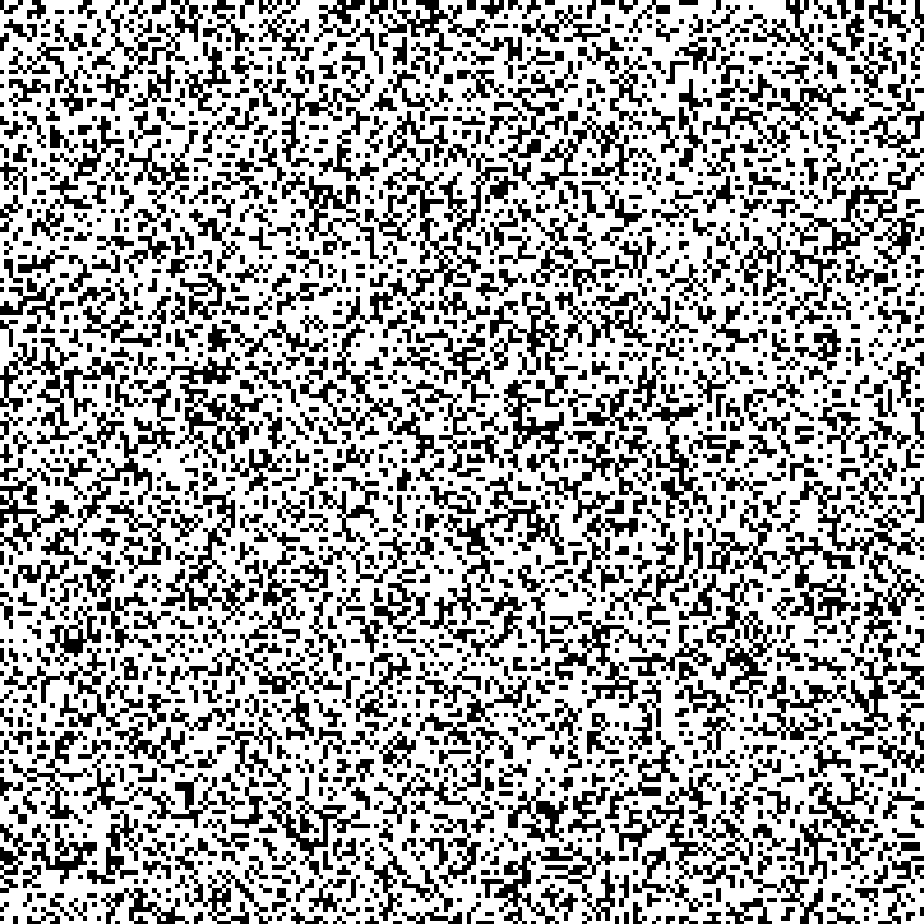}}\\
			{\footnotesize t=10}
		\end{minipage}
		\begin{minipage}{0.14\linewidth}
			\centering
			\fbox{\includegraphics[width=\linewidth]{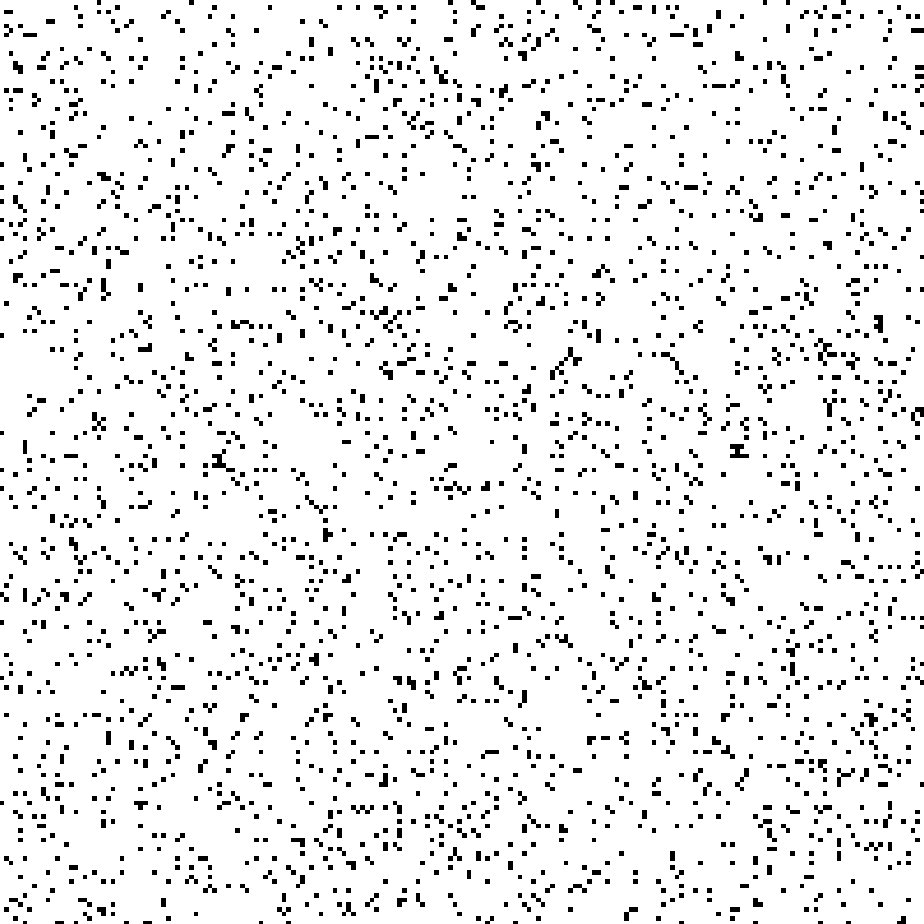}}\\
			{\footnotesize t=100}
		\end{minipage}
		\begin{minipage}{0.14\linewidth}
			\centering
			\fbox{\includegraphics[width=\linewidth]{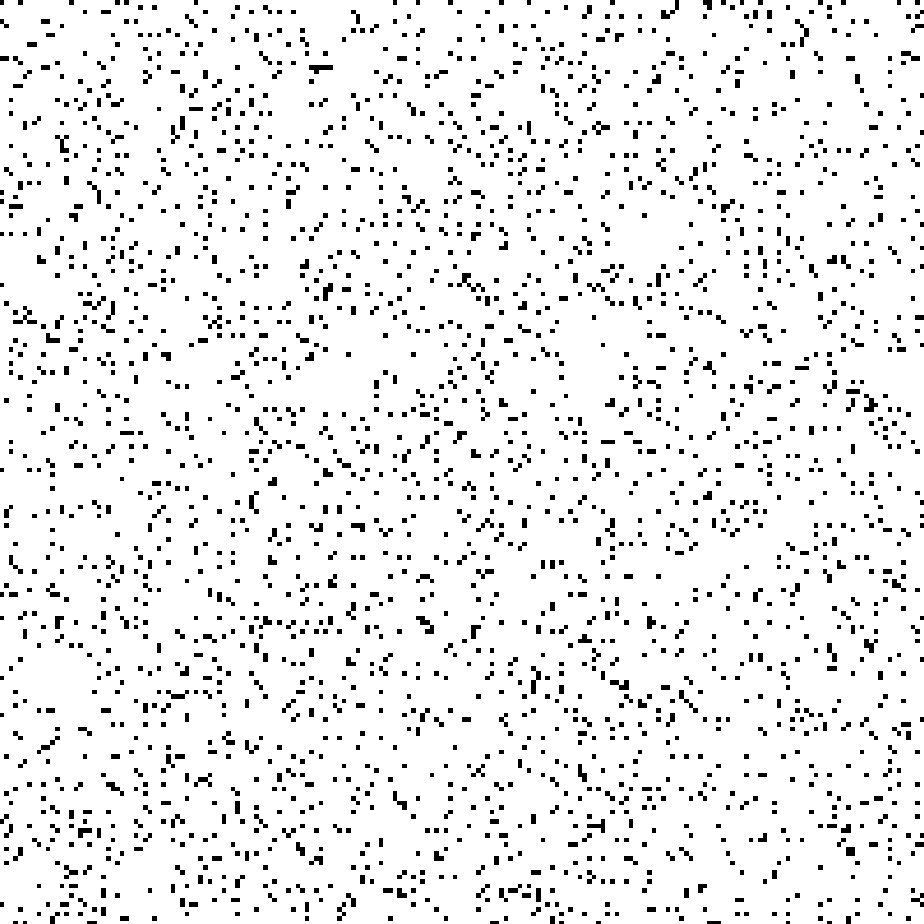}}\\
			{\footnotesize t=1000}
		\end{minipage}
\\
\centering
		{\footnotesize (a) GRPO-GCC}
	\end{minipage}
	\\[2mm]
	\begin{minipage}{\linewidth}
		\begin{minipage}{0.27\linewidth}
			\centering
			\includegraphics[width=\linewidth]{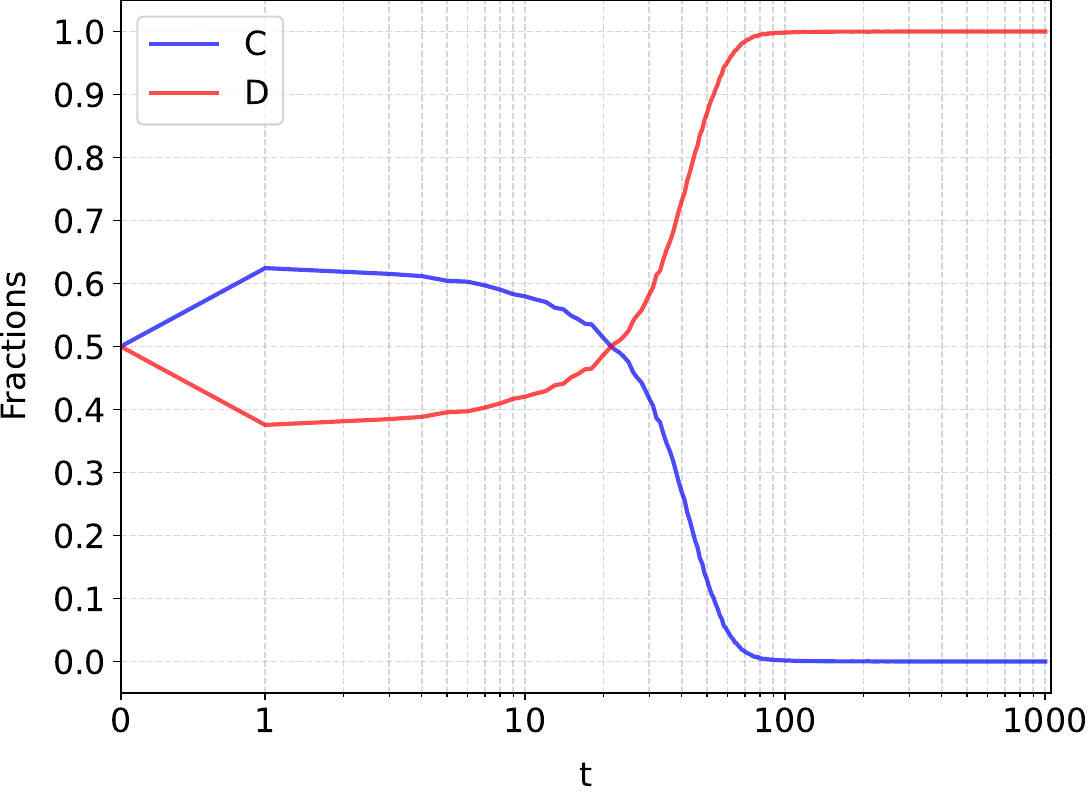}\\
		\end{minipage}
		\begin{minipage}{0.14\linewidth}
			\centering
			\fbox{\includegraphics[width=\linewidth]{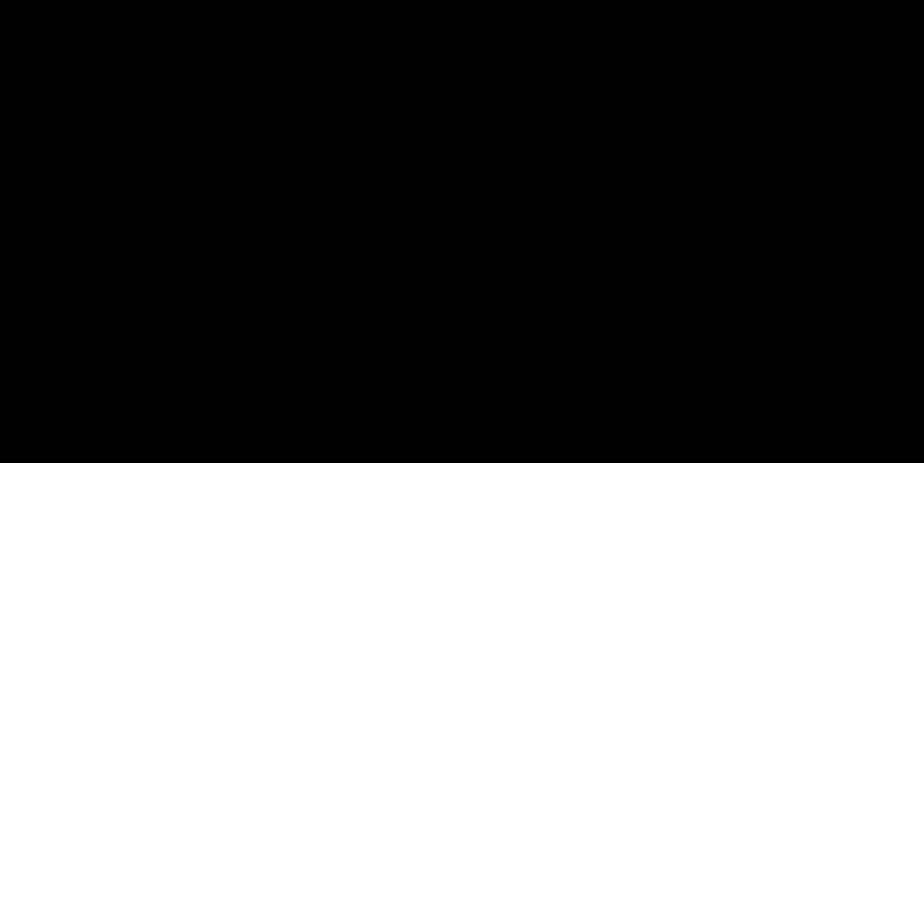}}\\
			{\footnotesize t=0}
		\end{minipage}
		\begin{minipage}{0.14\linewidth}
			\centering
			\fbox{\includegraphics[width=\linewidth]{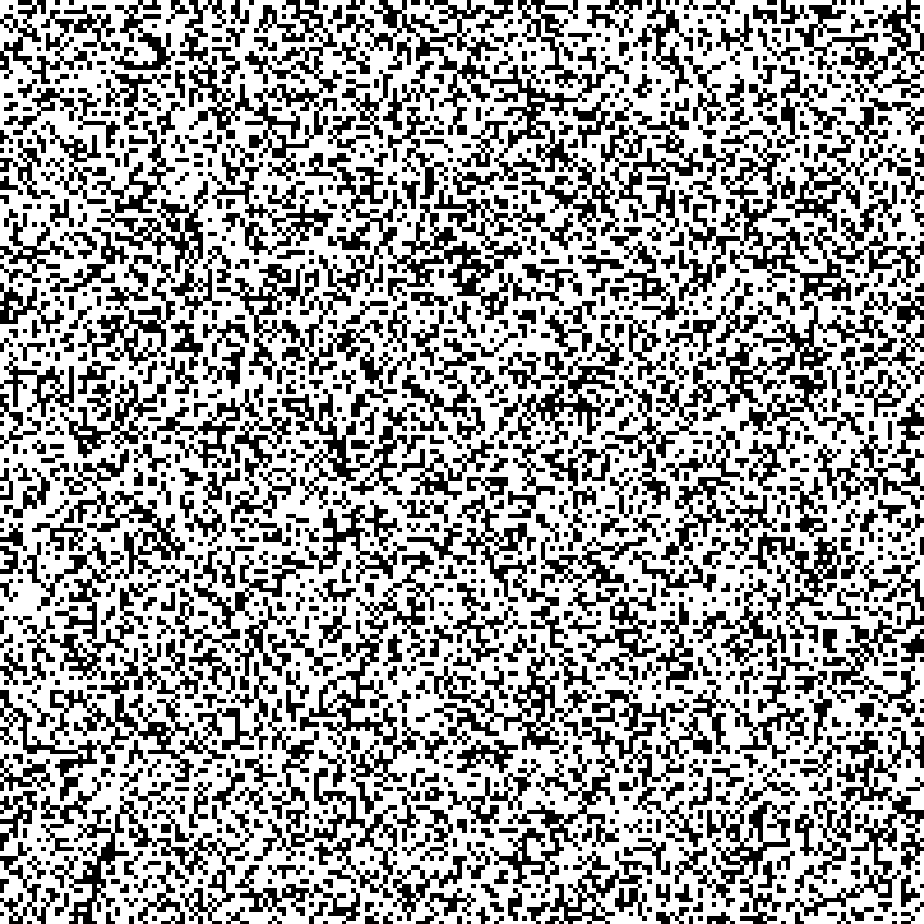}}\\
			{\footnotesize t=1}
		\end{minipage}
		\begin{minipage}{0.14\linewidth}
			\centering
			\fbox{\includegraphics[width=\linewidth]{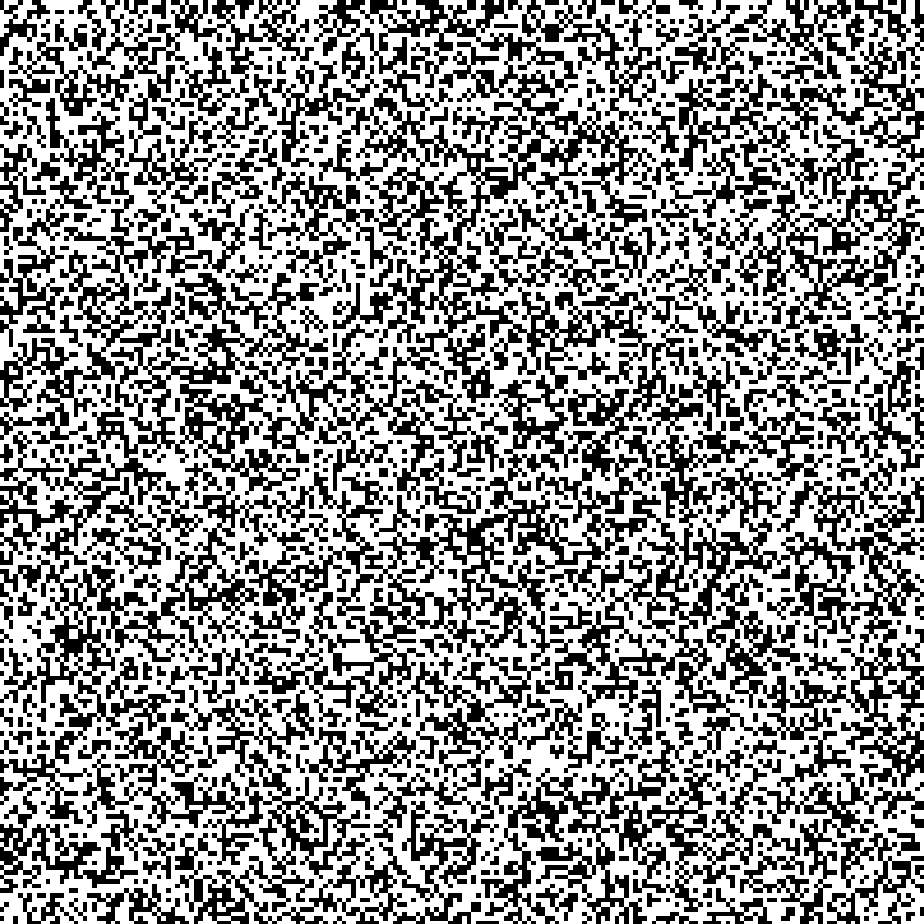}}\\
			{\footnotesize t=10}
		\end{minipage}
		\begin{minipage}{0.14\linewidth}
			\centering
			\fbox{\includegraphics[width=\linewidth]{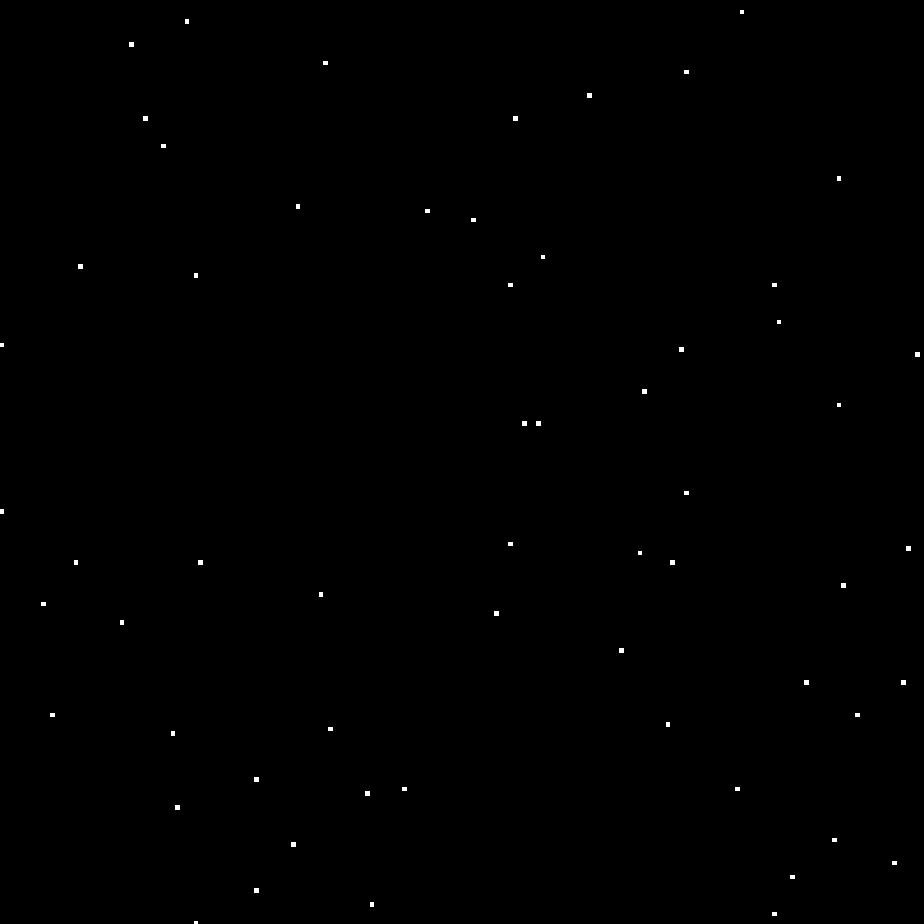}}\\
			{\footnotesize t=100}
		\end{minipage}
		\begin{minipage}{0.14\linewidth}
			\centering
			\fbox{\includegraphics[width=\linewidth]{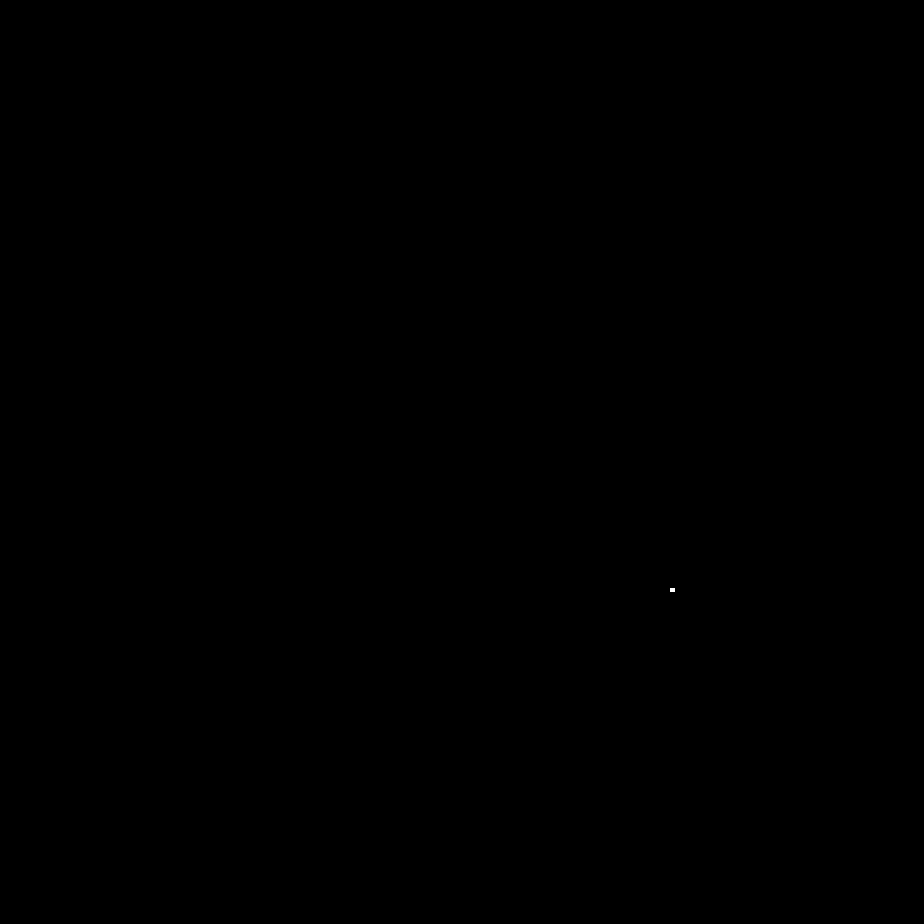}}\\
			{\footnotesize t=1000}
		\end{minipage}
\\
\centering
		{\footnotesize (b) GRPO}
	\end{minipage}
	\\[2mm]
	\begin{minipage}{\linewidth}
		\begin{minipage}{0.27\linewidth}
			\centering
			\includegraphics[width=\linewidth]{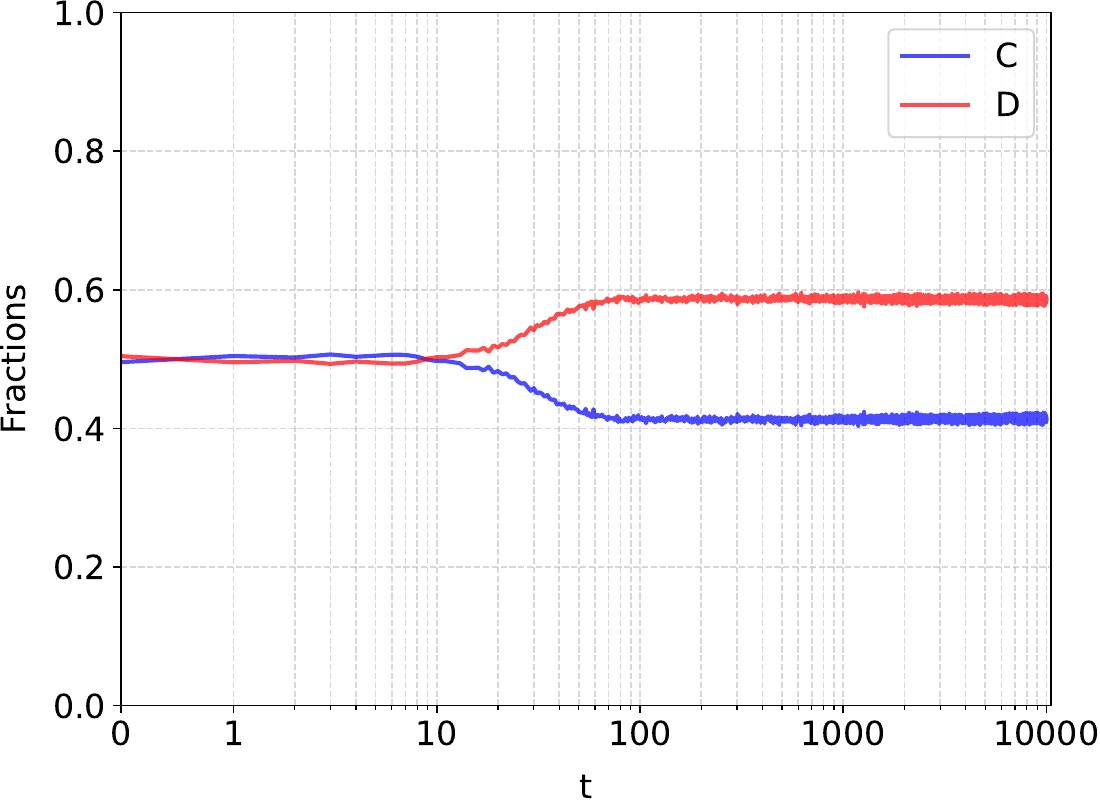}\\
		\end{minipage}
		\begin{minipage}{0.14\linewidth}
			\centering
			\fbox{\includegraphics[width=\linewidth]{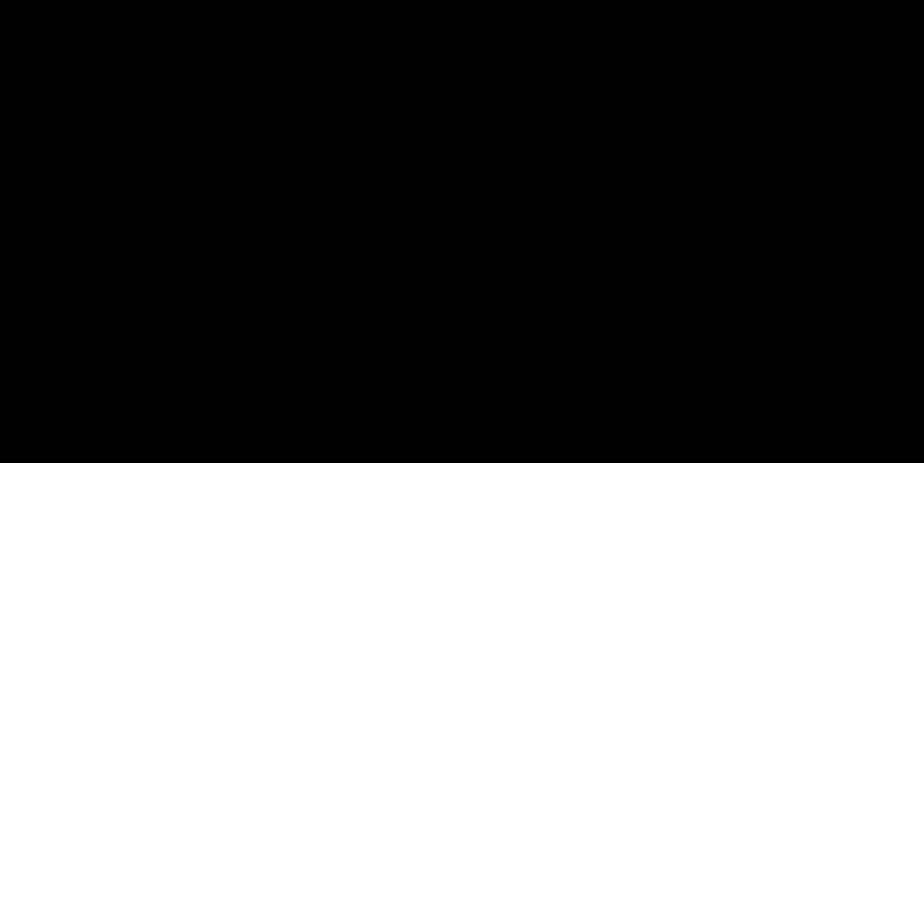}}\\
			{\footnotesize t=0}
		\end{minipage}
		\begin{minipage}{0.14\linewidth}
			\centering
			\fbox{\includegraphics[width=\linewidth]{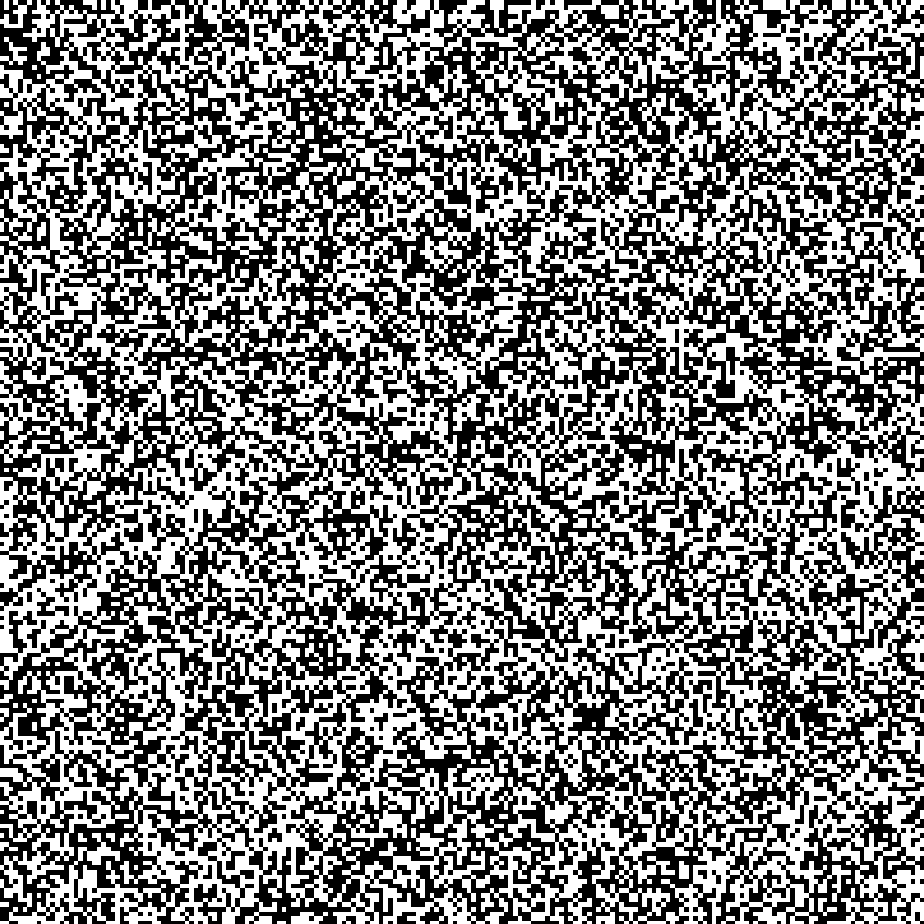}}\\
			{\footnotesize t=10}
		\end{minipage}
		\begin{minipage}{0.14\linewidth}
			\centering
			\fbox{\includegraphics[width=\linewidth]{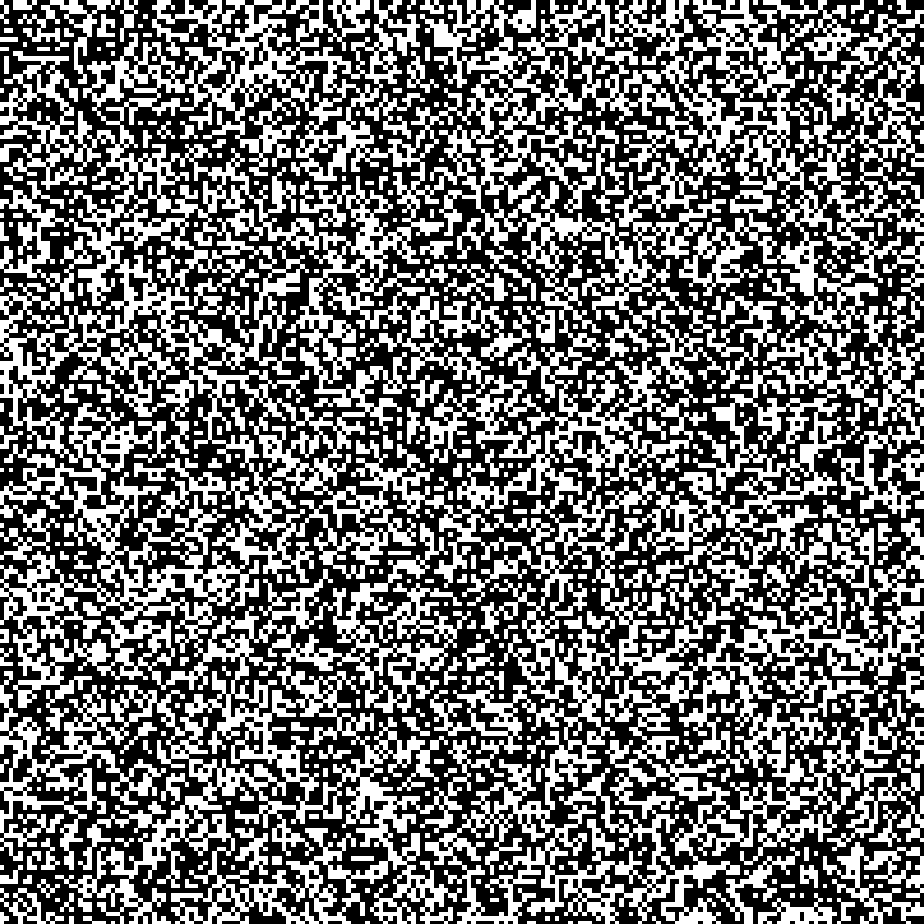}}\\
			{\footnotesize t=100}
		\end{minipage}
		\begin{minipage}{0.14\linewidth}
			\centering
			\fbox{\includegraphics[width=\linewidth]{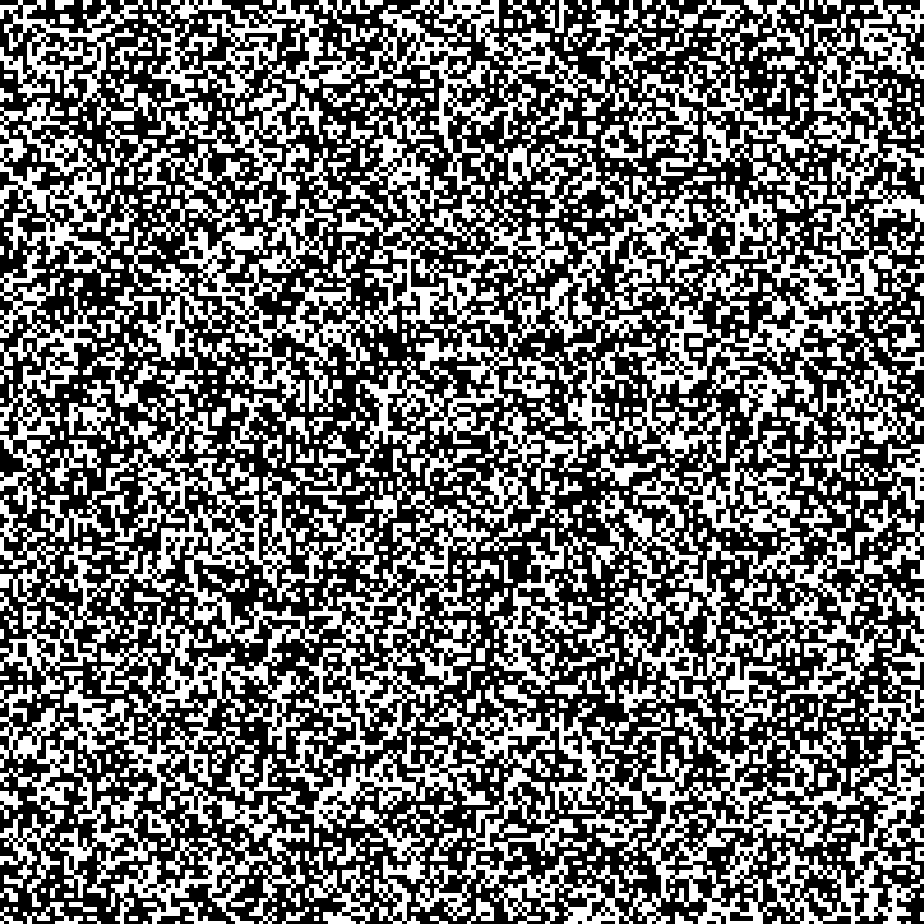}}\\
			{\footnotesize t=1000}
		\end{minipage}
		\begin{minipage}{0.14\linewidth}
			\centering
			\fbox{\includegraphics[width=\linewidth]{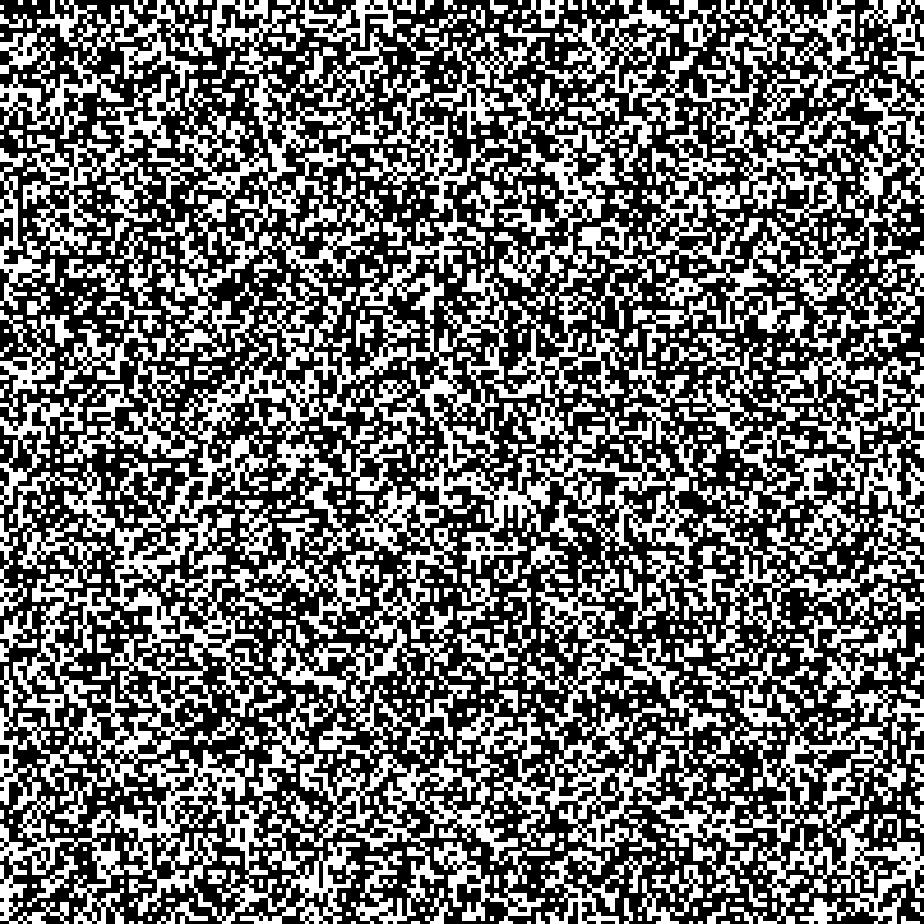}}\\
			{\footnotesize t=10000}
		\end{minipage}
\\
\centering
		{\footnotesize (c) Q-learning}
	\end{minipage}	
	\\[2mm]
	\begin{minipage}{\linewidth}
		\begin{minipage}{0.27\linewidth}
			\centering
			\includegraphics[width=\linewidth]{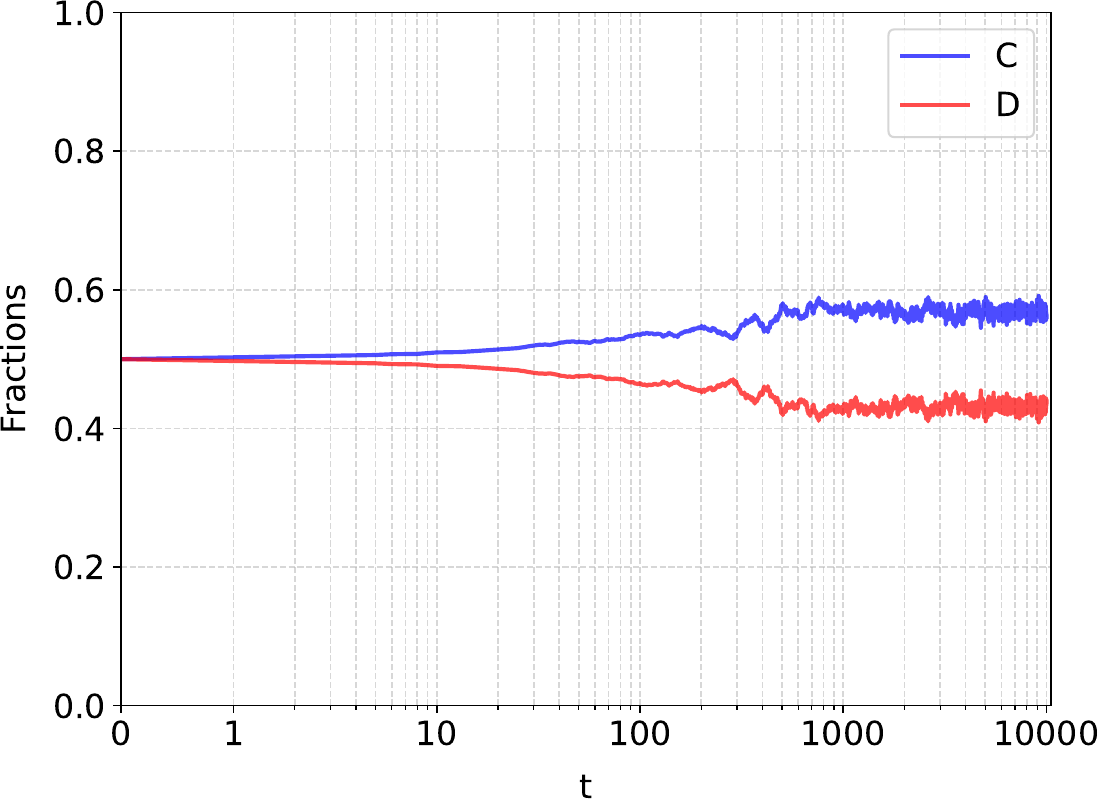}\\
		\end{minipage}
		\begin{minipage}{0.14\linewidth}
			\centering
			\fbox{\includegraphics[width=\linewidth]{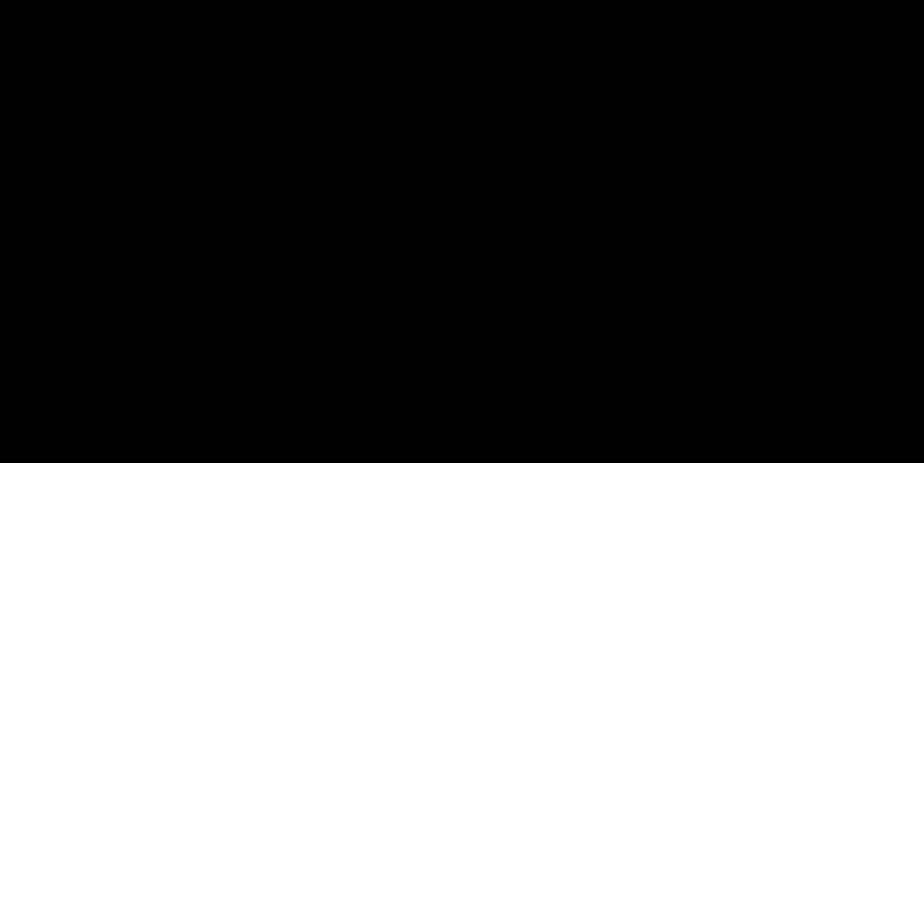}}\\
			{\footnotesize t=0}
		\end{minipage}
		\begin{minipage}{0.14\linewidth}
			\centering
			\fbox{\includegraphics[width=\linewidth]{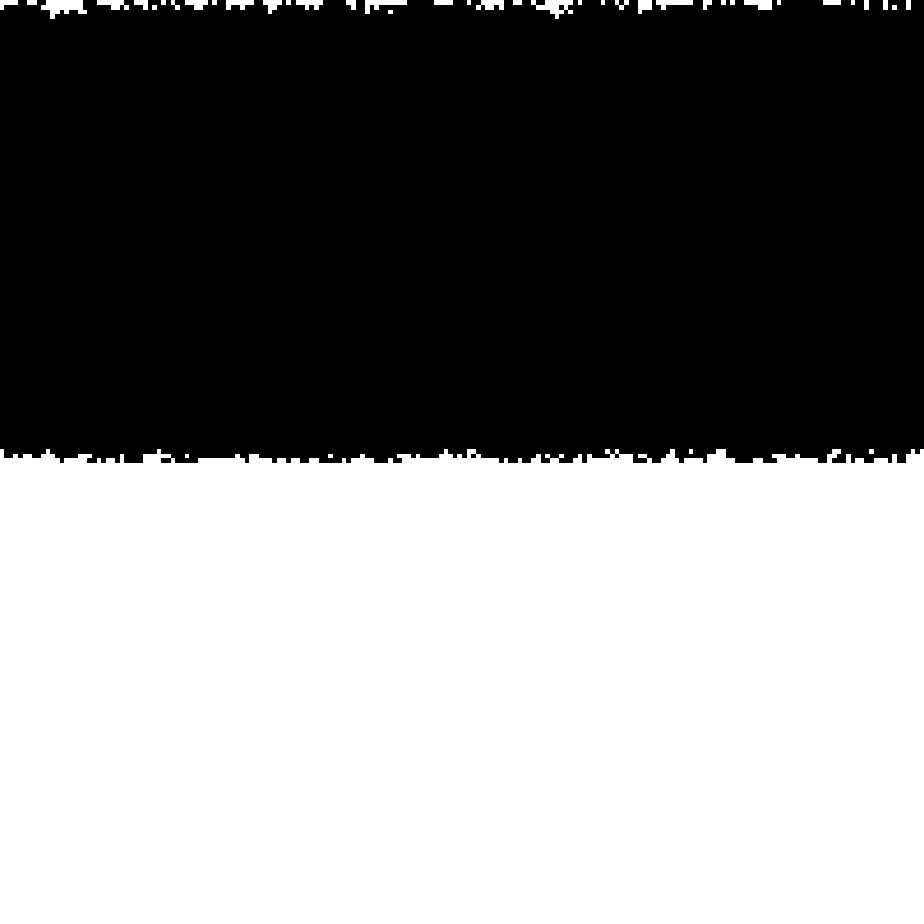}}\\
			{\footnotesize t=10}
		\end{minipage}
		\begin{minipage}{0.14\linewidth}
			\centering
			\fbox{\includegraphics[width=\linewidth]{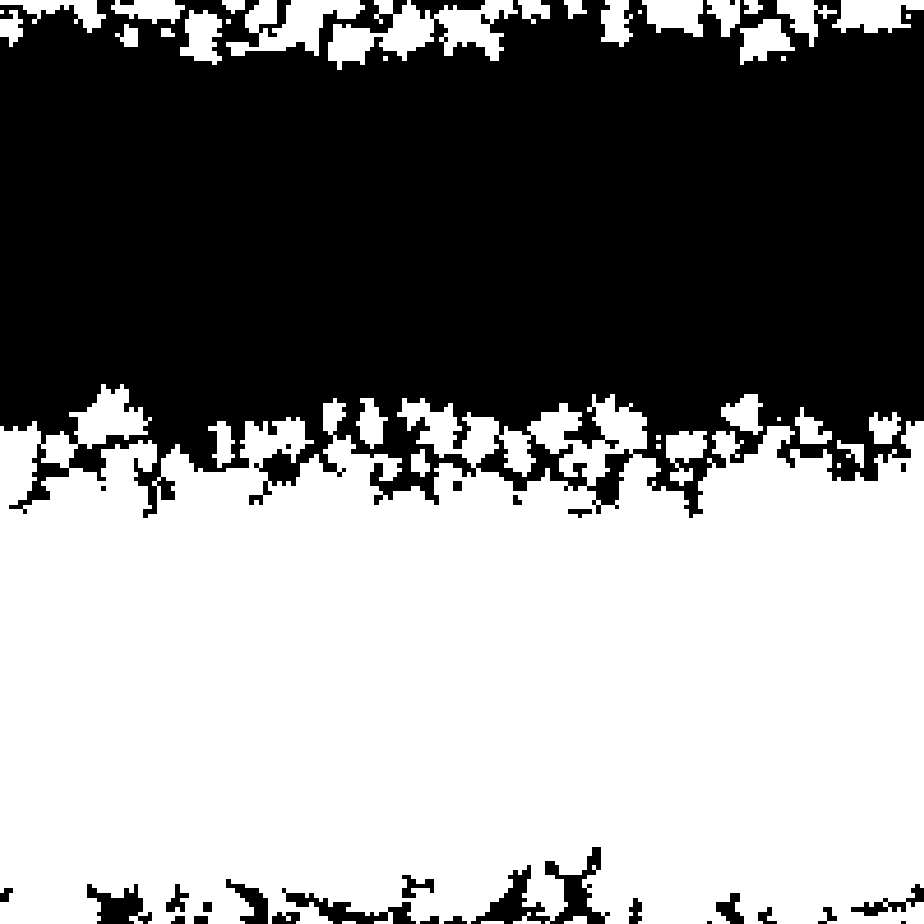}}\\
			{\footnotesize t=100}
		\end{minipage}
		\begin{minipage}{0.14\linewidth}
			\centering
			\fbox{\includegraphics[width=\linewidth]{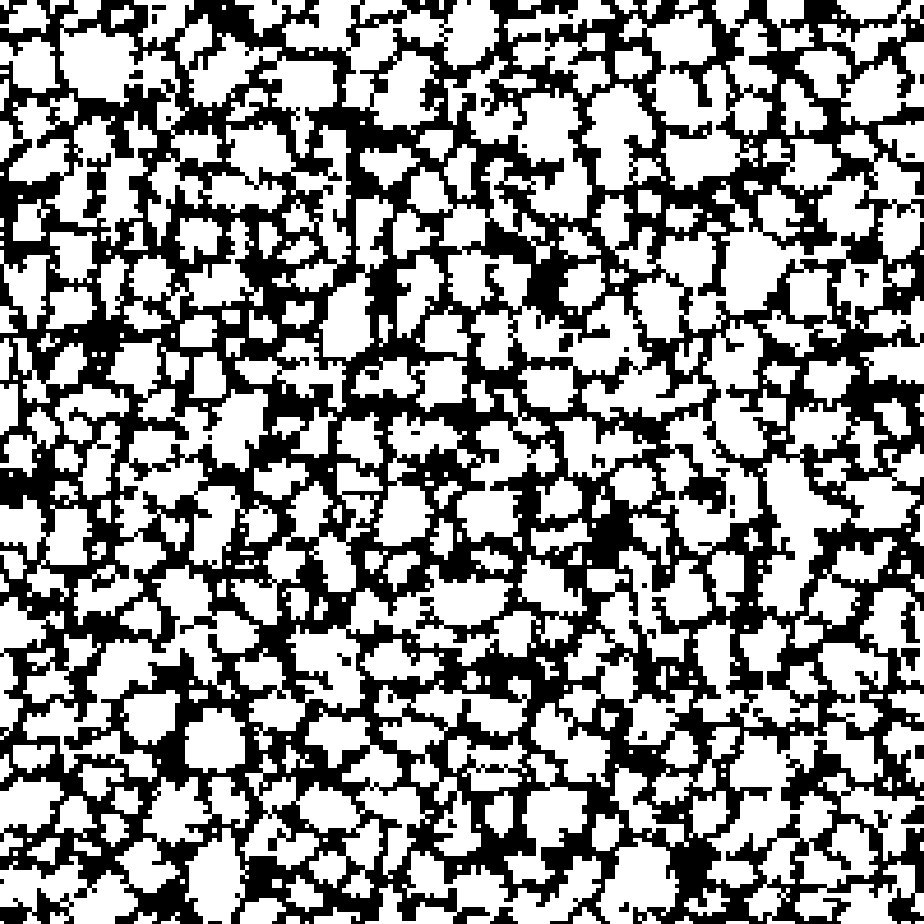}}\\
			{\footnotesize t=1000}
		\end{minipage}
		\begin{minipage}{0.14\linewidth}
			\centering
			\fbox{\includegraphics[width=\linewidth]{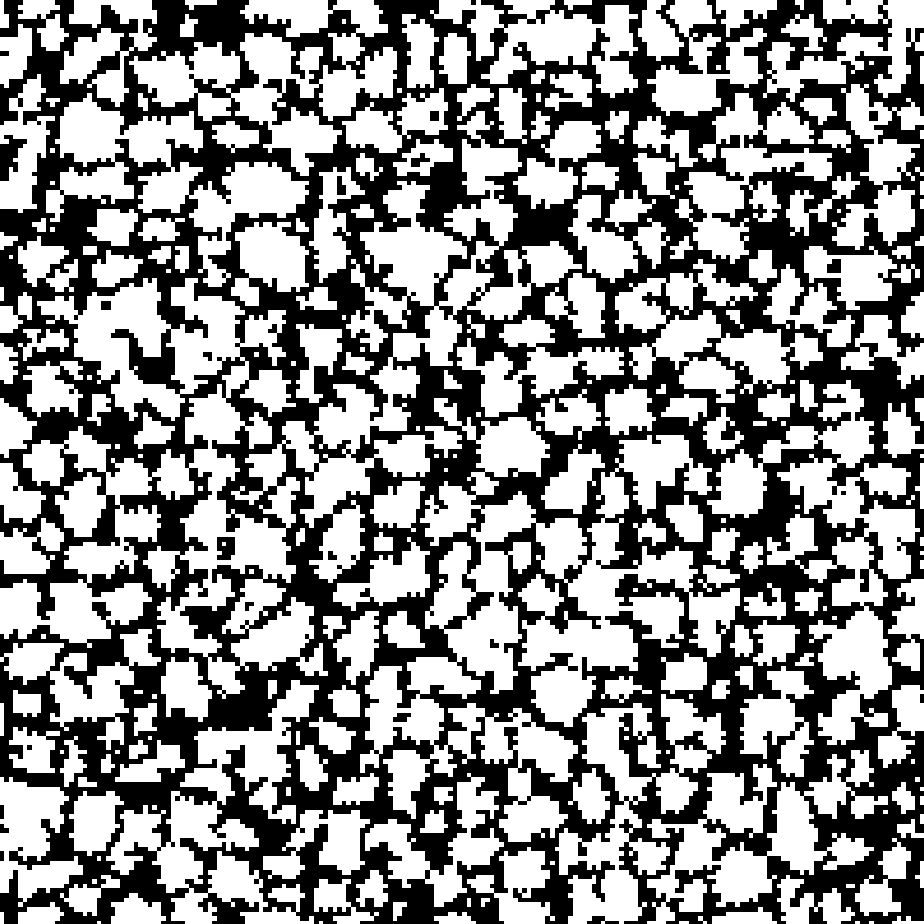}}\\
			{\footnotesize t=10000}
		\end{minipage}
\\
\centering
		{\footnotesize (d) Fermi update rule}
	\end{minipage}	
	\caption{Comparative performance of GRPO-GCC, GRPO, Q-learning, and the Fermi update rule at $r=4.0$. Subplots (a)--(d) correspond to GRPO-GCC, GRPO, Q-learning, and Fermi update rule respectively. Each subplot shows the temporal evolution of cooperation in blue and defection in red on the left. The right side presents representative state snapshots at different iterations. In snapshots, cooperators are shown as white dots and defectors as black dots.}
	\label{fig:GRPO_GCC_uDbC_compare}
\end{figure*}

We compare GRPO-GCC, GRPO, Q-learning, and the Fermi update rule at $r=4.0$. All algorithms are evaluated with identical spatial initialization and consistent experimental settings. At initialization, the upper half of the lattice is filled with defectors and the lower half with cooperators. The compared results include cooperation-defection dynamics and representative state snapshots at selected iterations. The experimental results are summarized in Fig.~\ref{fig:GRPO_GCC_uDbC_compare}, which contains four subplots. GRPO-GCC converges rapidly, exceeding $90\%$ cooperation within 100 iterations. Snapshots show initial random distribution of strategies gradually evolving toward dominant cooperation. Defectors do not form visible clusters and remain scattered across the lattice at equilibrium. This confirms that GCC significantly facilitates cooperative emergence in RL frameworks. GRPO without GCC exhibits a strong bias toward defection after 100 iterations. Snapshots illustrate nearly complete dominance of defectors, leaving only rare cooperators in the system. The absence of cooperative incentives explains why learning dynamics converge to universal defection. This result highlights the necessity of GCC to support stable cooperation under RL. Q-learning stabilizes near 41\% cooperation at $r=4.0$, without noticeable clustering. Snapshots display mixed distributions, with cooperators and defectors scattered throughout the grid. The limited cooperation stems from insufficient exploitation of global payoff structures. Thus, classical Q-learning struggles to achieve cooperation under weak enhancement conditions. The Fermi update rule reaches less than 60\% cooperation, even after extended iterations. Snapshots reveal cluster formation, where cooperators and defectors gradually occupy separate regions. Cooperator clusters are larger, yet defector groups persist due to local invasion dynamics. This illustrates how imitation-based rules favor spatial clustering but hinder global cooperation. In summary, RL-based models exhibit faster and more decisive convergence patterns than imitation-based ones. GRPO-GCC achieves higher cooperation than GRPO, validating the effectiveness of GCC. However, cooperation remains below 100\%, constrained by payoff structure rather than stochastic fluctuations. These results demonstrate distinct dynamics across learning paradigms and highlight the role of GCC.  

\subsection{Statistical analysis of GRPO-GCC and GRPO}
\label{exp:compare_stat}

To assess the effectiveness of the proposed GCC mechanism, we conducted comparative experiments between GRPO-GCC and the baseline GRPO. Each algorithm was independently trained for 50 runs under identical environmental configurations. The enhancement factor $r$ was varied from 3.0 to 6.0 in increments of 0.1, and the final cooperation rate after convergence was recorded. Two statistical visualization methods were employed: error-bar plots illustrating the mean and standard deviation of cooperation rates, and violin plots showing the distributional shape, mean, and median. The horizontal axis represents the final cooperation rate, while the vertical axis denotes the value of $r$.

\begin{figure*}[htbp!]
	\begin{minipage}{0.48\linewidth}
		\centering
		\includegraphics[width=\linewidth]{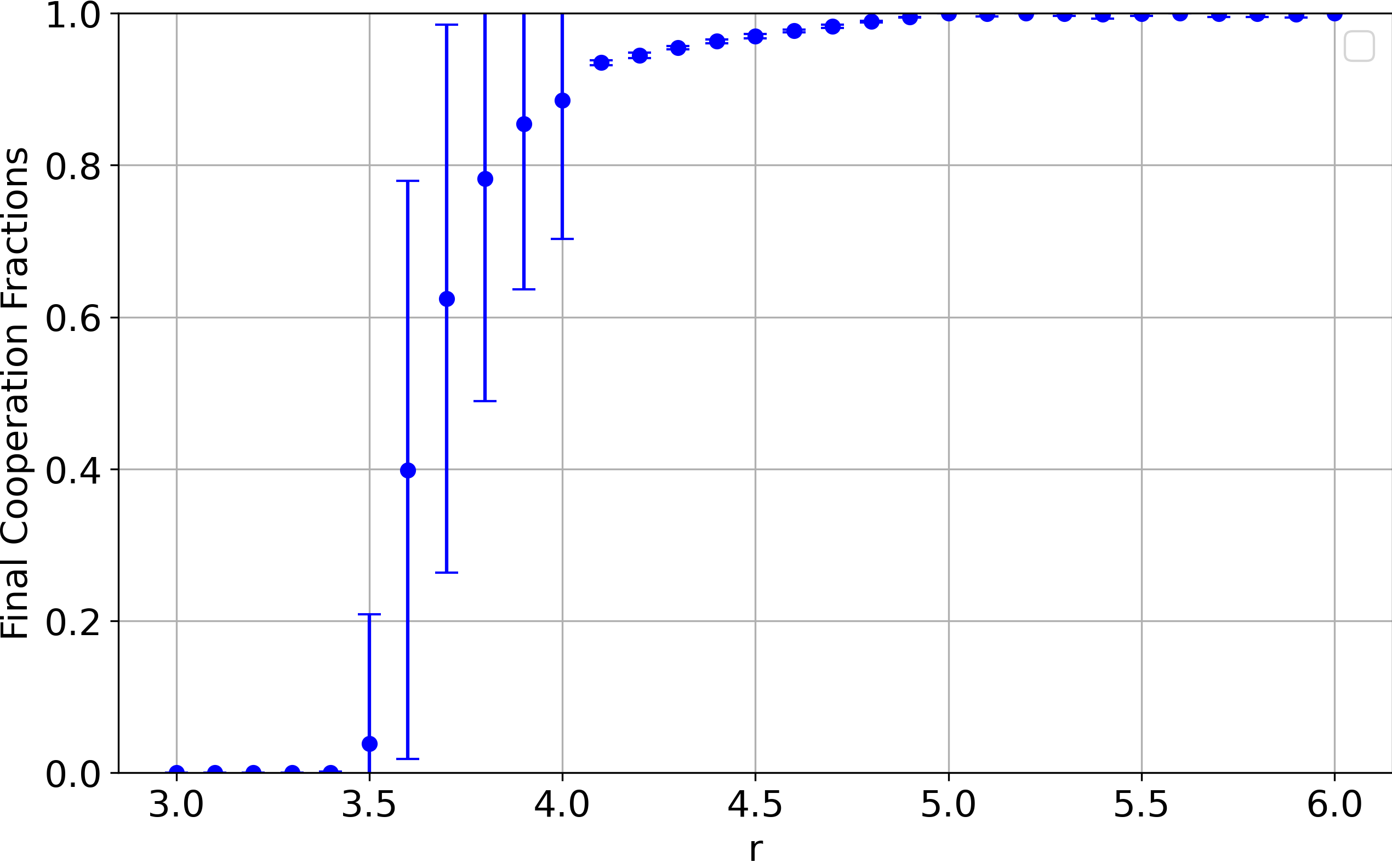}\\
		{\footnotesize (a) GRPO-GCC}
	\end{minipage}
	\hfill
	\begin{minipage}{0.48\linewidth}
		\centering
		\includegraphics[width=\linewidth]{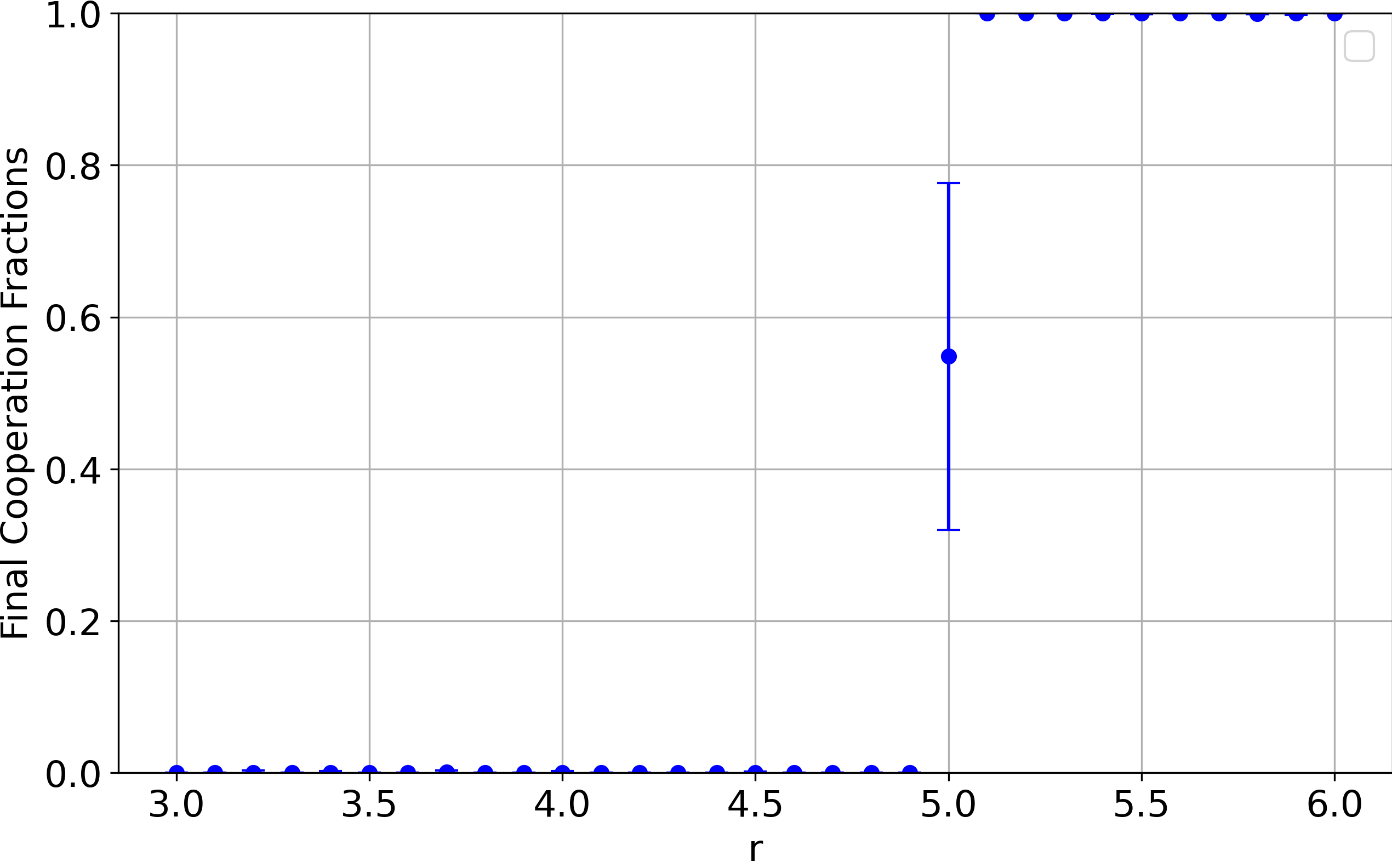}\\
		{\footnotesize (b) GRPO}
	\end{minipage}
	\caption{Error bar plots of final cooperation rates under varying $r$. GRPO-GCC achieves higher cooperation with lower $r$ values and shows smaller variance across runs.}		
	\label{fig:errorbar}
\end{figure*}

\begin{figure*}[htbp!]
	\begin{minipage}{0.48\linewidth}
		\centering
		\includegraphics[width=\linewidth]{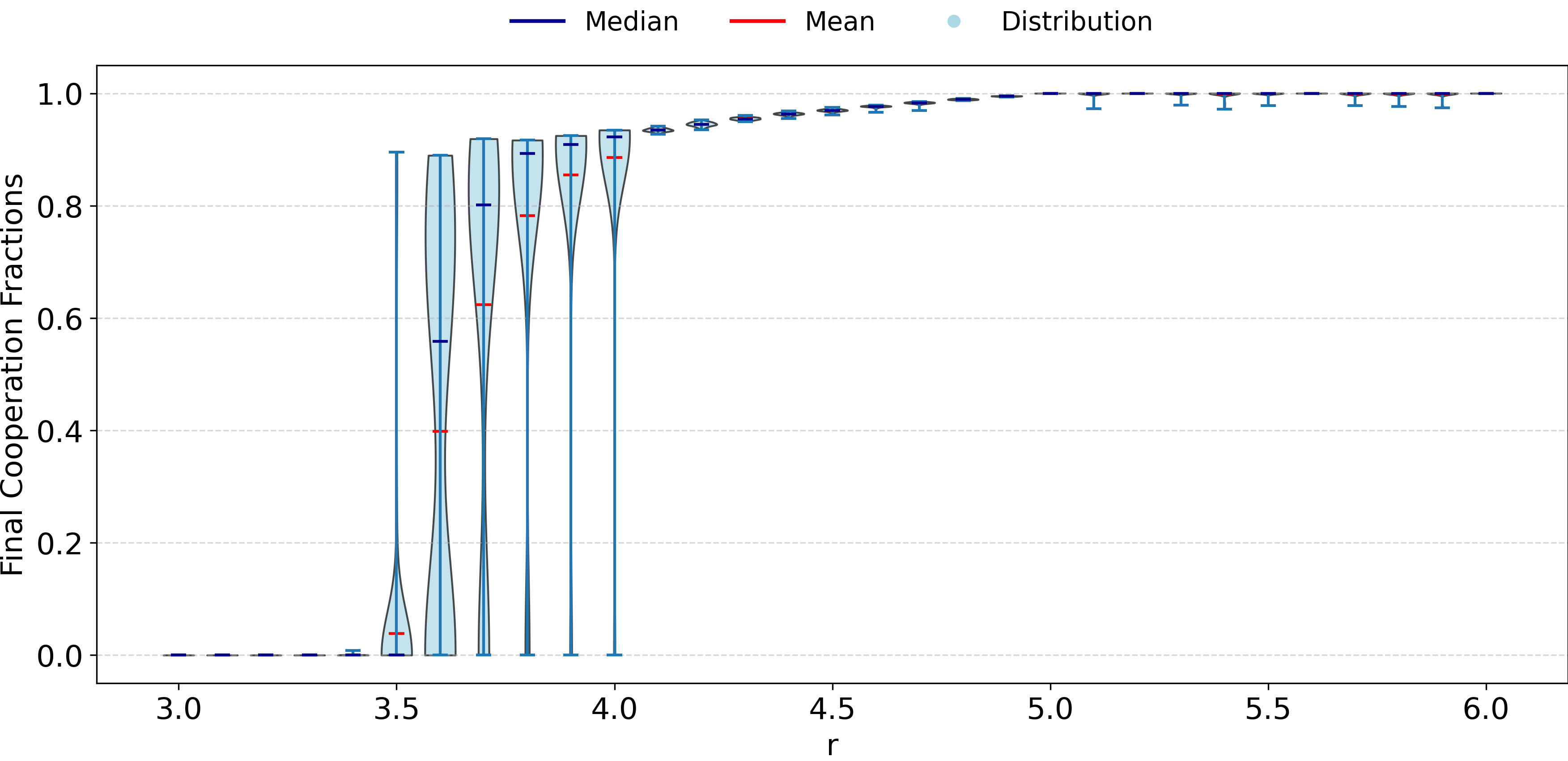}\\
		{\footnotesize (a) GRPO-GCC}
	\end{minipage}
	\hfill
	\begin{minipage}{0.48\linewidth}
		\centering
		\includegraphics[width=\linewidth]{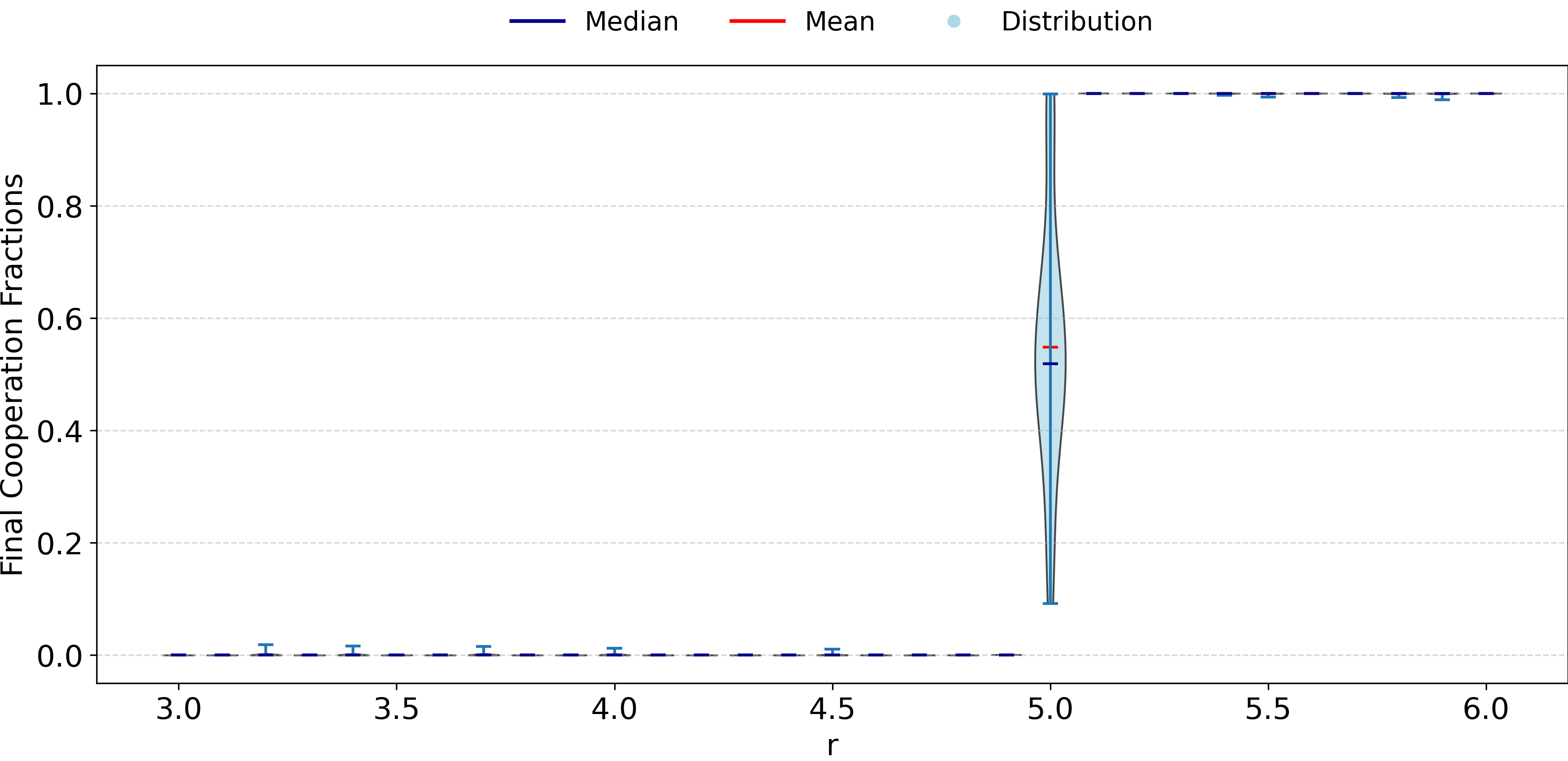}\\
		{\footnotesize (b) GRPO}
	\end{minipage}
	\caption{Violin plots of cooperation distributions. GRPO-GCC produces more concentrated and symmetric distributions while GRPO remains skewed toward defection.}		
	\label{fig:violin}
\end{figure*}

The error bar plots in Fig.~\ref{fig:errorbar} demonstrate that GRPO-GCC significantly outperforms GRPO across the examined $r$ values. When $r<3.5$, both algorithms exhibit negligible cooperation, reflecting the unfavorable incentive structure. From $r=3.6$ onward, GRPO-GCC shows a rapid increase in cooperation and achieves stable convergence beyond $r=4.0$ with mean cooperation exceeding 0.85 and narrow variance. In contrast, GRPO exhibits almost no cooperation when $r<5.0$. Cooperation begins to emerge at $r=5.0$ but with a large standard deviation. When $r>5.0$, the algorithm converges stably to a fully cooperative state where all agents choose cooperation. The violin plots in Fig.~\ref{fig:violin} further reveal that GRPO-GCC produces highly concentrated and symmetric distributions, reducing extreme outliers and ensuring robustness across runs. These results confirm that GCC lowers the critical threshold for cooperation emergence and enhances stability across repeated trials.

In addition, the $95\%$ confidence interval analysis presented in Table~\ref{tab:CI_comparison} corroborates the visual findings from Figs~\ref{fig:errorbar} and~\ref{fig:violin}. The intervals for GRPO-GCC consistently narrow and shift upward at lower $r$ values compared with GRPO. This provides statistical confirmation that the GCC mechanism reduces uncertainty and enhances robustness. This joint evidence from figures and tables underscores the dual role of GCC in both accelerating cooperation emergence and ensuring stable convergence.

\begin{table*}[h]
	\centering
	\footnotesize
	\caption{$95\%$ confidence intervals comparison for cooperation fractions}
	\label{tab:CI_comparison}
	\resizebox{\textwidth}{!}{
		\begin{tabular}{@{}c*{7}{S[table-format=1.2]@{\,--\,}S[table-format=1.2]}@{}}
			\toprule
			$r$ & \multicolumn{2}{c}{3.5} & \multicolumn{2}{c}{3.6} & \multicolumn{2}{c}{3.7} & \multicolumn{2}{c}{3.8} & \multicolumn{2}{c}{3.9} & \multicolumn{2}{c}{4.0} & \multicolumn{2}{c}{4.1} \\
			\cmidrule(lr){2-3} \cmidrule(lr){4-5} \cmidrule(lr){6-7} \cmidrule(lr){8-9} \cmidrule(lr){10-11} \cmidrule(lr){12-13} \cmidrule(lr){14-15}
			GRPO-GCC & 0.00 & 0.00 & 0.05 & 0.22 & 0.12 & 0.35 & 0.18 & 0.42 & 0.27 & 0.53 & 0.41 & 0.67 & 0.58 & 0.82 \\
			GRPO     & 0.00 & 0.00 & 0.00 & 0.00 & 0.00 & 0.00 & 0.00 & 0.00 & 0.00 & 0.00 & 0.00 & 0.00 & 0.00 & 0.00 \\
			\midrule
			
			$r$ & \multicolumn{2}{c}{4.2} & \multicolumn{2}{c}{4.3} & \multicolumn{2}{c}{4.4} & \multicolumn{2}{c}{4.5} & \multicolumn{2}{c}{4.6} & \multicolumn{2}{c}{4.7} & \multicolumn{2}{c}{4.8} \\
			\cmidrule(lr){2-3} \cmidrule(lr){4-5} \cmidrule(lr){6-7} \cmidrule(lr){8-9} \cmidrule(lr){10-11} \cmidrule(lr){12-13} \cmidrule(lr){14-15}
			GRPO-GCC & 0.72 & 0.94 & 0.85 & 0.98 & 0.91 & 1.00 & 0.93 & 1.00 & 0.95 & 1.00 & 0.97 & 1.00 & 0.99 & 1.00 \\
			GRPO     & 0.00 & 0.00 & 0.00 & 0.12 & 0.00 & 0.21 & 0.00 & 0.36 & 0.00 & 0.49 & 0.00 & 0.63 & 0.15 & 0.77 \\
			\midrule
			
			$r$ & \multicolumn{2}{c}{4.9} & \multicolumn{2}{c}{5.0} & \multicolumn{2}{c}{5.1} & \multicolumn{2}{c}{5.2} & \multicolumn{2}{c}{5.3} & \multicolumn{2}{c}{5.4} & \multicolumn{2}{c}{5.5} \\
			\cmidrule(lr){2-3} \cmidrule(lr){4-5} \cmidrule(lr){6-7} \cmidrule(lr){8-9} \cmidrule(lr){10-11} \cmidrule(lr){12-13} \cmidrule(lr){14-15}
			GRPO-GCC & 1.00 & 1.00 & 1.00 & 1.00 & 1.00 & 1.00 & 1.00 & 1.00 & 1.00 & 1.00 & 1.00 & 1.00 & 1.00 & 1.00 \\
			GRPO     & 0.35 & 0.81 & 0.52 & 0.88 & 0.61 & 0.92 & 0.68 & 0.94 & 0.72 & 0.96 & 0.79 & 0.98 & 0.85 & 0.99 \\
			\bottomrule
		\end{tabular}
	}
\end{table*}

\subsection{GRPO-GCC with half-and-half initialization}
\label{exp_hh}

We further examine the robustness of GRPO-GCC under a half-and-half initialization setting. Agents are arranged on a $200 \times 200$ lattice, with the upper half initialized as defectors and the lower half as cooperators. The experiment is conducted for two representative enhancement factors, $r=3.6$ and $r=4.6$. For each case, the top panel shows the temporal evolution of cooperation and defection fractions over iterations, while the bottom panel presents spatial snapshots at $t=0,1,10,100,1000$. In addition, payoff heatmaps are provided to illustrate the spatial distribution of accumulated returns at the same timesteps. The color scale ranges from yellow (high payoff) through green and blue to purple (low payoff).

\begin{figure*}[htbp!]
	\begin{minipage}{0.45\linewidth}
		\begin{minipage}{\linewidth}
			\centering
			\includegraphics[width=\linewidth]{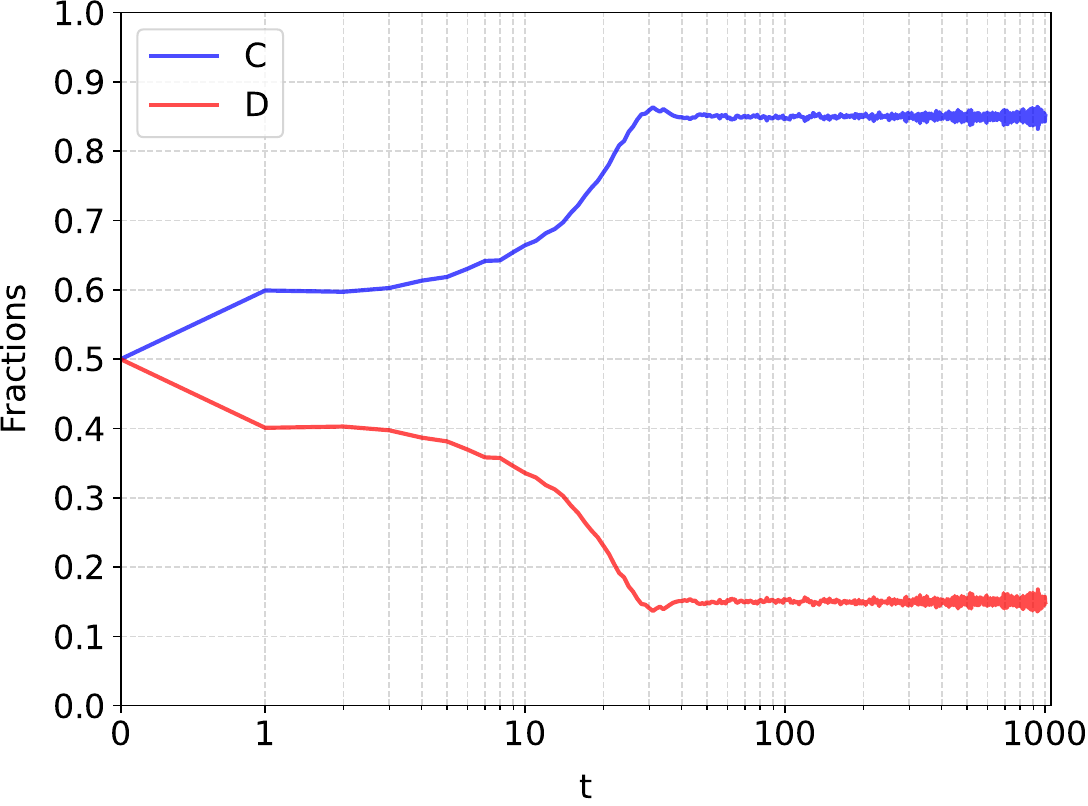}\\
		\end{minipage}
		\vspace{2mm}
		\\
		\begin{minipage}{0.188\linewidth}
			\centering
			\fbox{\includegraphics[width=\linewidth]{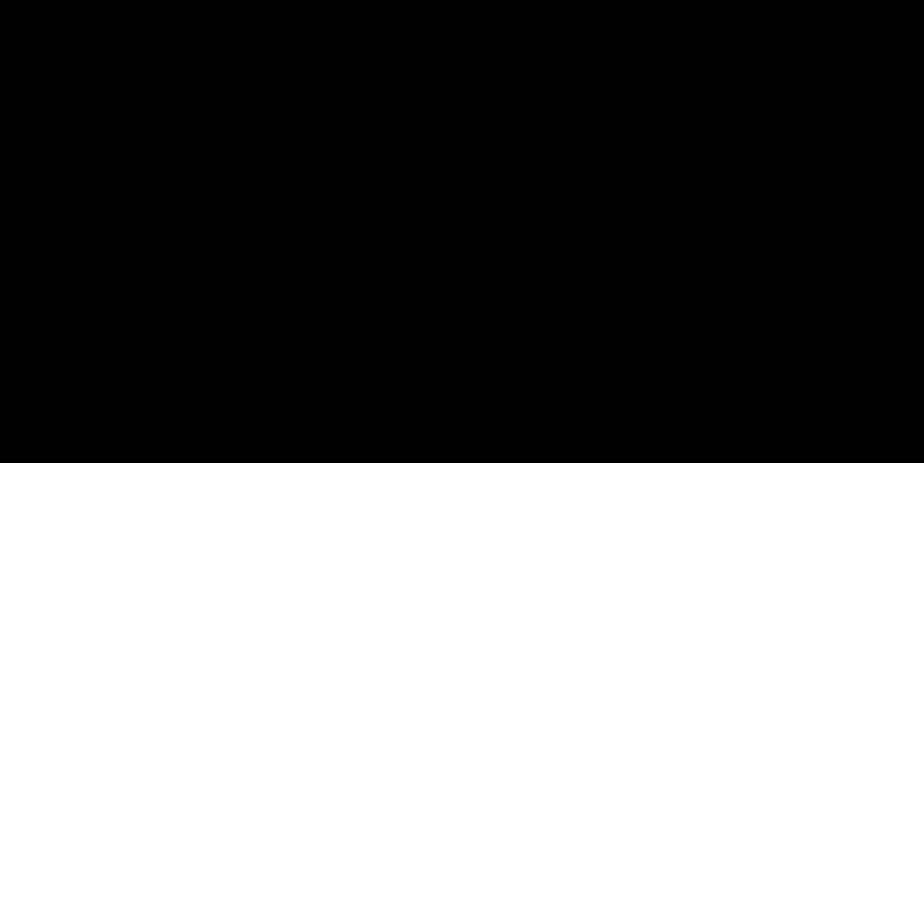}}\\
			{\footnotesize t=0}
		\end{minipage}
		\begin{minipage}{0.188\linewidth}
			\centering
			\fbox{\includegraphics[width=\linewidth]{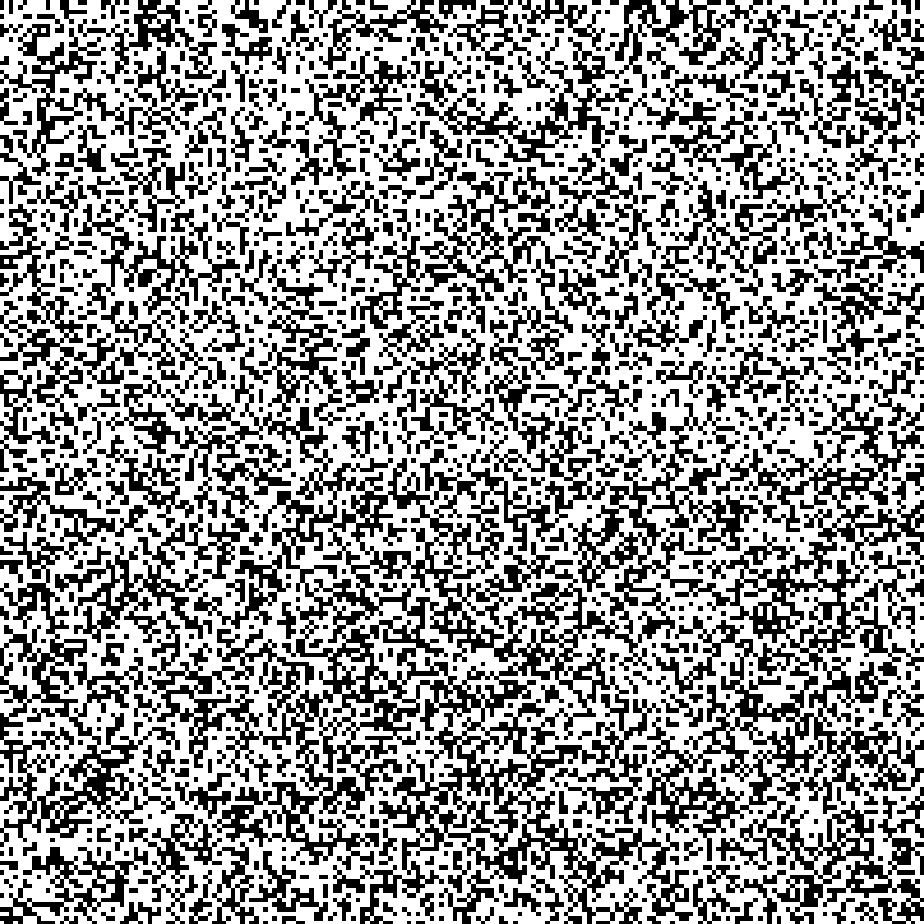}}\\
			{\footnotesize t=1}
		\end{minipage}
		\begin{minipage}{0.188\linewidth}
			\centering
			\fbox{\includegraphics[width=\linewidth]{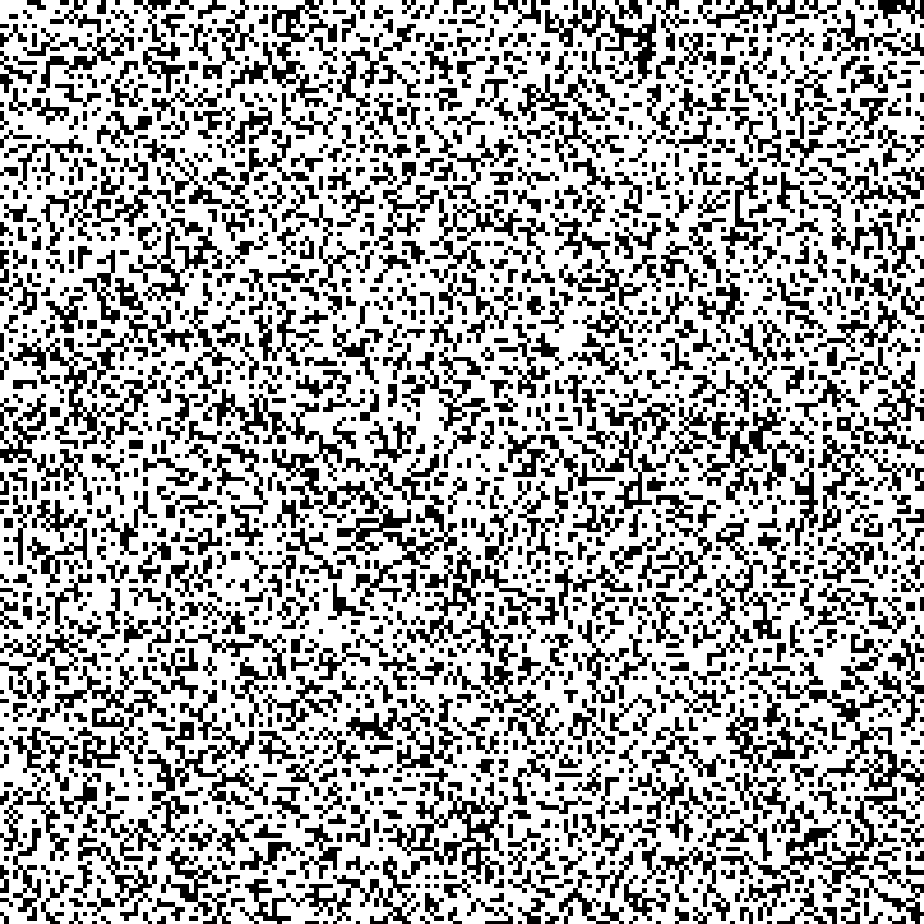}}\\
			{\footnotesize t=10}
		\end{minipage}
		\begin{minipage}{0.188\linewidth}
			\centering
			\fbox{\includegraphics[width=\linewidth]{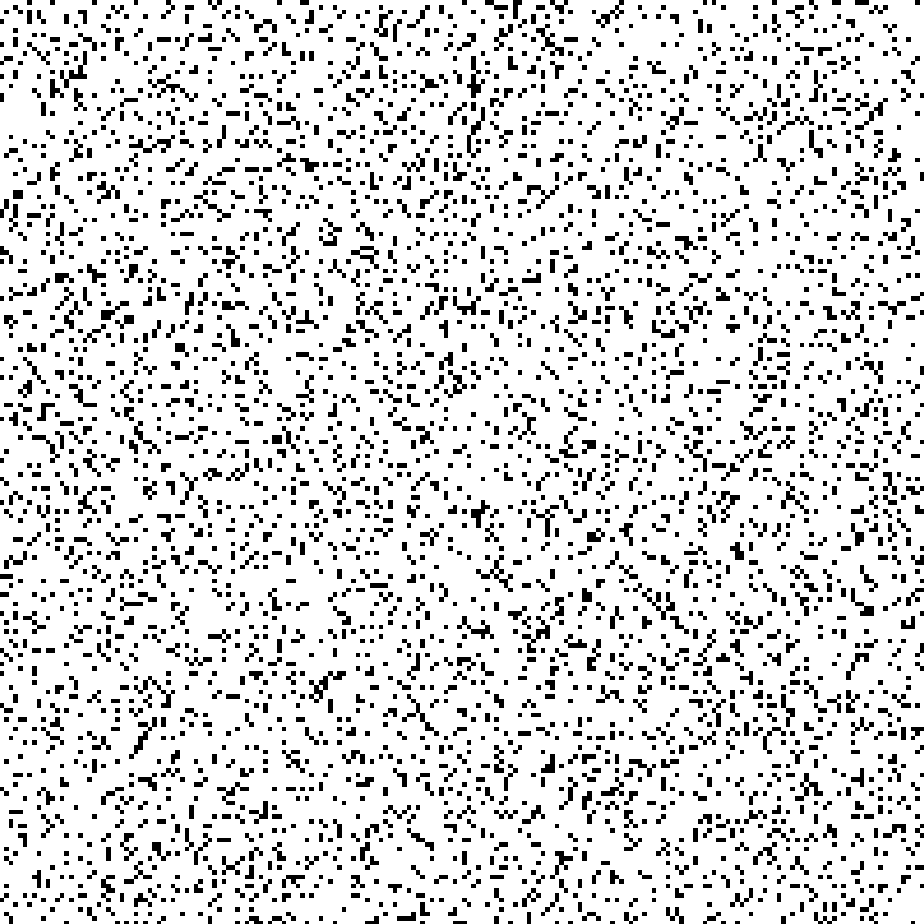}}\\
			{\footnotesize t=100}
		\end{minipage}
		\begin{minipage}{0.188\linewidth}
			\centering
			\fbox{\includegraphics[width=\linewidth]{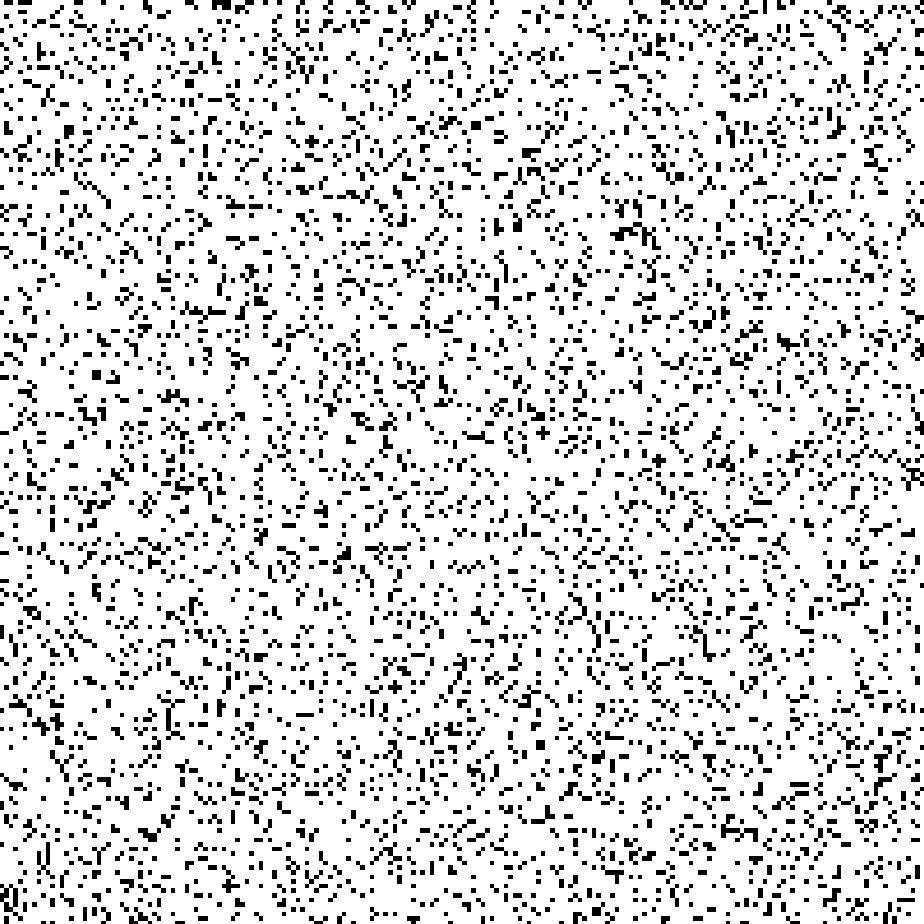}}\\
			{\footnotesize t=1000}
		\end{minipage}
	\centering
	\\
	[2mm]
		{\footnotesize (a) r=3.6}
	\end{minipage}
	\hfill
	\begin{minipage}{0.45\linewidth}
		\begin{minipage}{\linewidth}
			\centering
			\includegraphics[width=\linewidth]{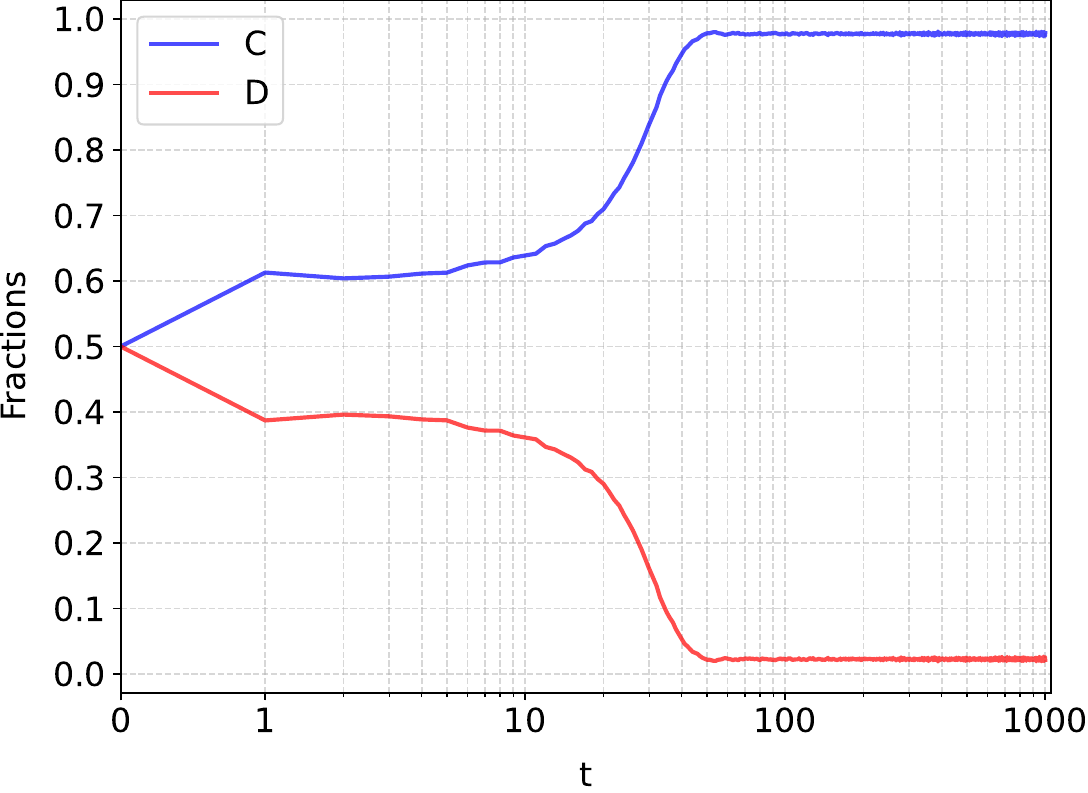}\\
		\end{minipage}
		\vspace{2mm}
		\\
		\begin{minipage}{0.188\linewidth}
			\centering
			\fbox{\includegraphics[width=\linewidth]{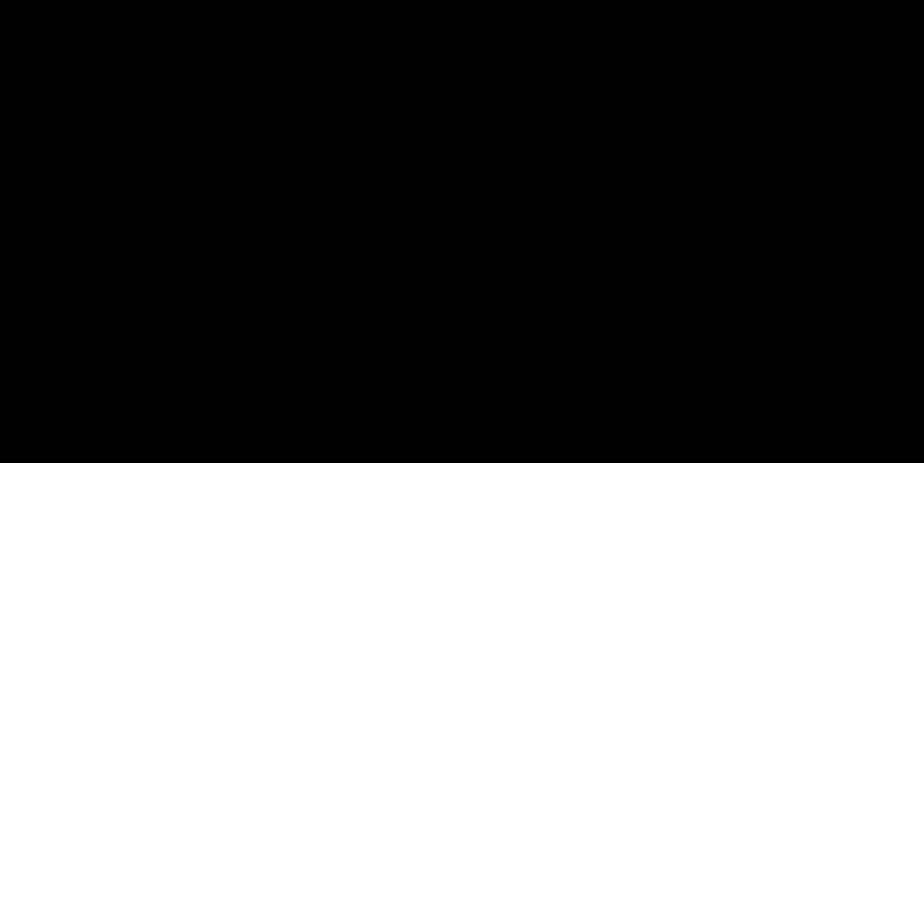}}\\
			{\footnotesize t=0}
		\end{minipage}
		\begin{minipage}{0.188\linewidth}
			\centering
			\fbox{\includegraphics[width=\linewidth]{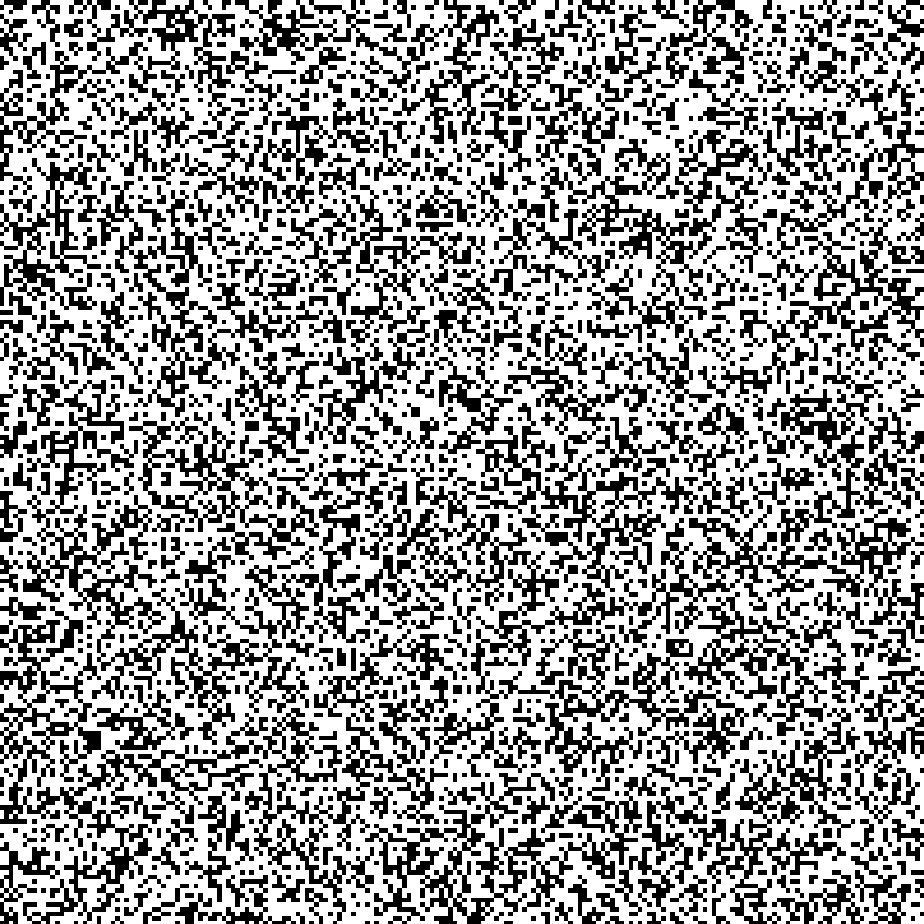}}\\
			{\footnotesize t=1}
		\end{minipage}
		\begin{minipage}{0.188\linewidth}
			\centering
			\fbox{\includegraphics[width=\linewidth]{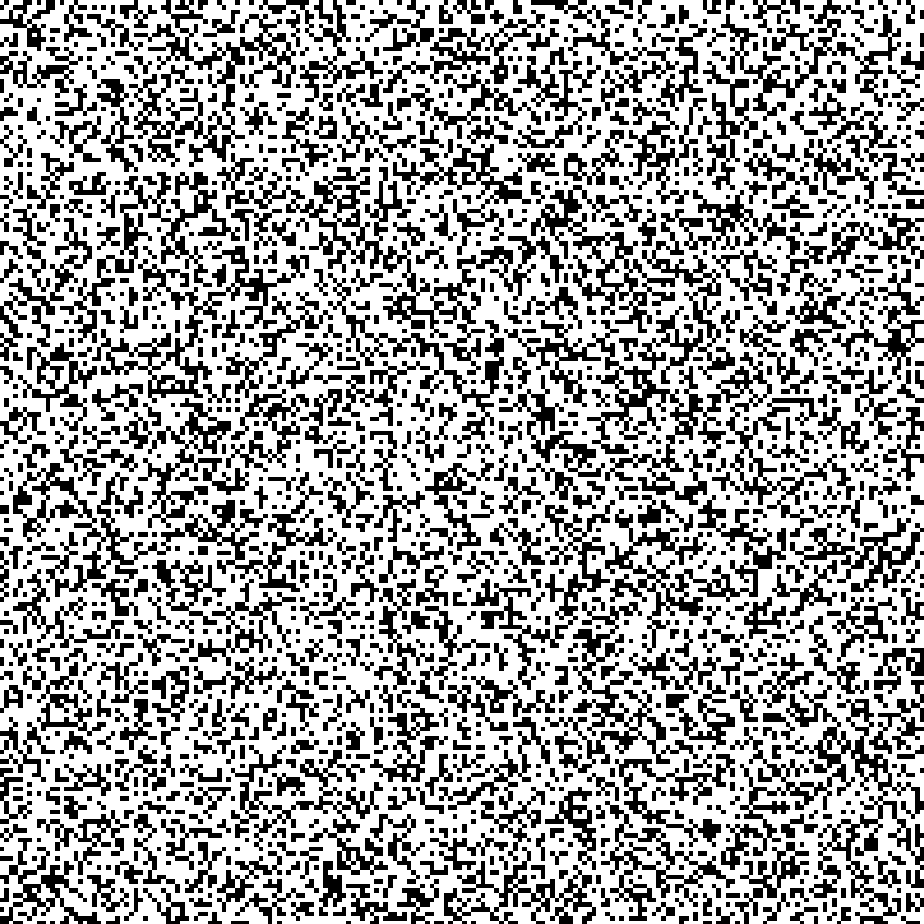}}\\
			{\footnotesize t=10}
		\end{minipage}
		\begin{minipage}{0.188\linewidth}
			\centering
			\fbox{\includegraphics[width=\linewidth]{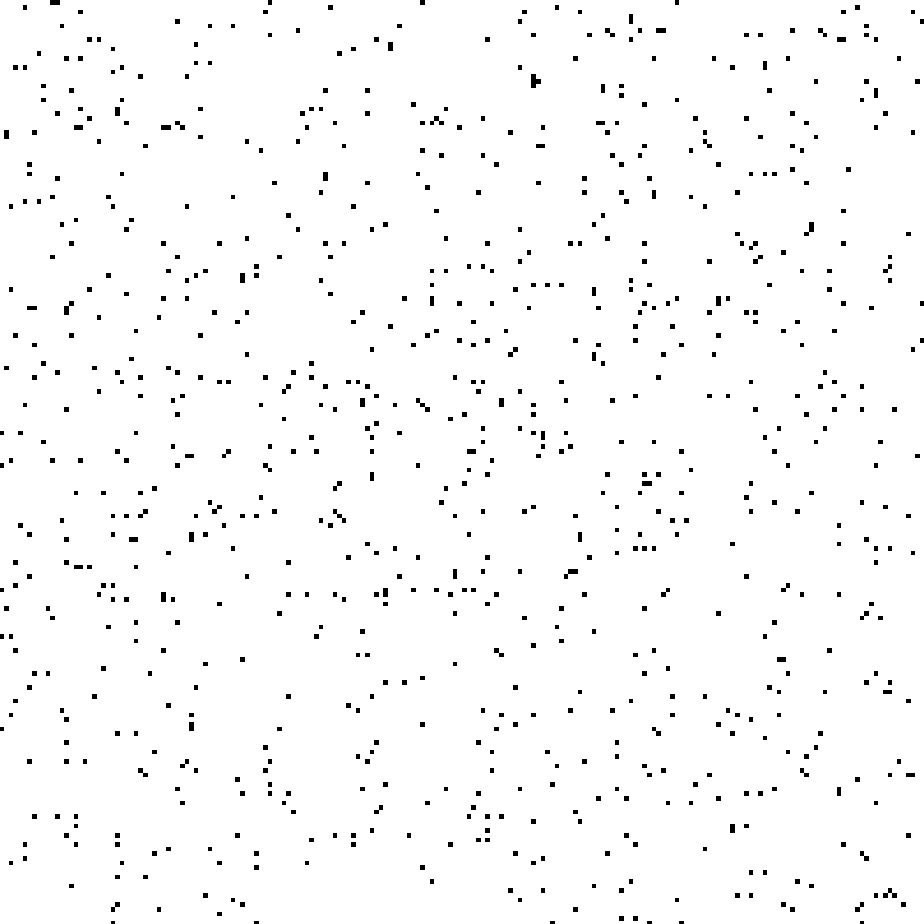}}\\
			{\footnotesize t=100}
		\end{minipage}
		\begin{minipage}{0.188\linewidth}
			\centering
			\fbox{\includegraphics[width=\linewidth]{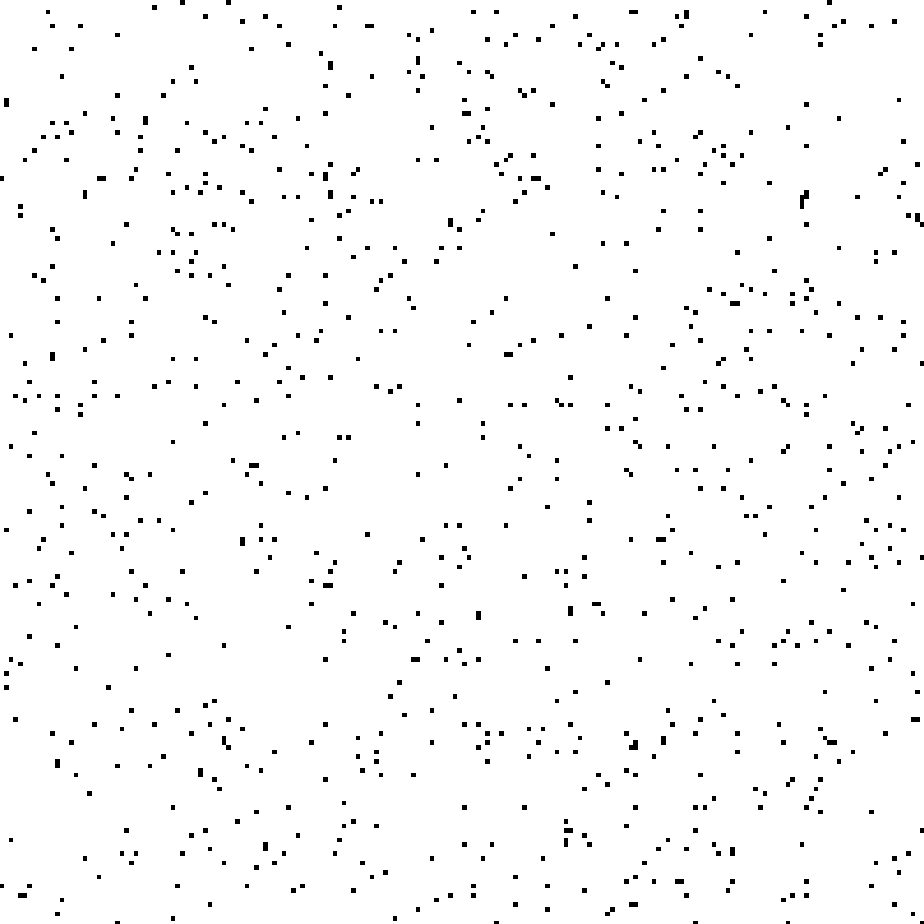}}\\
			{\footnotesize t=1000}
		\end{minipage}
	\centering
\\
	[2mm]
		{\footnotesize (b) r=4.6}
	\end{minipage}
	\\
	[2mm]
	\begin{minipage}{\linewidth}
		\begin{minipage}{0.188\linewidth}
			\centering
			\includegraphics[width=\linewidth]{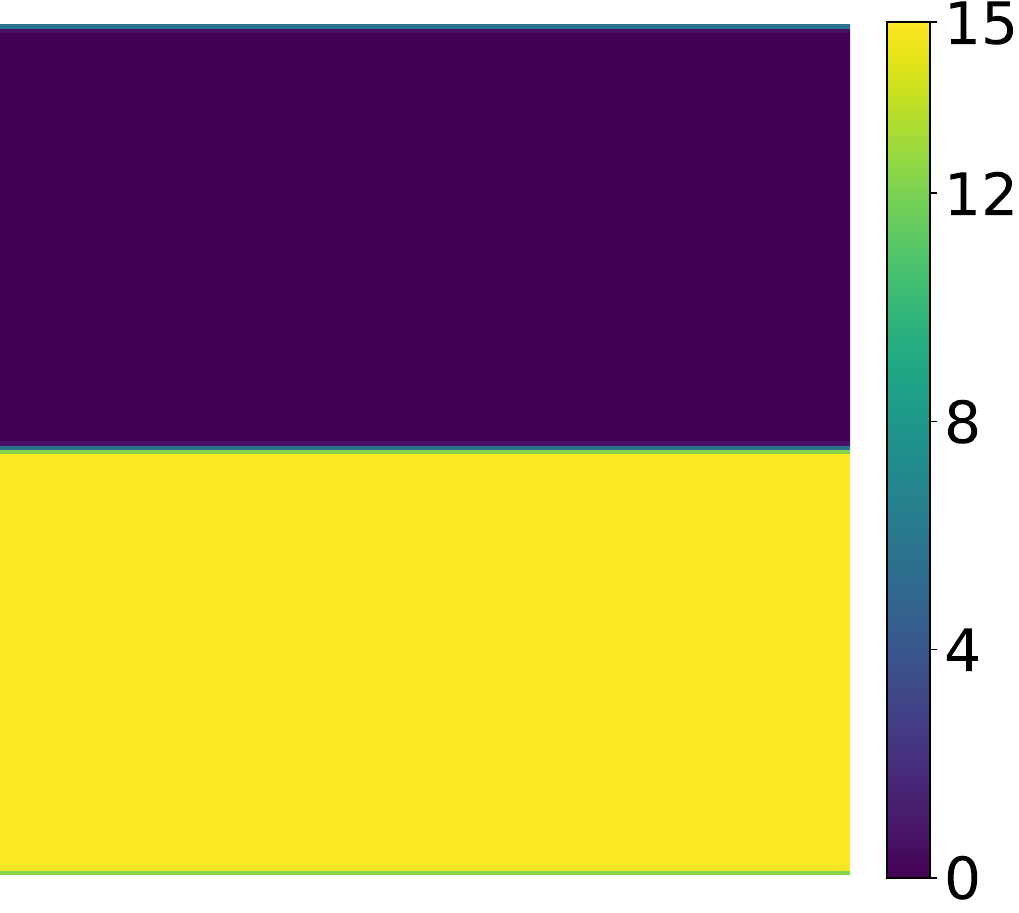}\\
			{\footnotesize t=0}
		\end{minipage}
		\hfill
		\begin{minipage}{0.188\linewidth}
			\centering
			\includegraphics[width=\linewidth]{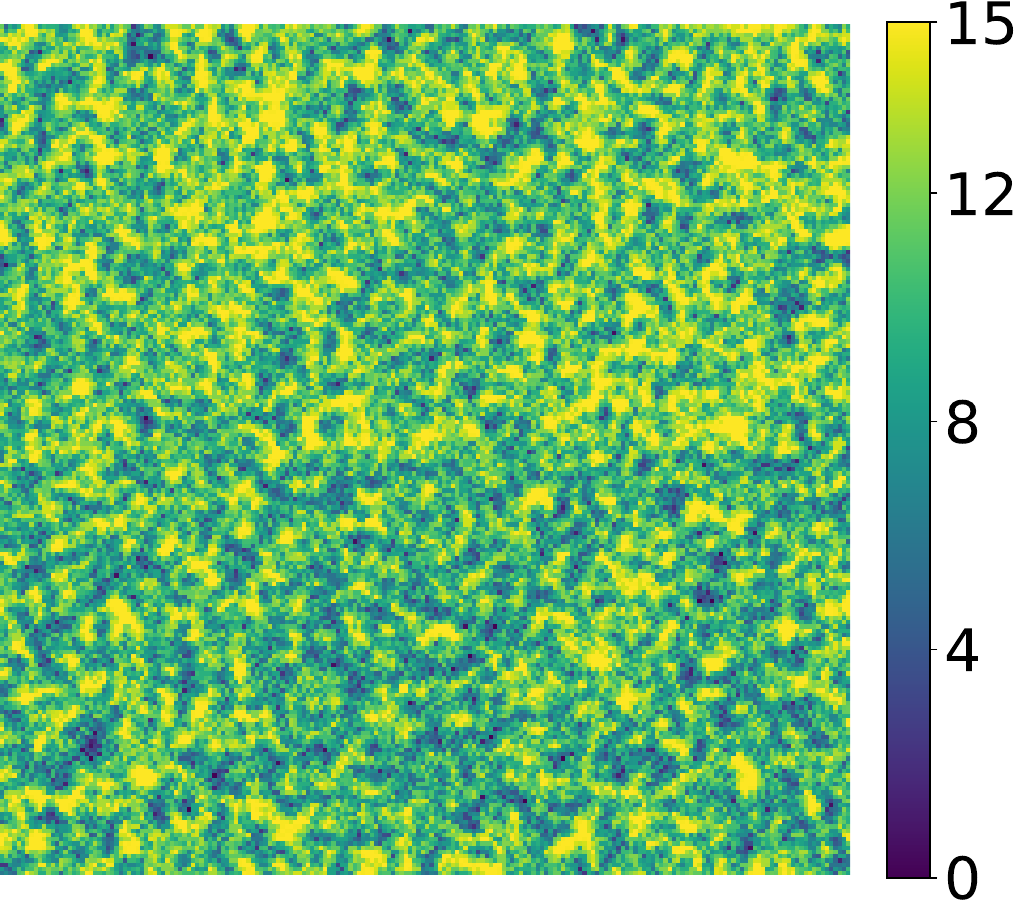}\\
			{\footnotesize t=1}
		\end{minipage}
		\hfill
		\begin{minipage}{0.188\linewidth}
			\centering
			\includegraphics[width=\linewidth]{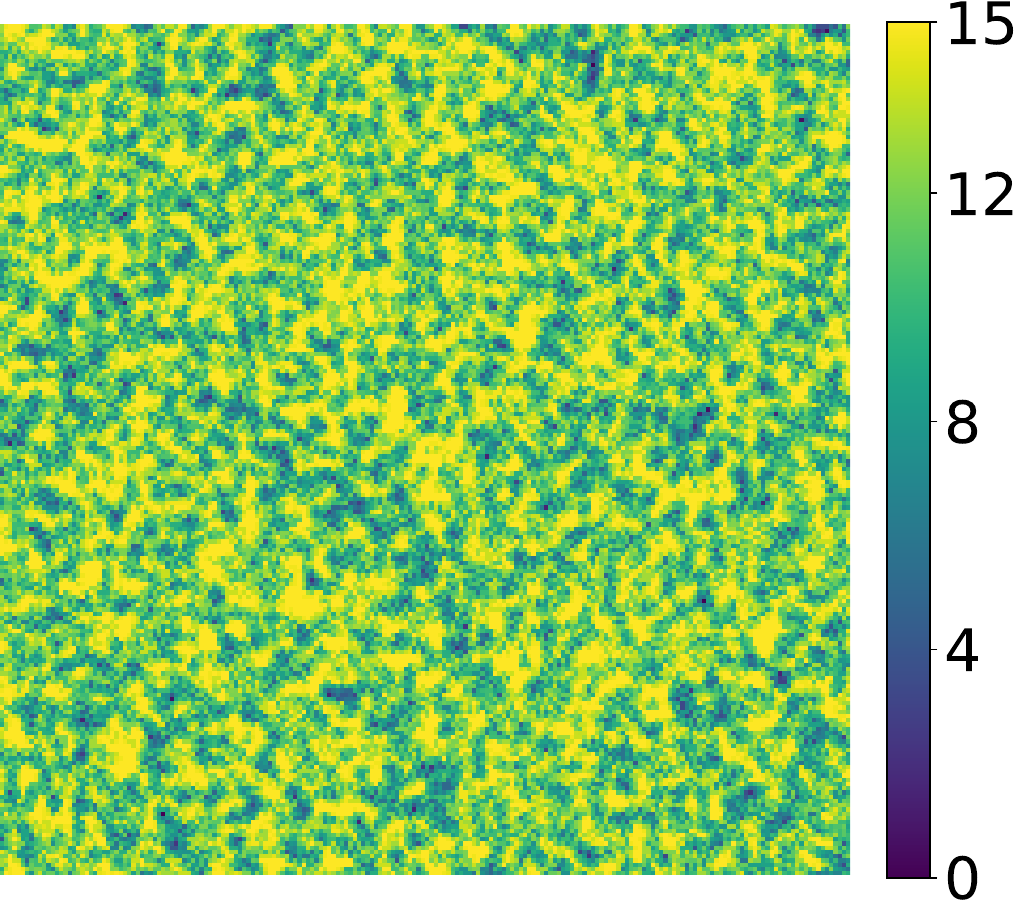}\\
			{\footnotesize t=10}
		\end{minipage}
		\hfill
		\begin{minipage}{0.188\linewidth}
			\centering
			\includegraphics[width=\linewidth]{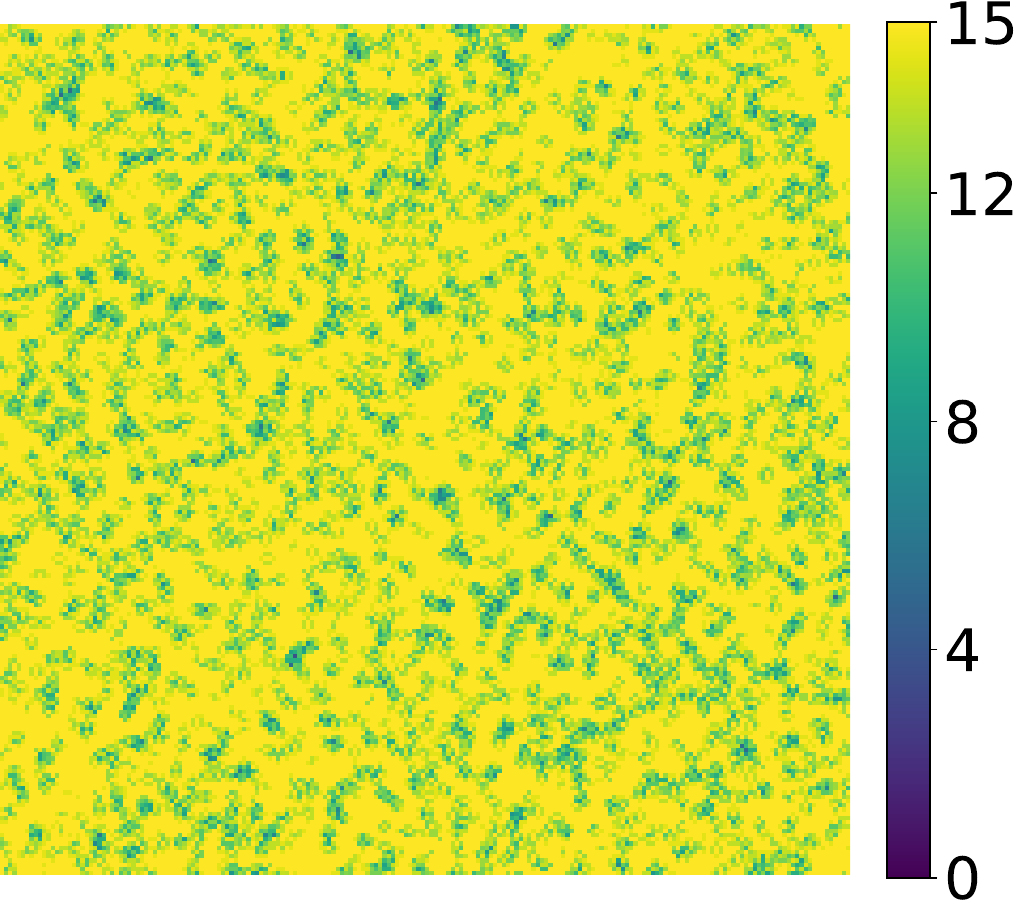}\\
			{\footnotesize t=100}
		\end{minipage}
		\hfill
		\begin{minipage}{0.188\linewidth}
			\centering
			\includegraphics[width=\linewidth]{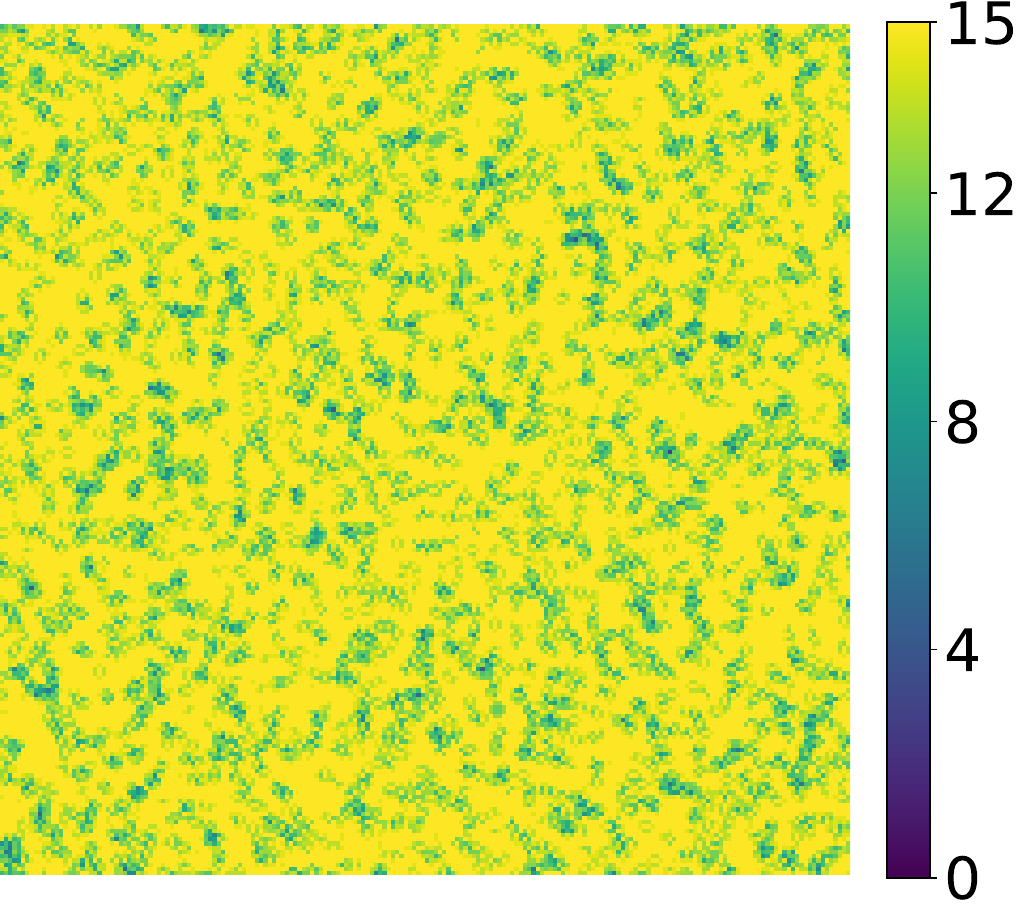}\\
			{\footnotesize t=1000}
		\end{minipage}
		\\
		[2mm]
			\centering
		{\footnotesize (c) r=3.6 (Payoff heatmaps)}
	\end{minipage}
	\\
	[2mm]
	\begin{minipage}{\linewidth}
		\begin{minipage}{0.188\linewidth}
			\centering
			\includegraphics[width=\linewidth]{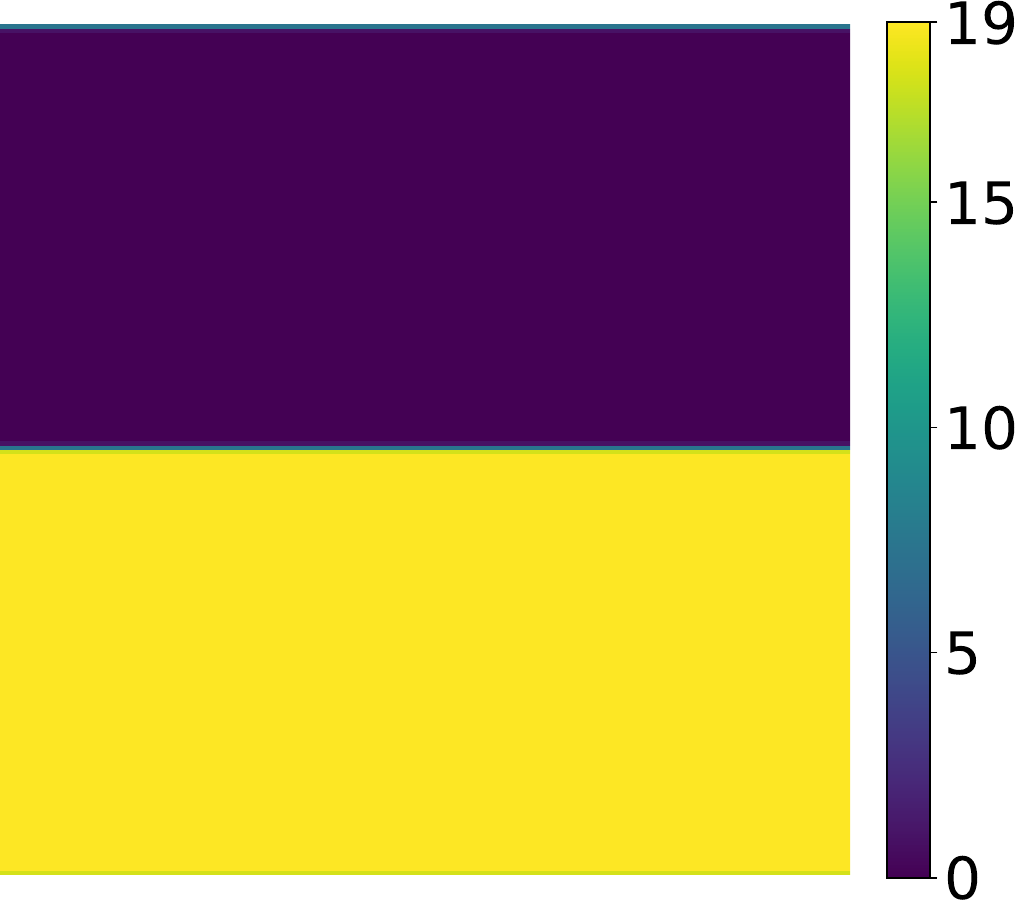}\\
			{\footnotesize t=0}
		\end{minipage}
		\hfill
		\begin{minipage}{0.188\linewidth}
			\centering
			\includegraphics[width=\linewidth]{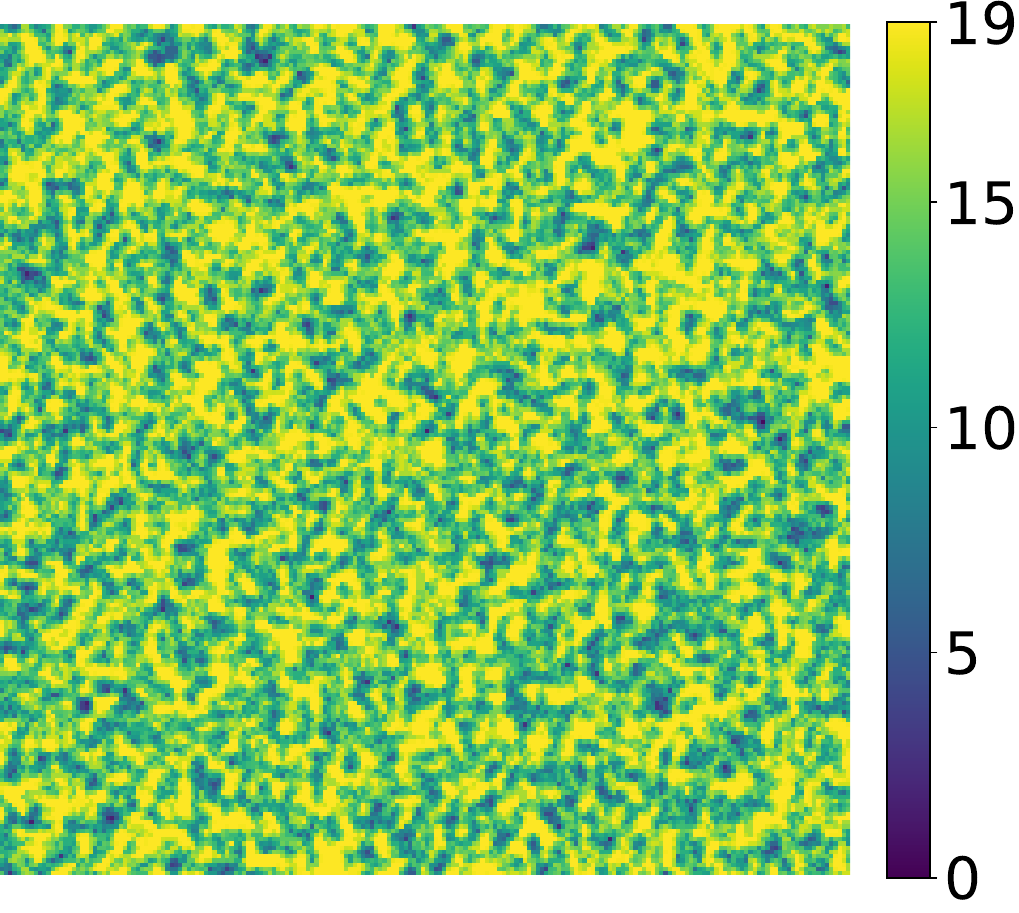}\\
			{\footnotesize t=1}
		\end{minipage}
		\hfill
		\begin{minipage}{0.188\linewidth}
			\centering
			\includegraphics[width=\linewidth]{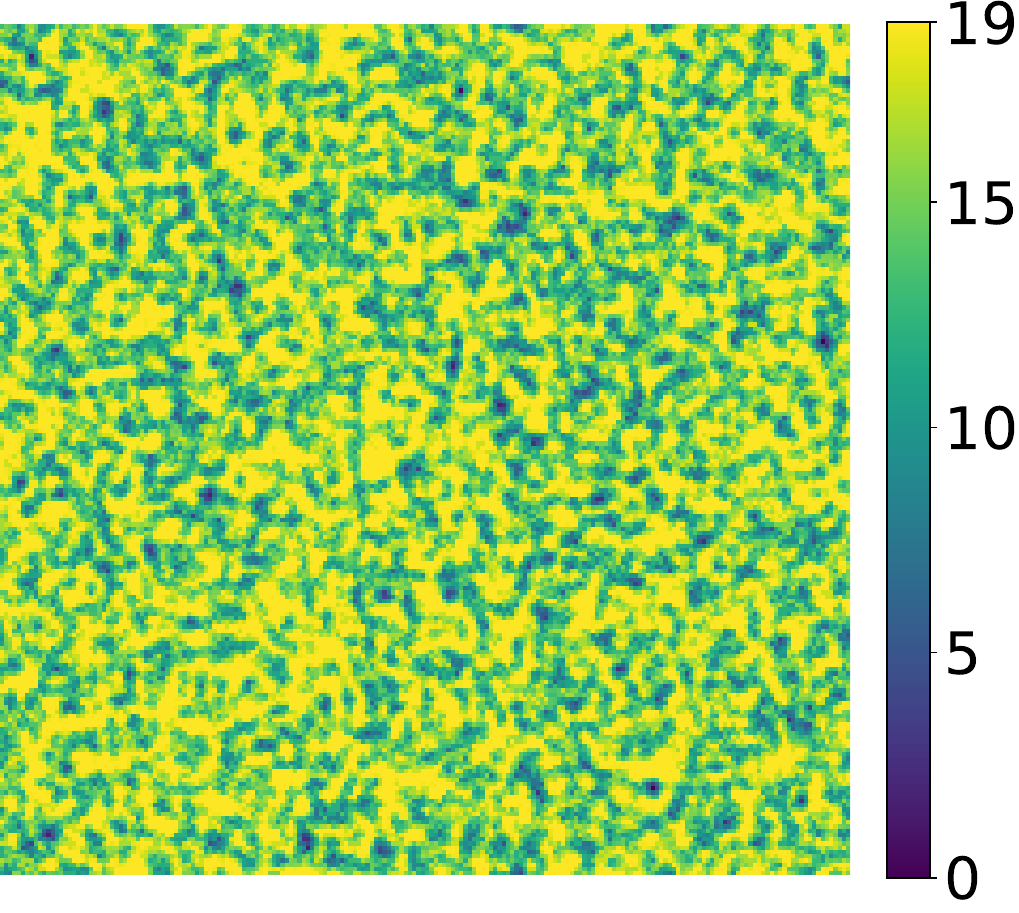}\\
			{\footnotesize t=10}
		\end{minipage}
		\hfill
		\begin{minipage}{0.188\linewidth}
			\centering
			\includegraphics[width=\linewidth]{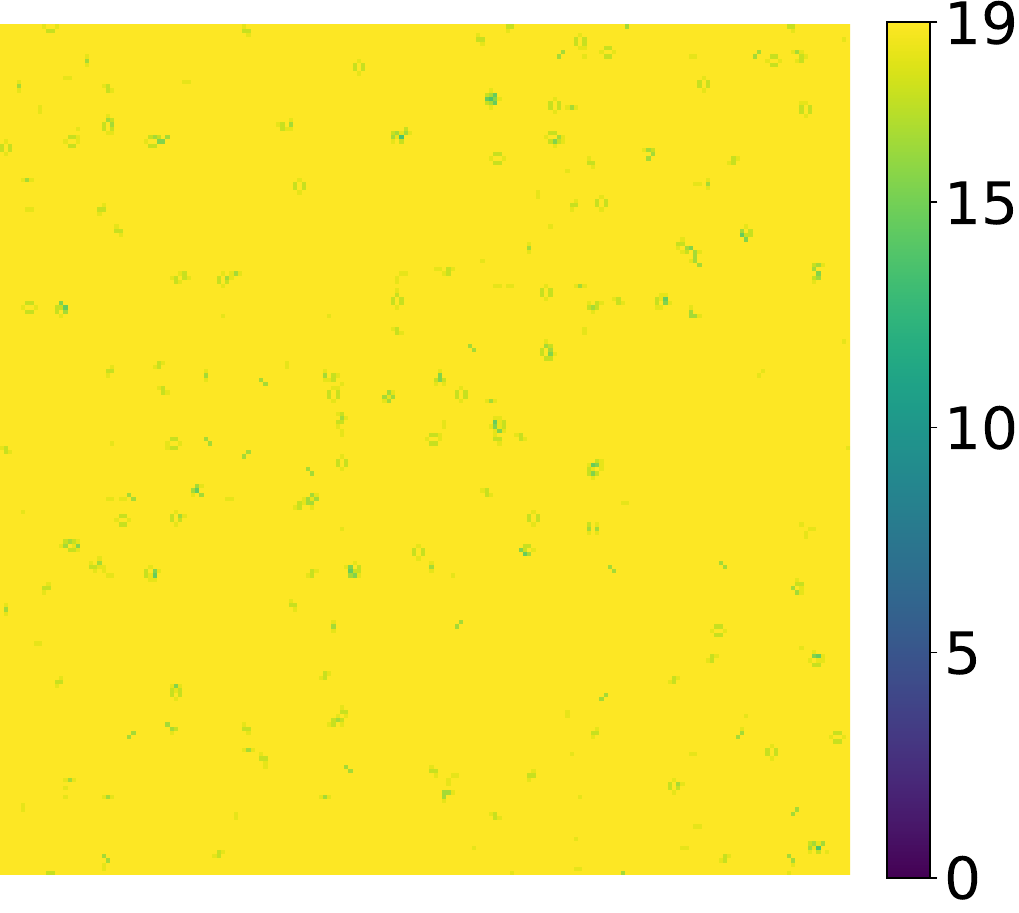}\\
			{\footnotesize t=100}
		\end{minipage}
		\hfill
		\begin{minipage}{0.188\linewidth}
			\centering
			\includegraphics[width=\linewidth]{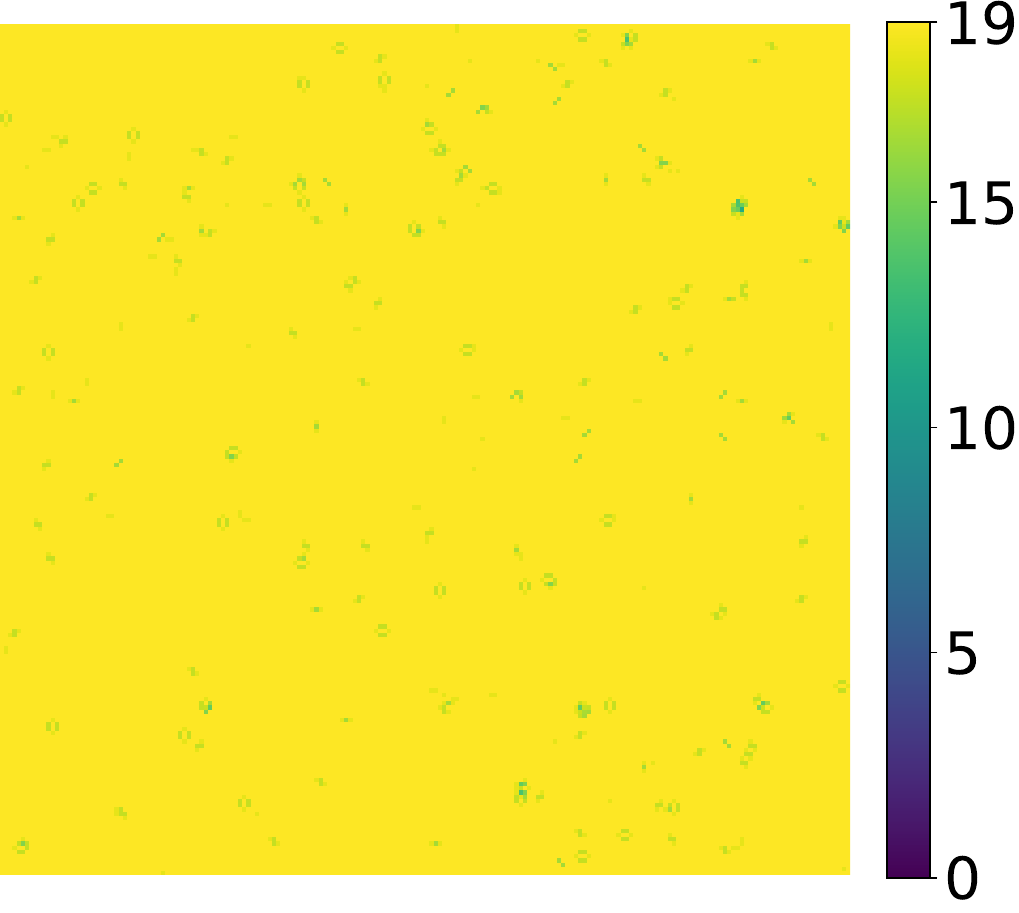}\\
			{\footnotesize t=1000}
		\end{minipage}
		\\
		[2mm]
	\centering
		{\footnotesize (d) r=4.6 (Payoff heatmaps)}
	\end{minipage}
	\caption{GRPO-GCC with half-and-half initialization on a $200 \times 200$ lattice. (a) $r=3.6$ cooperation dynamics and state snapshots. (b) $r=4.6$ cooperation dynamics and state snapshots. (c) $r=3.6$ payoff heatmaps at $t=0,1,10,100,1000$. (d) $r=4.6$ payoff heatmaps at $t=0,1,10,100,1000$. Cooperation expands robustly in both cases, with higher $r$ producing faster and more complete convergence.}
	\label{fig:GRPO-GCC_uDbC}
\end{figure*}

The results are summarized in Fig.~\ref{fig:GRPO-GCC_uDbC}. In subfigure (a) with $r=3.6$, the cooperation fraction increases rapidly and stabilizes at approximately $85\%$ after about 40 iterations. The evolution curve shows a monotonic increase in the number of cooperators. The spatial snapshots reveal that from the first iteration onward, cooperators and defectors mix across the lattice, dissolving the initial half-and-half boundary. The payoff heatmaps in subfigure (c) display heterogeneous regions where higher returns cluster around cooperative areas. In contrast, defectors experience comparatively lower payoffs, confirming that cooperation becomes dominant despite the polarized start. In subfigure (b) with $r=4.6$, cooperation rises more sharply and converges to about $98\%$ after roughly 60 iterations. The evolution curve highlights the accelerated dominance of cooperation. The state snapshots illustrate that by $t=100$, defectors persist only as scattered individuals dispersed across the lattice. The payoff heatmaps in subfigure (d) corroborate this observation. They show widespread high-payoff regions consistent with near-complete cooperation, while defectors remain confined to isolated pockets of low returns. These findings demonstrate that GRPO-GCC sustains cooperative emergence under half-and-half initialization. Higher $r$ values drive faster convergence and a more complete dominance of cooperation.

\subsection{GRPO-GCC with bernoulli random initialization}
\label{exp_b}

To further investigate the adaptability of GRPO-GCC, we consider a Bernoulli random initialization setting. Agents are arranged on a $200 \times 200$ lattice, and each agent is independently assigned as a cooperator or defector with equal probability $p=0.5$. The experiment is conducted for $r=3.6$ and $r=4.6$. For each case, the top panel shows the temporal evolution of cooperation and defection fractions, while the bottom panel presents spatial snapshots at $t=0,1,10,100,1000$. In addition, payoff heatmaps are provided at the same timesteps, with color ranging from yellow (high payoff) through green and blue to purple (low payoff).

\begin{figure*}[htbp!]
	\begin{minipage}{0.45\linewidth}
		\begin{minipage}{\linewidth}
			\centering
			\includegraphics[width=\linewidth]{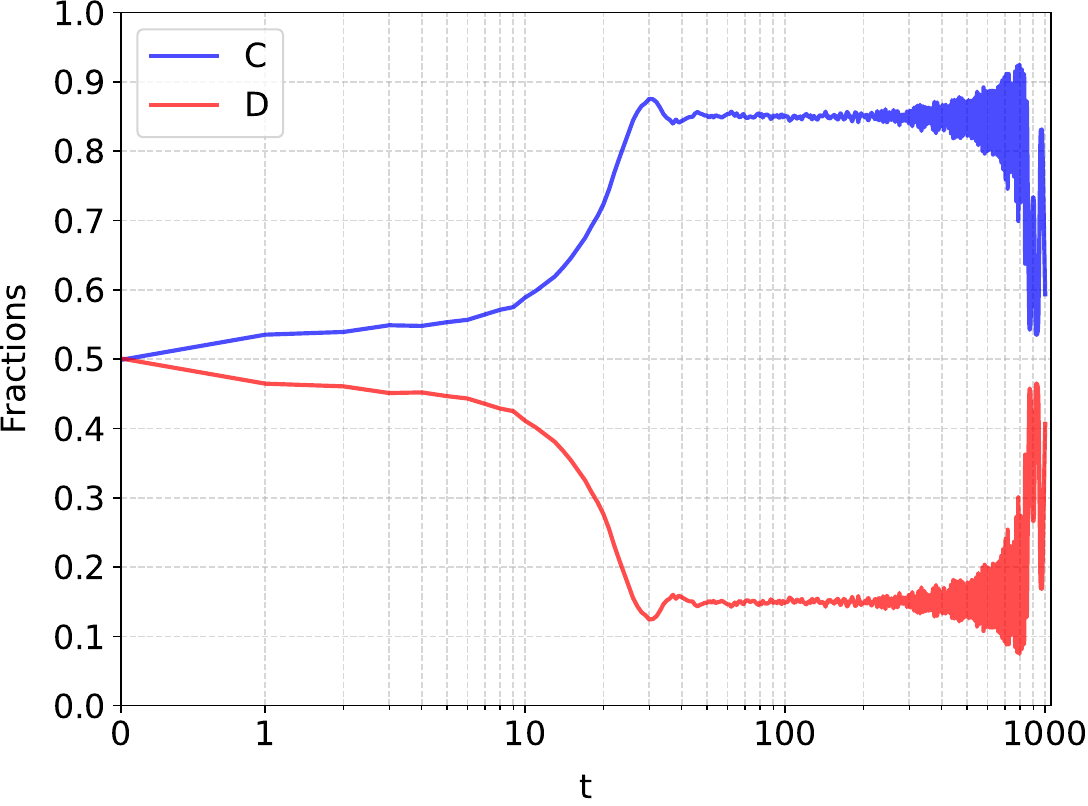}\\
		\end{minipage}
		\vspace{2mm}
		\\
		\begin{minipage}{0.188\linewidth}
			\centering
			\fbox{\includegraphics[width=\linewidth]{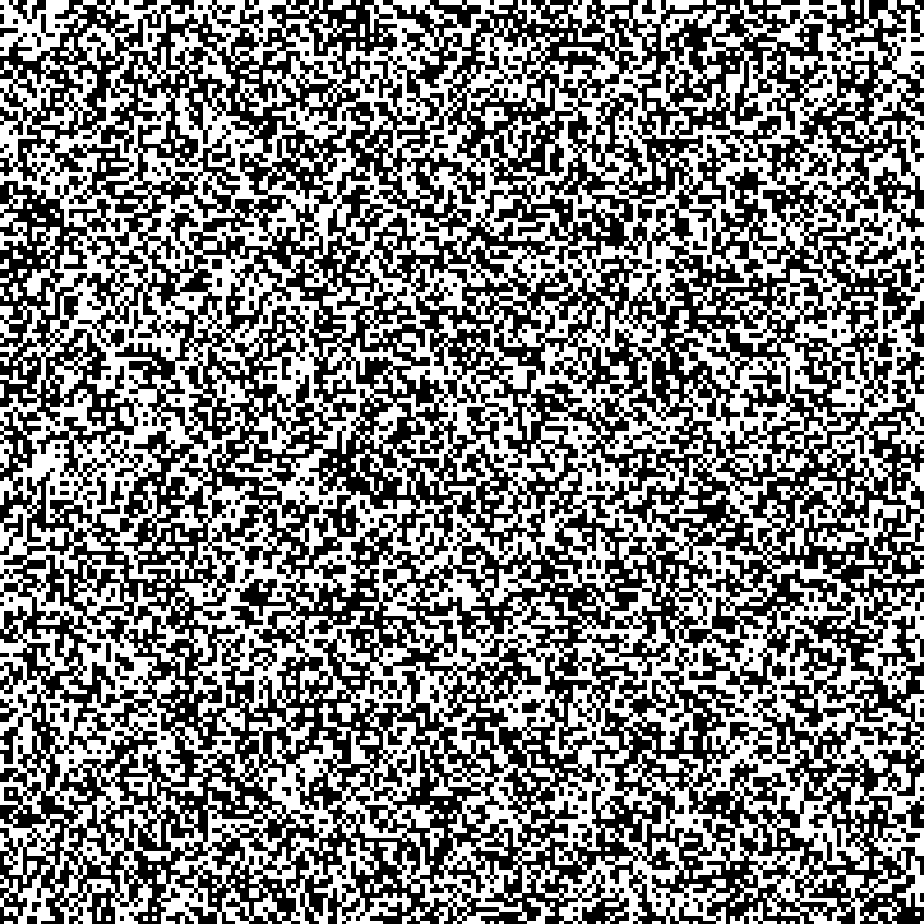}}\\
			{\footnotesize t=0}
		\end{minipage}
		\begin{minipage}{0.188\linewidth}
			\centering
			\fbox{\includegraphics[width=\linewidth]{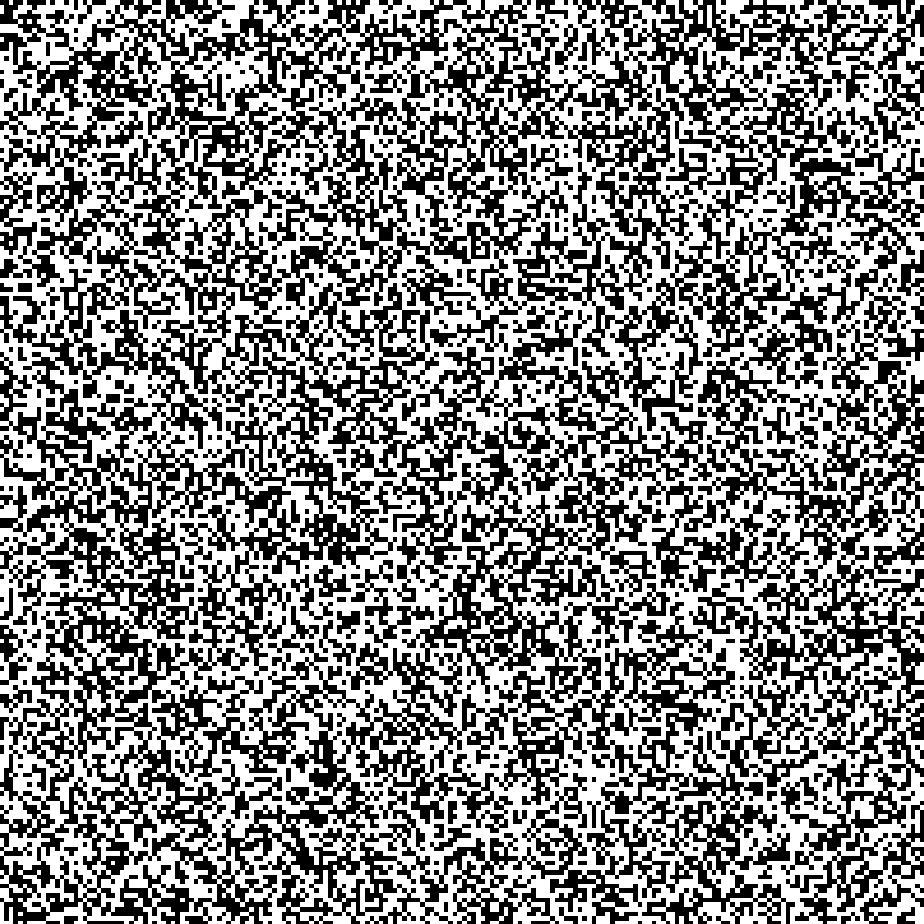}}\\
			{\footnotesize t=1}
		\end{minipage}
		\begin{minipage}{0.188\linewidth}
			\centering
			\fbox{\includegraphics[width=\linewidth]{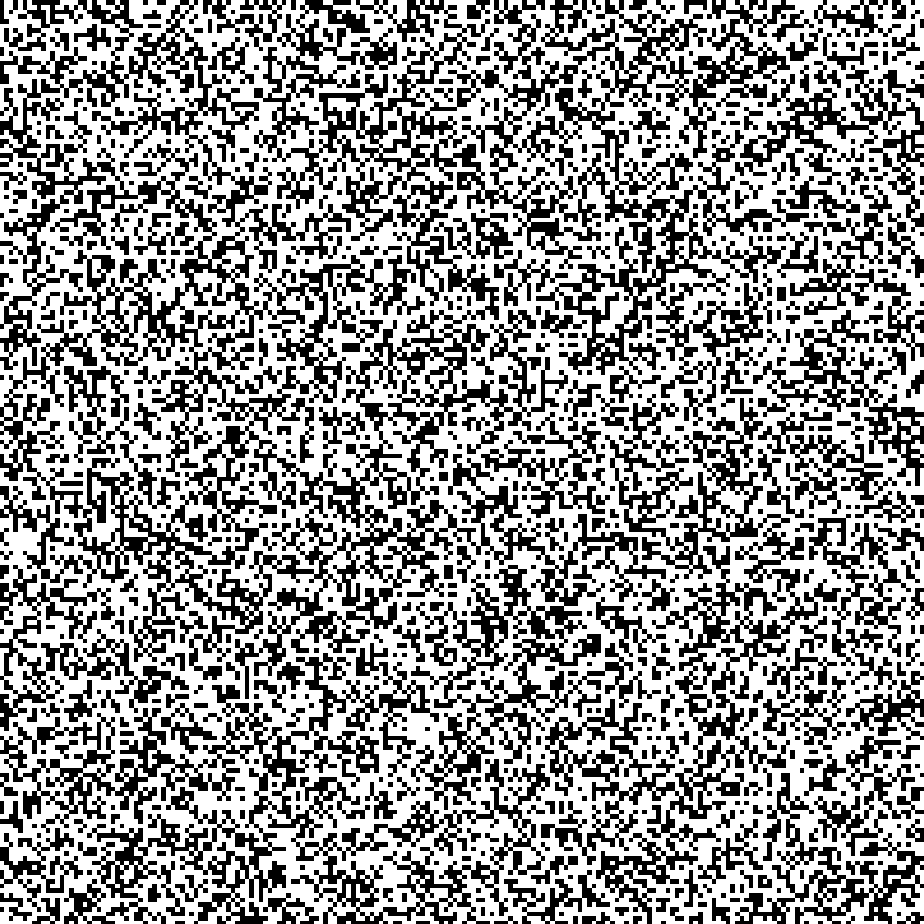}}\\
			{\footnotesize t=10}
		\end{minipage}
		\begin{minipage}{0.188\linewidth}
			\centering
			\fbox{\includegraphics[width=\linewidth]{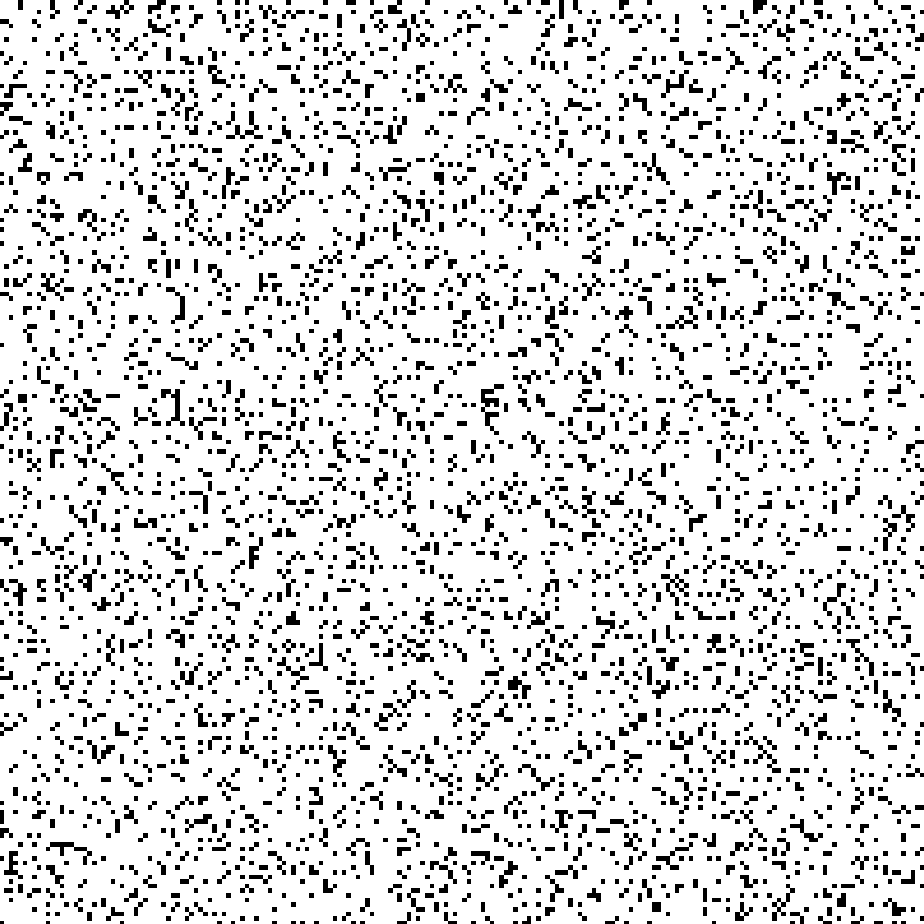}}\\
			{\footnotesize t=100}
		\end{minipage}
		\begin{minipage}{0.188\linewidth}
			\centering
			\fbox{\includegraphics[width=\linewidth]{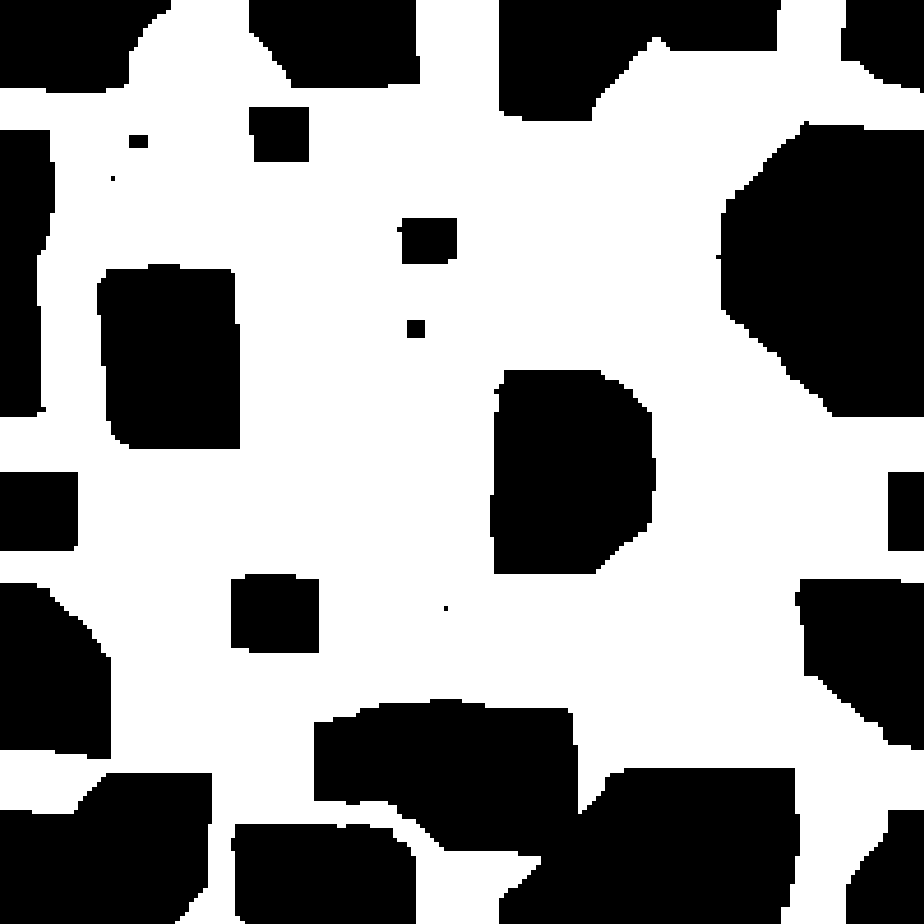}}\\
			{\footnotesize t=1000}
		\end{minipage}
		\\
[2mm]
\centering
		{\footnotesize (a) r=3.6}
	\end{minipage}
	\hfill
	\begin{minipage}{0.45\linewidth}
		\begin{minipage}{\linewidth}
			\centering
			\includegraphics[width=\linewidth]{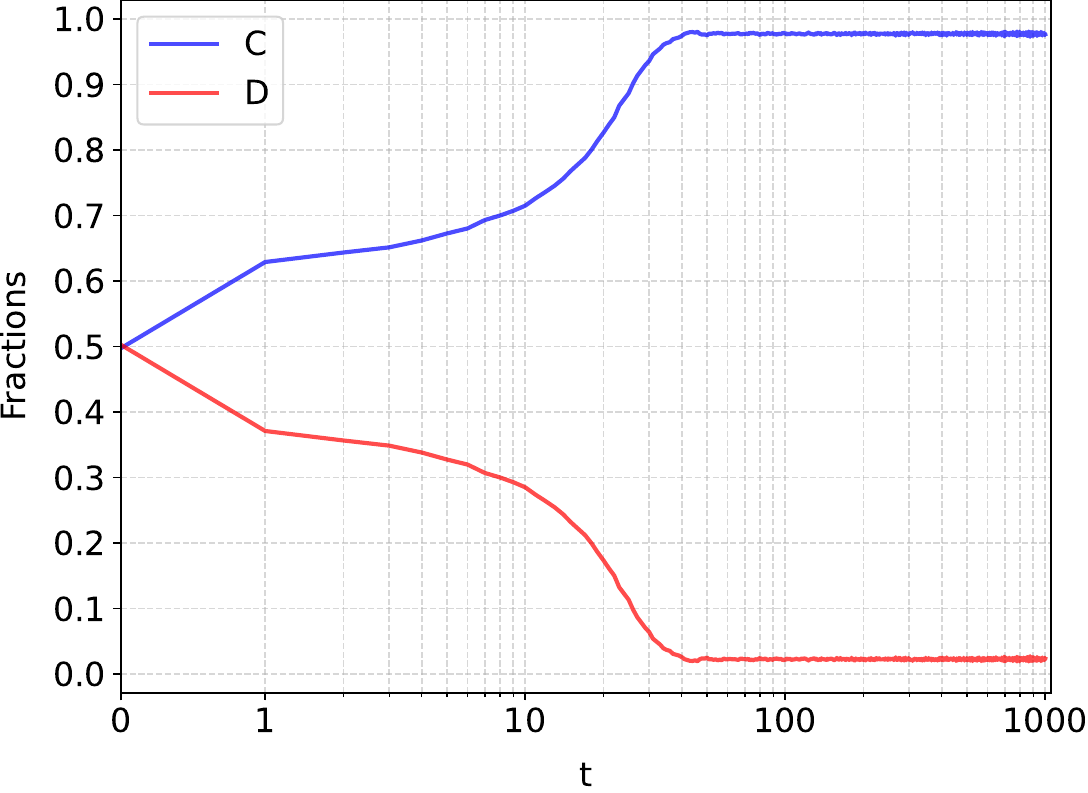}\\
		\end{minipage}
		\vspace{2mm}
		\\
		\begin{minipage}{0.188\linewidth}
			\centering
			\fbox{\includegraphics[width=\linewidth]{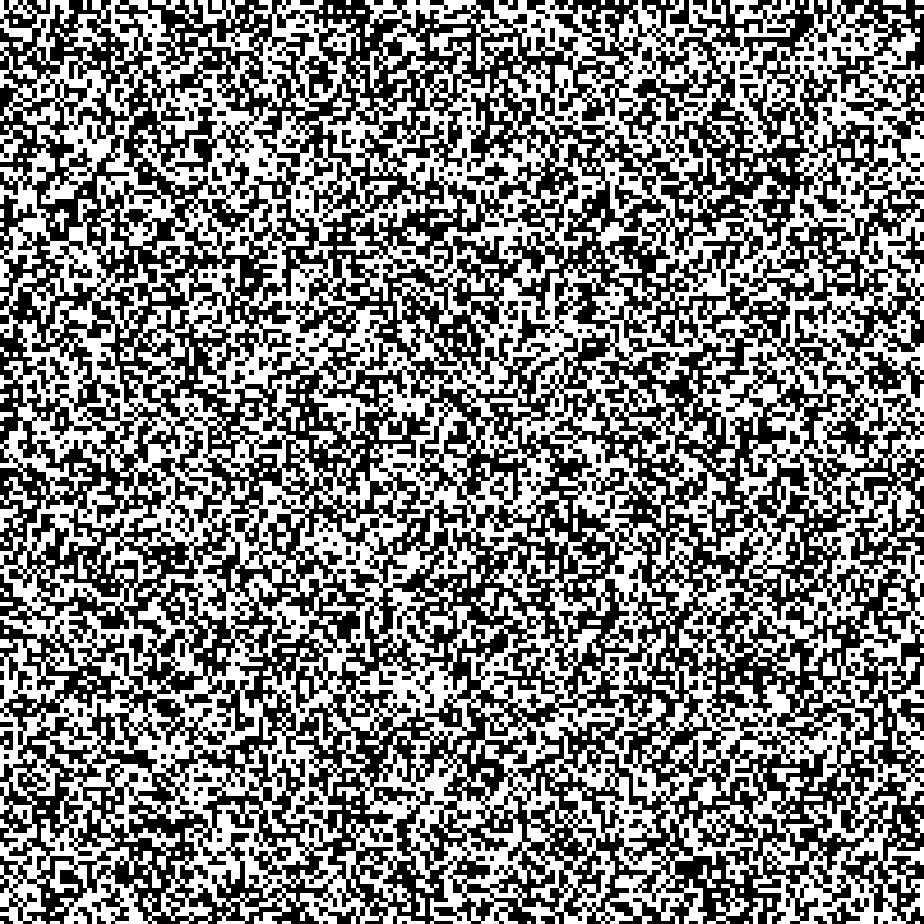}}\\
			{\footnotesize t=0}
		\end{minipage}
		\begin{minipage}{0.188\linewidth}
			\centering
			\fbox{\includegraphics[width=\linewidth]{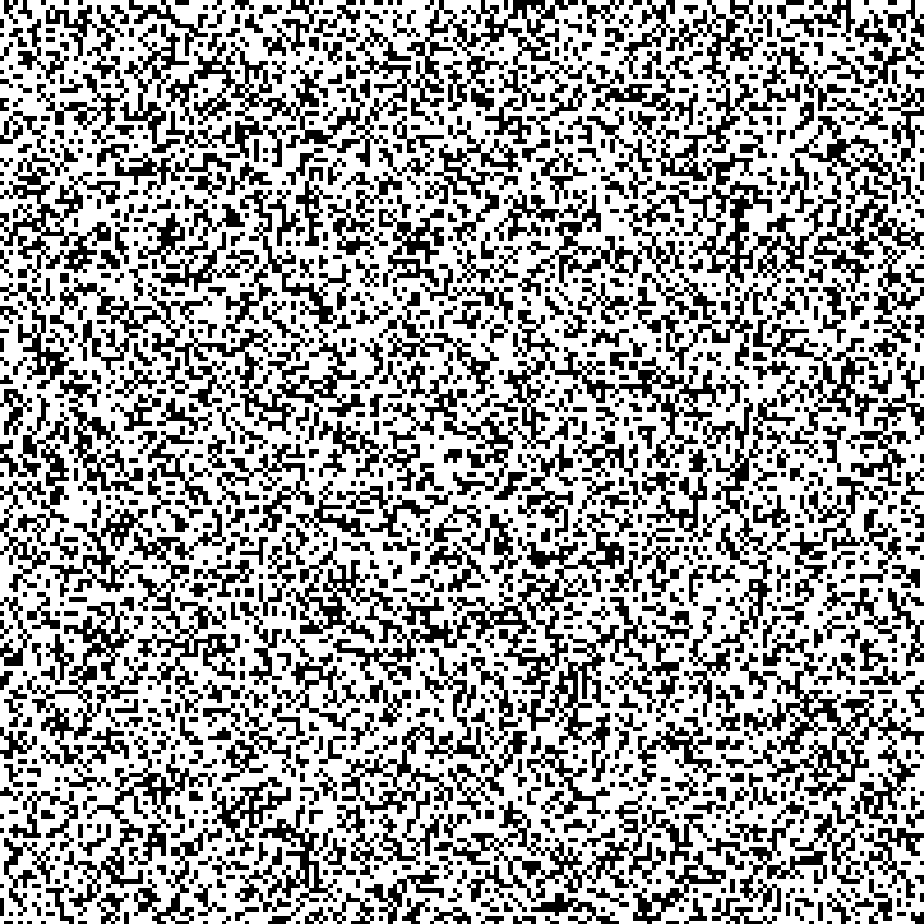}}\\
			{\footnotesize t=1}
		\end{minipage}
		\begin{minipage}{0.188\linewidth}
			\centering
			\fbox{\includegraphics[width=\linewidth]{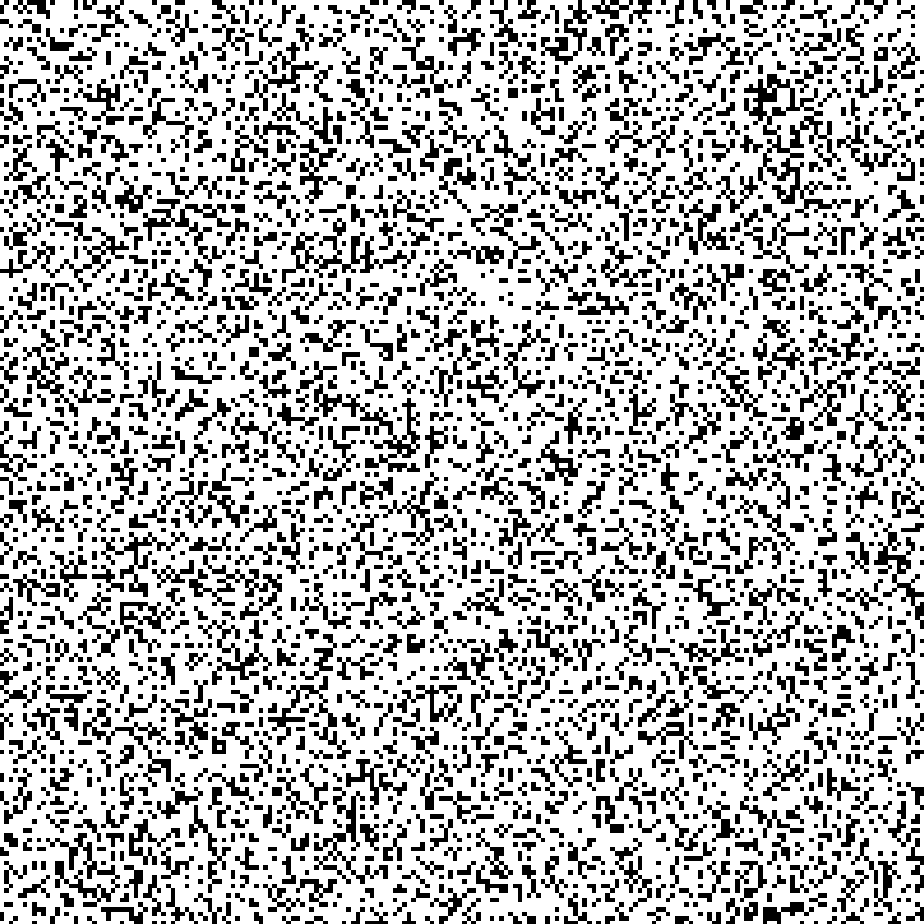}}\\
			{\footnotesize t=10}
		\end{minipage}
		\begin{minipage}{0.188\linewidth}
			\centering
			\fbox{\includegraphics[width=\linewidth]{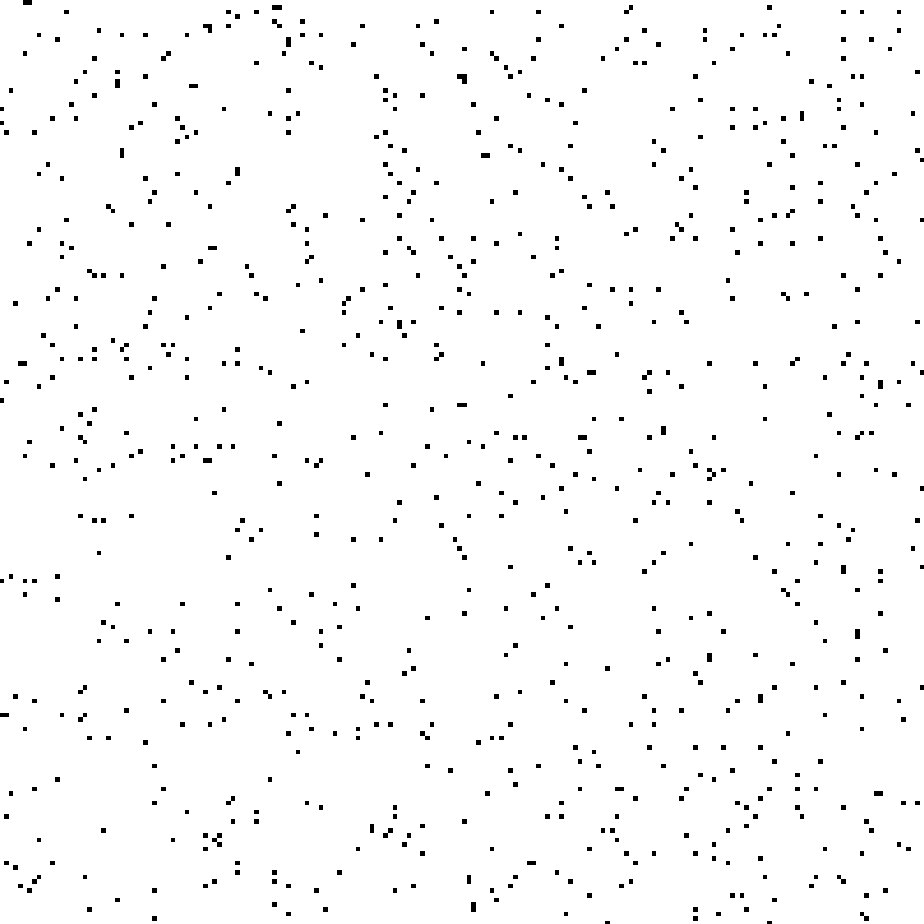}}\\
			{\footnotesize t=100}
		\end{minipage}
		\begin{minipage}{0.188\linewidth}
			\centering
			\fbox{\includegraphics[width=\linewidth]{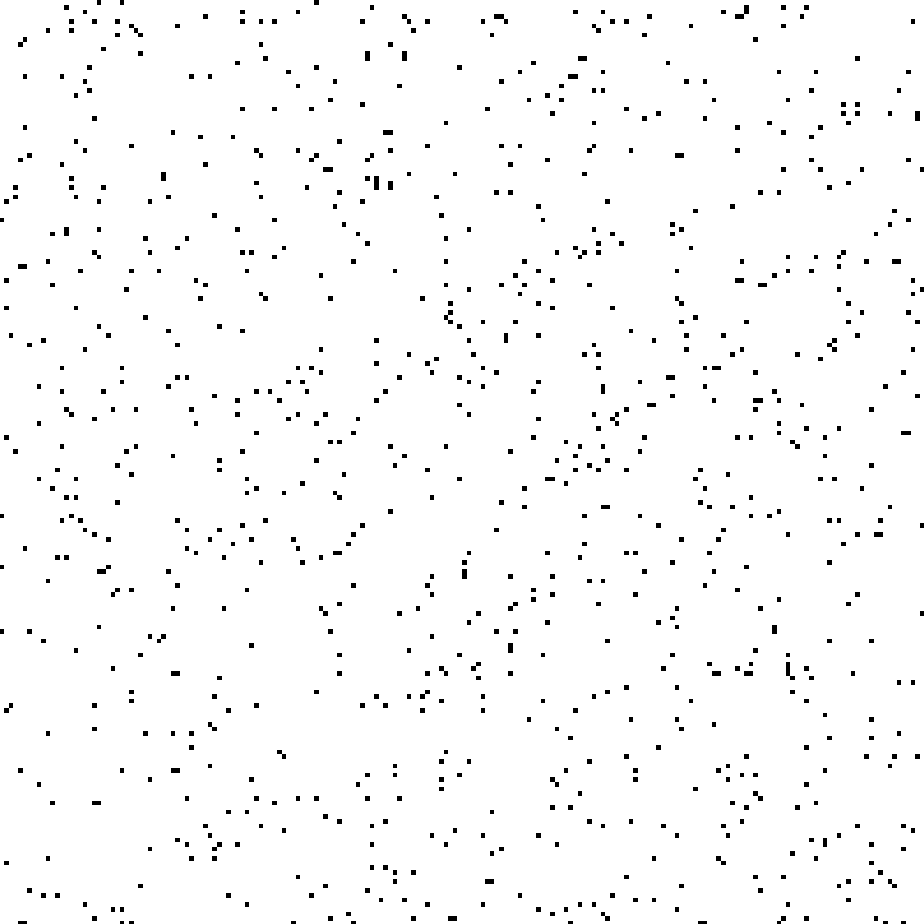}}\\
			{\footnotesize t=1000}
		\end{minipage}
		\\
[2mm]
\centering
		{\footnotesize (b) r=4.6}
	\end{minipage}
	\\
	[2mm]
	\begin{minipage}{\linewidth}
		\begin{minipage}{0.188\linewidth}
			\centering
			\includegraphics[width=\linewidth]{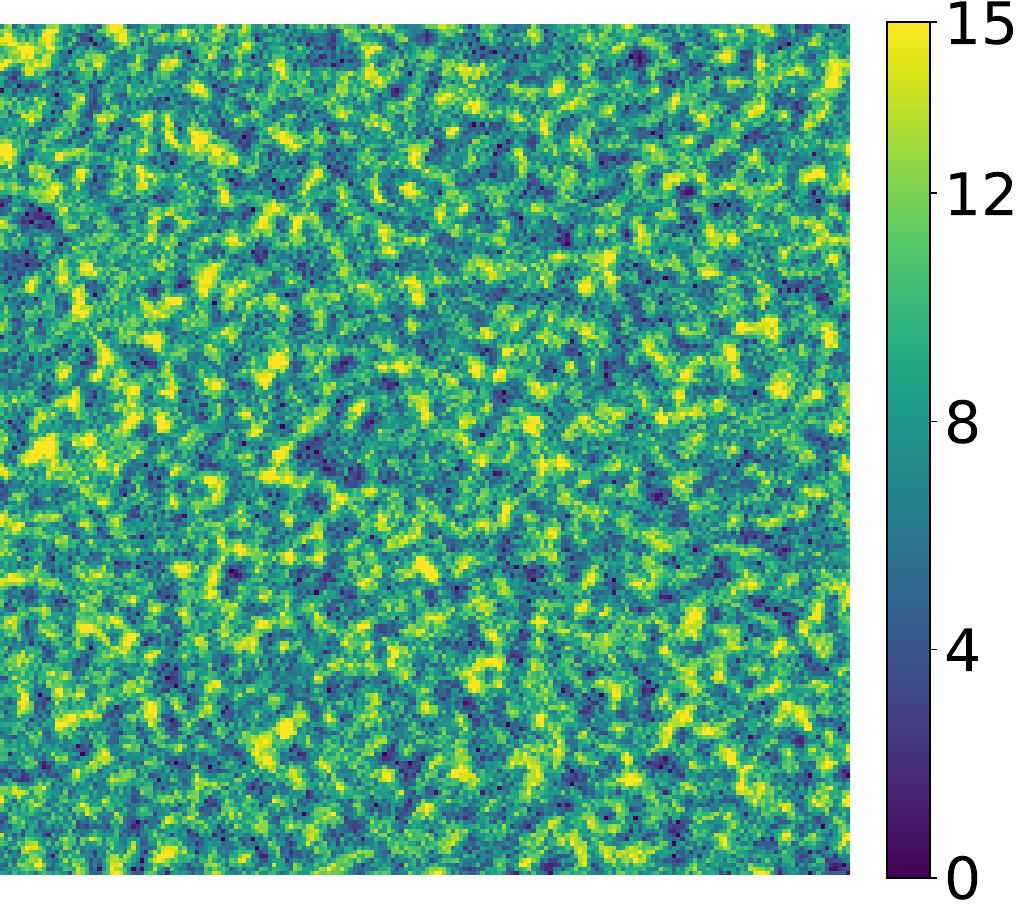}\\
			{\footnotesize t=0}
		\end{minipage}
		\hfill
		\begin{minipage}{0.188\linewidth}
			\centering
			\includegraphics[width=\linewidth]{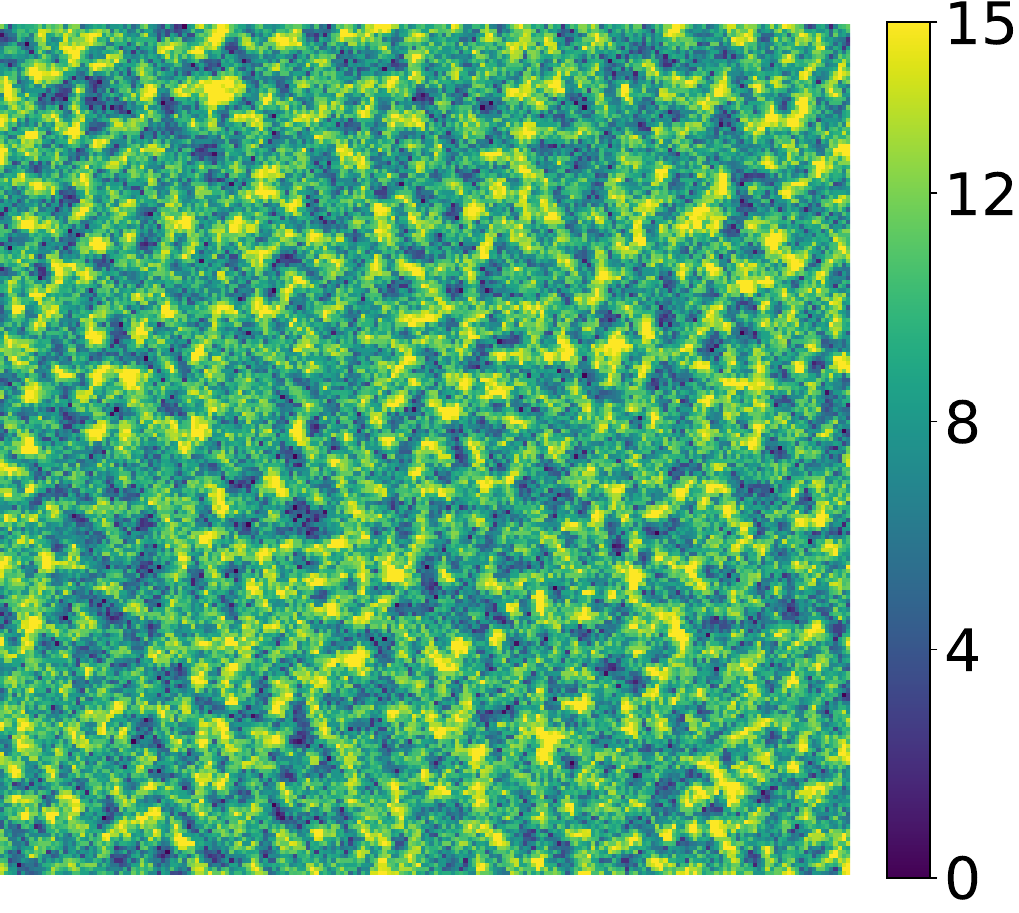}\\
			{\footnotesize t=1}
		\end{minipage}
		\hfill
		\begin{minipage}{0.188\linewidth}
			\centering
			\includegraphics[width=\linewidth]{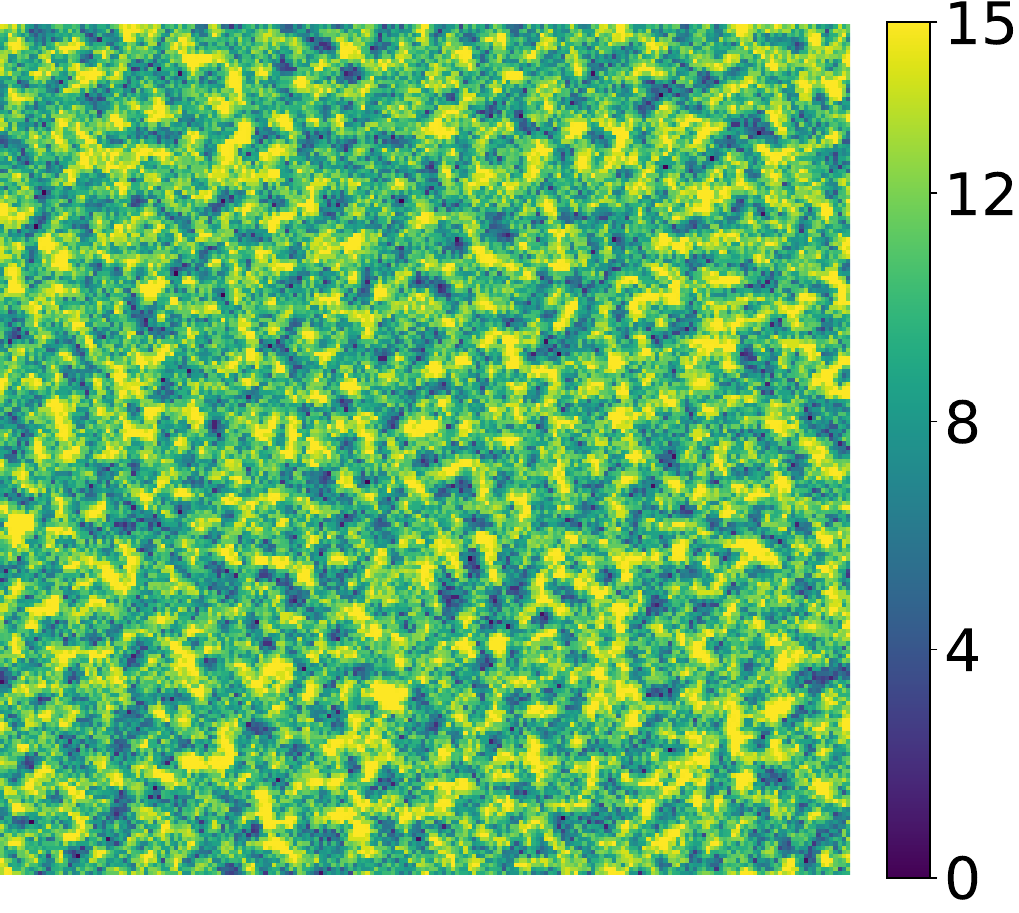}\\
			{\footnotesize t=10}
		\end{minipage}
		\hfill
		\begin{minipage}{0.188\linewidth}
			\centering
			\includegraphics[width=\linewidth]{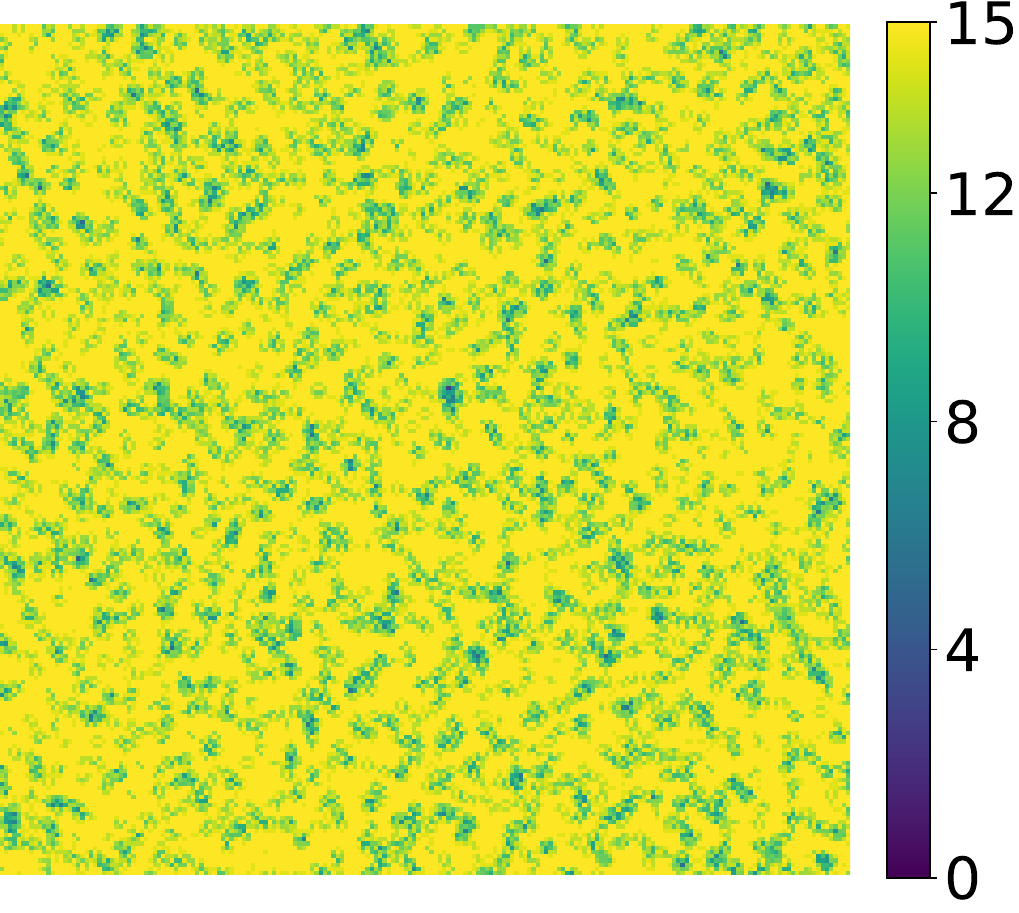}\\
			{\footnotesize t=100}
		\end{minipage}
		\hfill
		\begin{minipage}{0.188\linewidth}
			\centering
			\includegraphics[width=\linewidth]{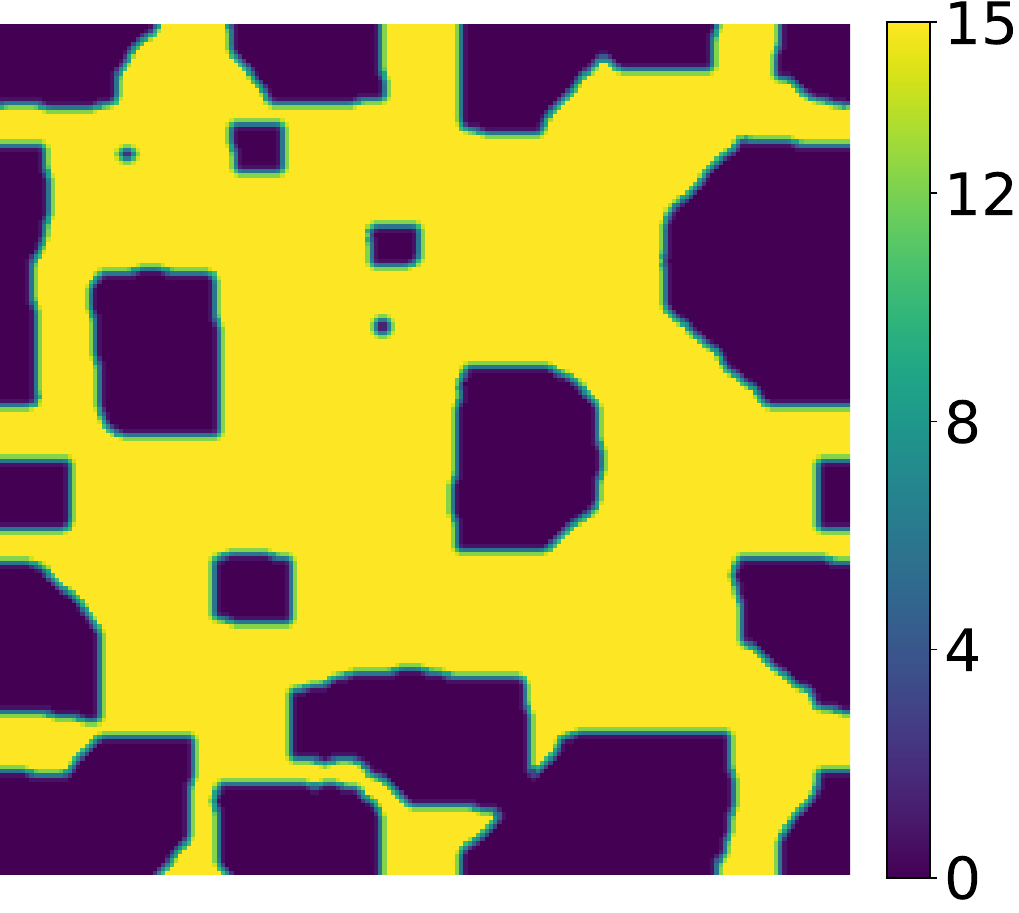}\\
			{\footnotesize t=1000}
		\end{minipage}
		\\
[2mm]
\centering
		{\footnotesize (c) r=3.6 (Payoff heatmaps)}
	\end{minipage}
	\\
	[2mm]
	\begin{minipage}{\linewidth}
		\begin{minipage}{0.188\linewidth}
			\centering
			\includegraphics[width=\linewidth]{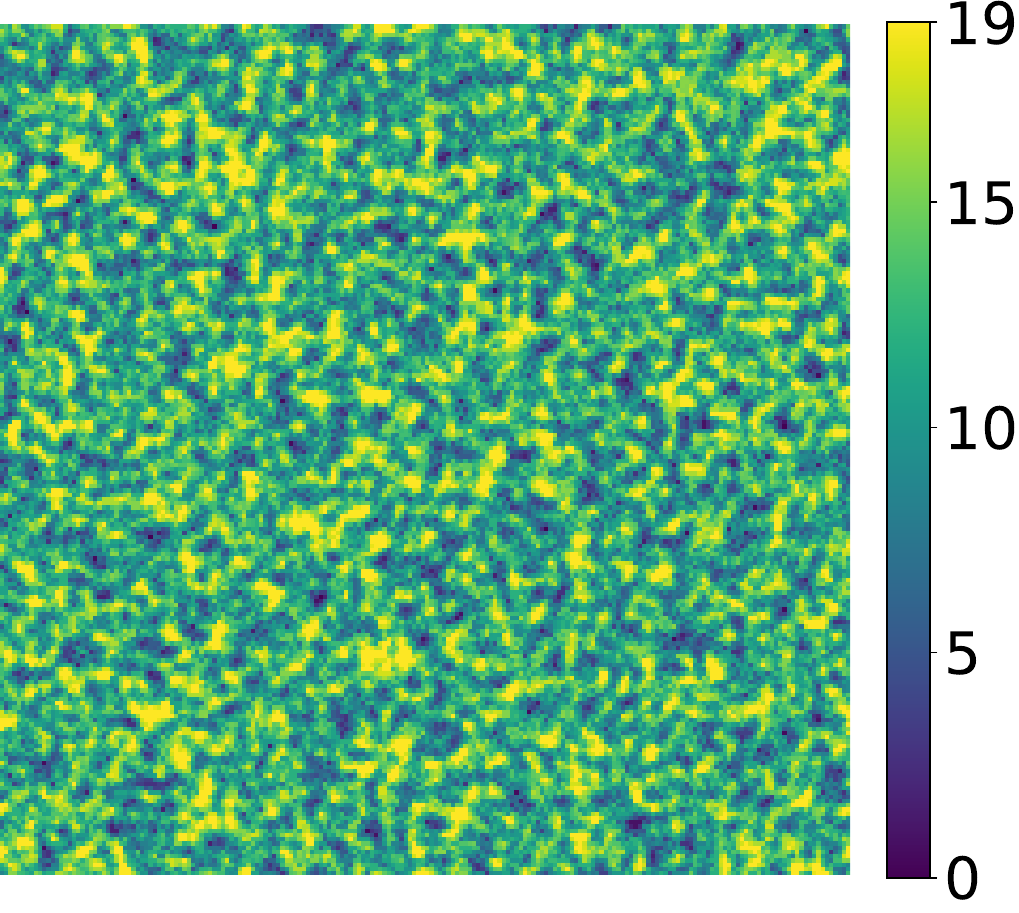}\\
			{\footnotesize t=0}
		\end{minipage}
		\hfill
		\begin{minipage}{0.188\linewidth}
			\centering
			\includegraphics[width=\linewidth]{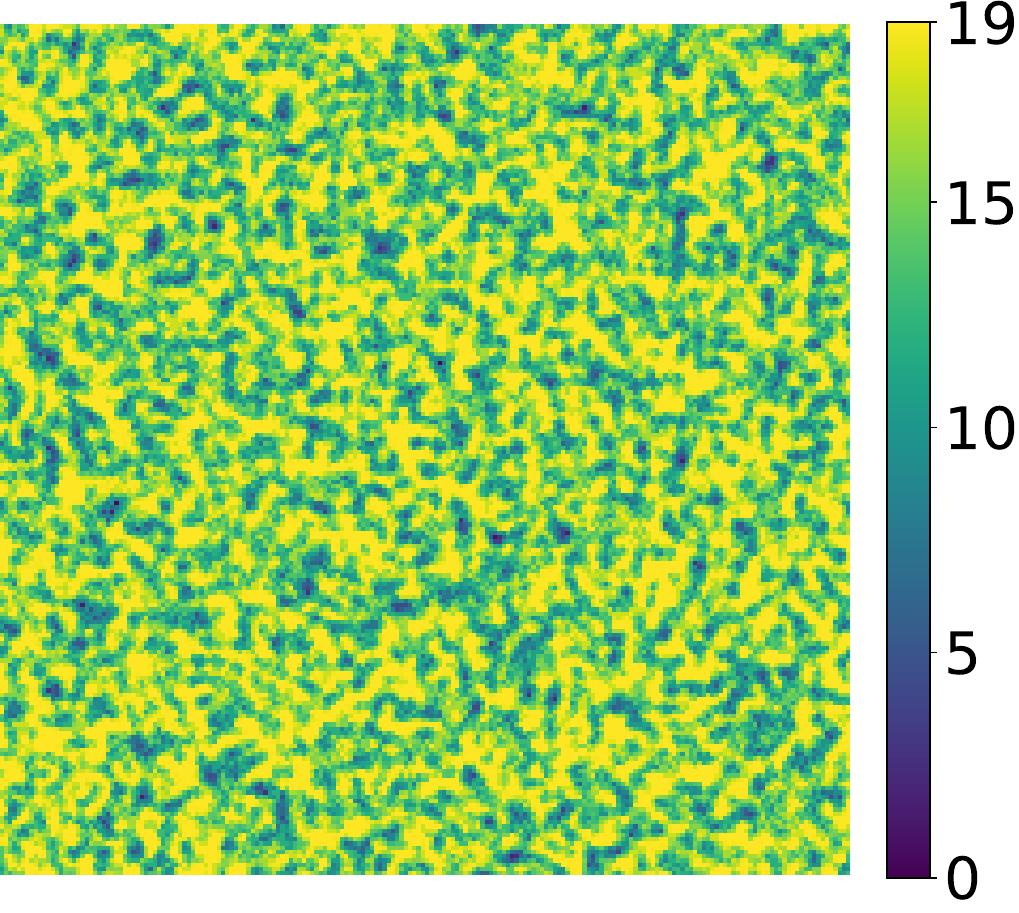}\\
			{\footnotesize t=1}
		\end{minipage}
		\hfill
		\begin{minipage}{0.188\linewidth}
			\centering
			\includegraphics[width=\linewidth]{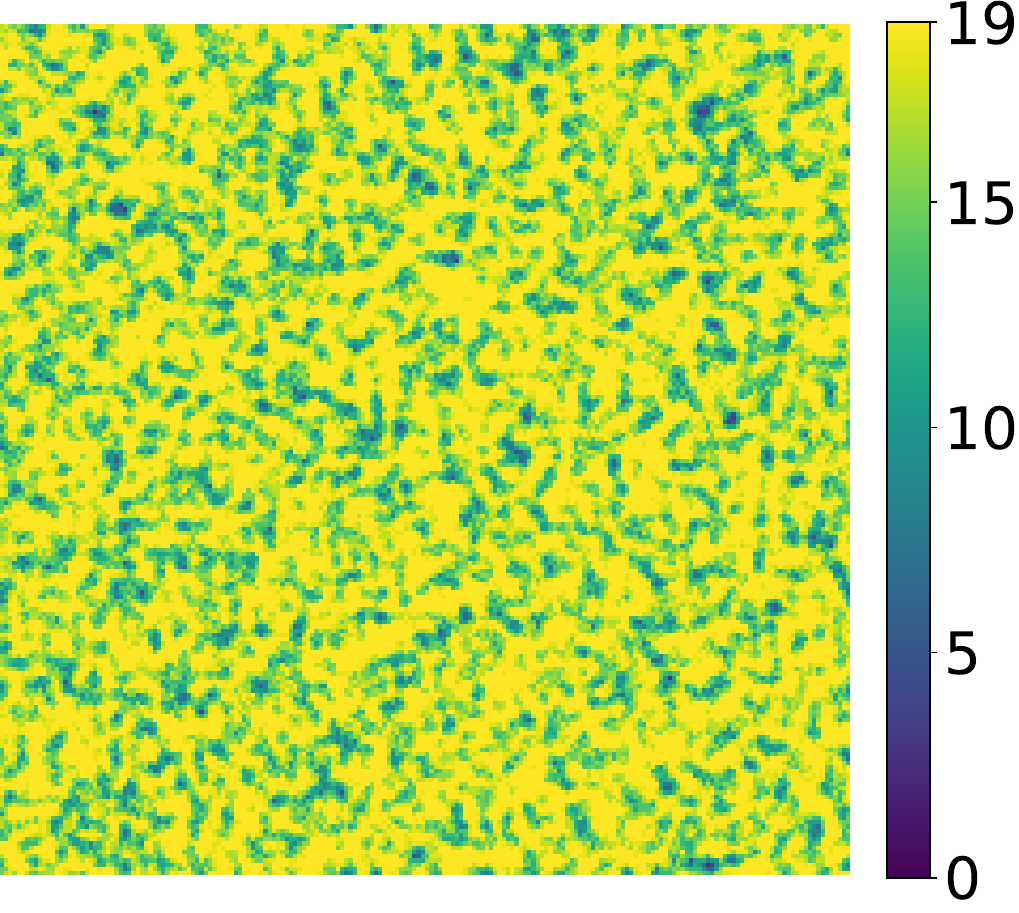}\\
			{\footnotesize t=10}
		\end{minipage}
		\hfill
		\begin{minipage}{0.188\linewidth}
			\centering
			\includegraphics[width=\linewidth]{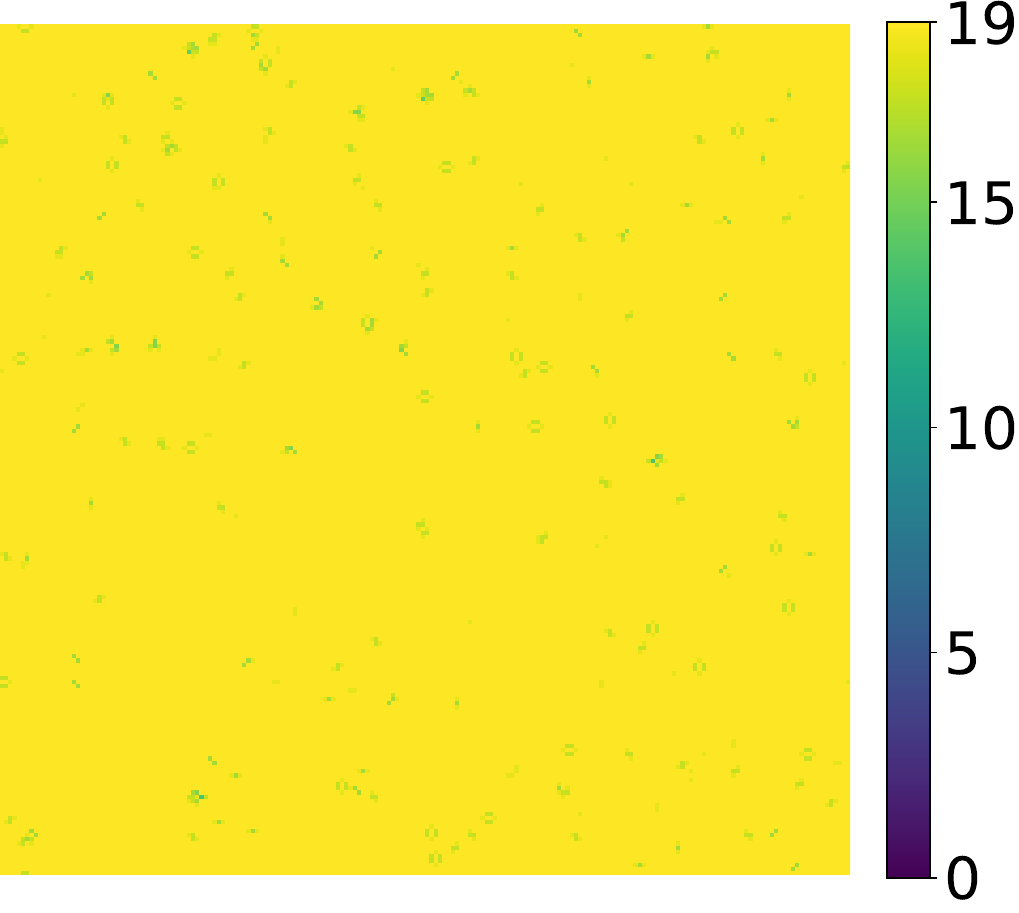}\\
			{\footnotesize t=100}
		\end{minipage}
		\hfill
		\begin{minipage}{0.188\linewidth}
			\centering
			\includegraphics[width=\linewidth]{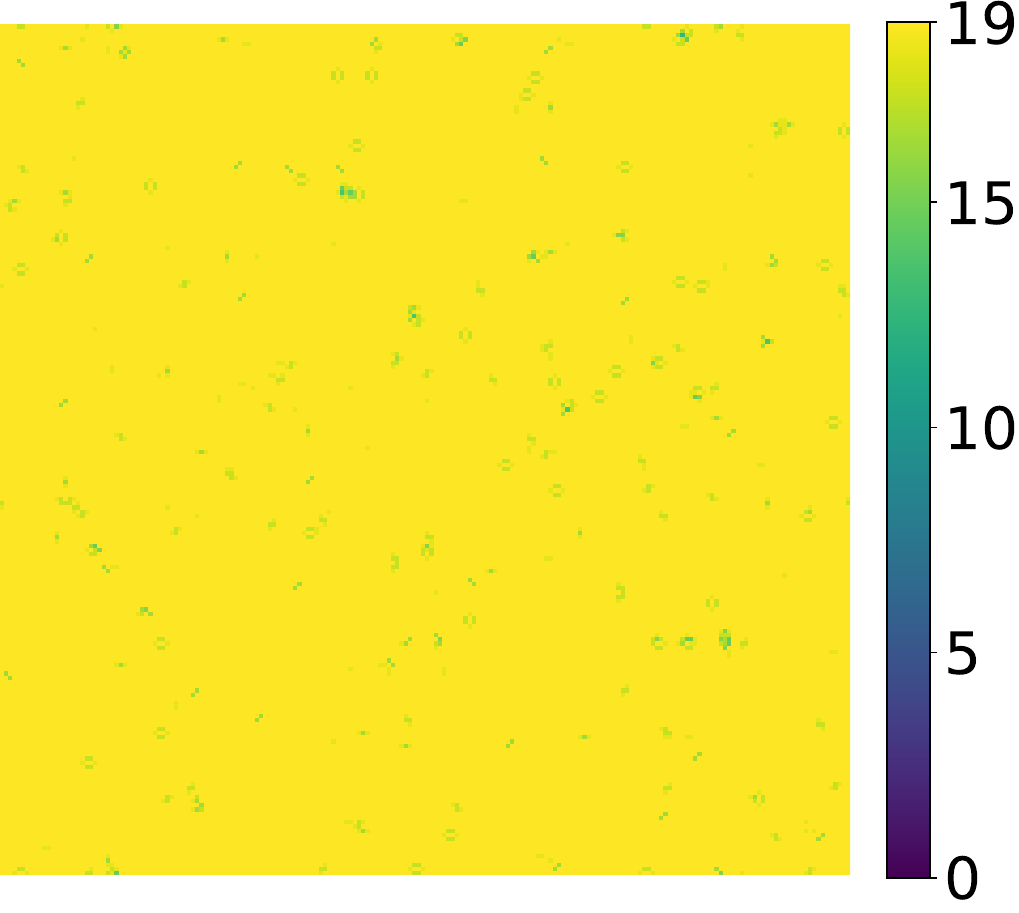}\\
			{\footnotesize t=1000}
		\end{minipage}
		\\
[2mm]
\centering
		{\footnotesize (d) r=4.6 (Payoff heatmaps)}
	\end{minipage}
	\caption{GRPO-GCC with Bernoulli random initialization on a $200 \times 200$ lattice. (a) $r=3.6$ cooperation dynamics and state snapshots. (b) $r=4.6$ cooperation dynamics and state snapshots. (c) $r=3.6$ payoff heatmaps at $t=0,1,10,100,1000$. (d) $r=4.6$ payoff heatmaps at $t=0,1,10,100,1000$. At $r=3.6$ cooperation dominates but defectors survive as clusters due to the GCC constraint, whereas at $r=4.6$ cooperation rapidly converges to near-complete dominance.}
	\label{fig:GRPO-GCC_Bernoulli}
\end{figure*}

The results are presented in Fig.~\ref{fig:GRPO-GCC_Bernoulli}. In subfigure (a) with $r=3.6$, the cooperation fraction initially rises but later exhibits oscillations around an intermediate level, with cooperators consistently outnumbering defectors. The oscillatory behavior reflects the influence of the GCC mechanism: when global cooperation becomes too high, the self-limiting term reduces incentives, allowing defectors to reemerge. The spatial snapshots confirm this interpretation, showing that defectors gradually form compact clusters despite their minority status. This cluster formation under RL dynamics is unusual and highlights the distinctive impact of GCC. The corresponding payoff heatmaps in subfigure (c) display heterogeneous distributions where clustered defectors persist as localized low-payoff zones, while cooperators dominate high-payoff areas. In subfigure (b) with $r=4.6$, the cooperation fraction rises rapidly and stabilizes near $98\%$ after about 50 iterations. The snapshots show that by $t=100$, defectors are scattered sparsely across the lattice, unable to sustain clusters. The payoff heatmap in subfigure (d) corroborates this observation. It shows widespread high-reward regions consistent with almost complete cooperation. Defectors are relegated to isolated pockets of low returns. Overall, these results demonstrate that under random initialization, GRPO-GCC maintains cooperative dominance. At moderate $r$, GCC introduces dynamic balance that permits defectors to cluster, while at higher $r$, cooperation prevails almost universally.

\subsection{GRPO-GCC with all-defectors initialization}
\label{exp_ad}

We further test GRPO-GCC under the most unfavorable starting condition, where all agents in a $200 \times 200$ lattice are initialized as defectors. The experiment is conducted for $r=3.6$ and $r=4.6$. For each case, the top panel shows the temporal evolution of cooperation and defection fractions, while the bottom panel presents spatial snapshots at $t=0,1,10,100,1000$. Payoff heatmaps are also provided for the same timesteps, with color ranging from yellow (high payoff) through green and blue to purple (low payoff).

\begin{figure*}[htbp!]
	\begin{minipage}{0.45\linewidth}
		\begin{minipage}{\linewidth}
			\centering
			\includegraphics[width=\linewidth]{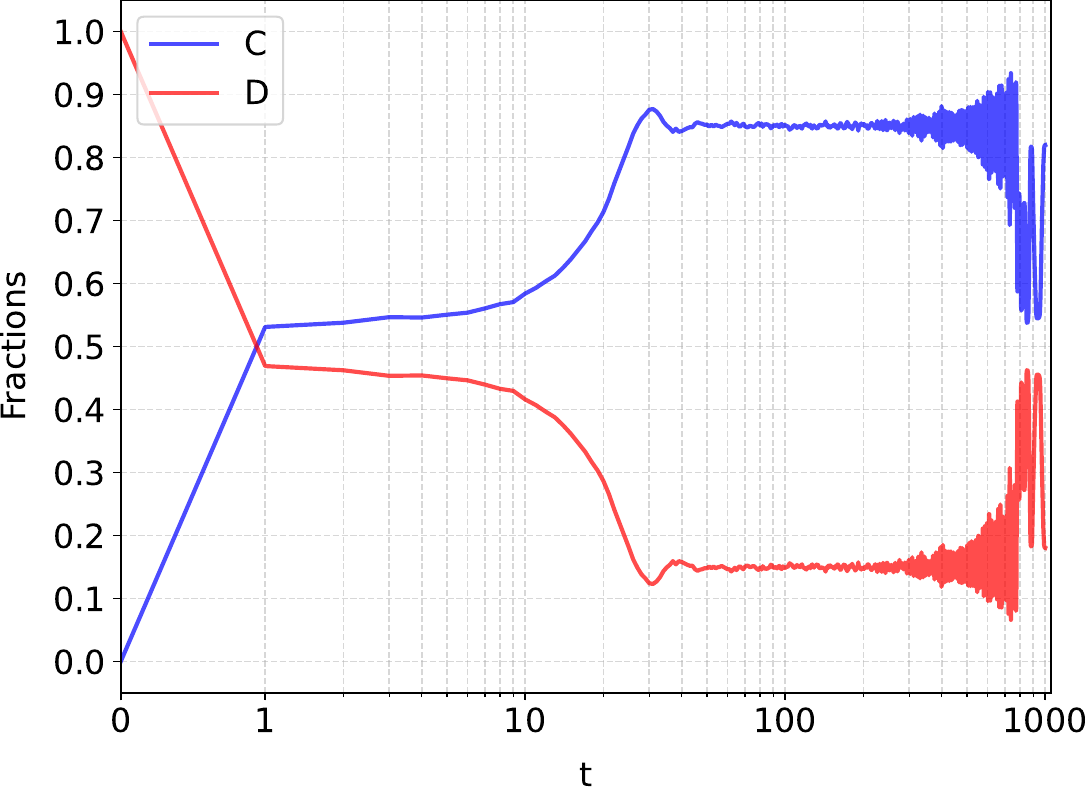}\\
		\end{minipage}
		\vspace{2mm}
		\\
		\begin{minipage}{0.188\linewidth}
			\centering
			\fbox{\includegraphics[width=\linewidth]{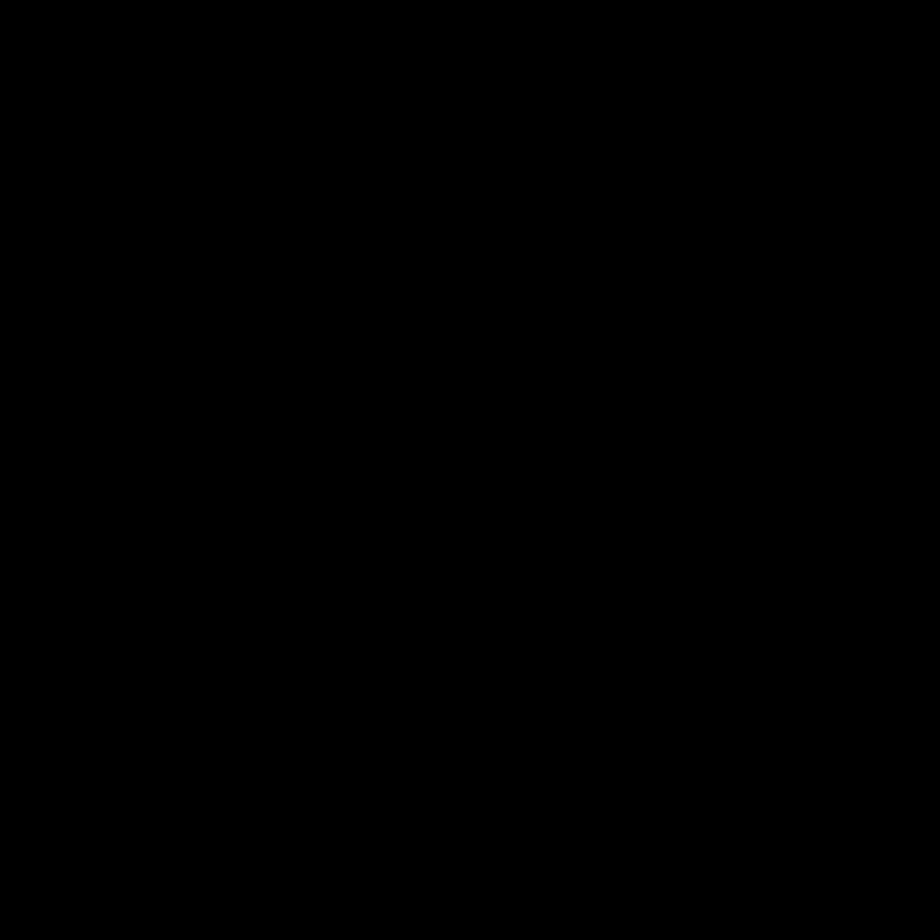}}\\
			{\footnotesize t=0}
		\end{minipage}
		\begin{minipage}{0.188\linewidth}
			\centering
			\fbox{\includegraphics[width=\linewidth]{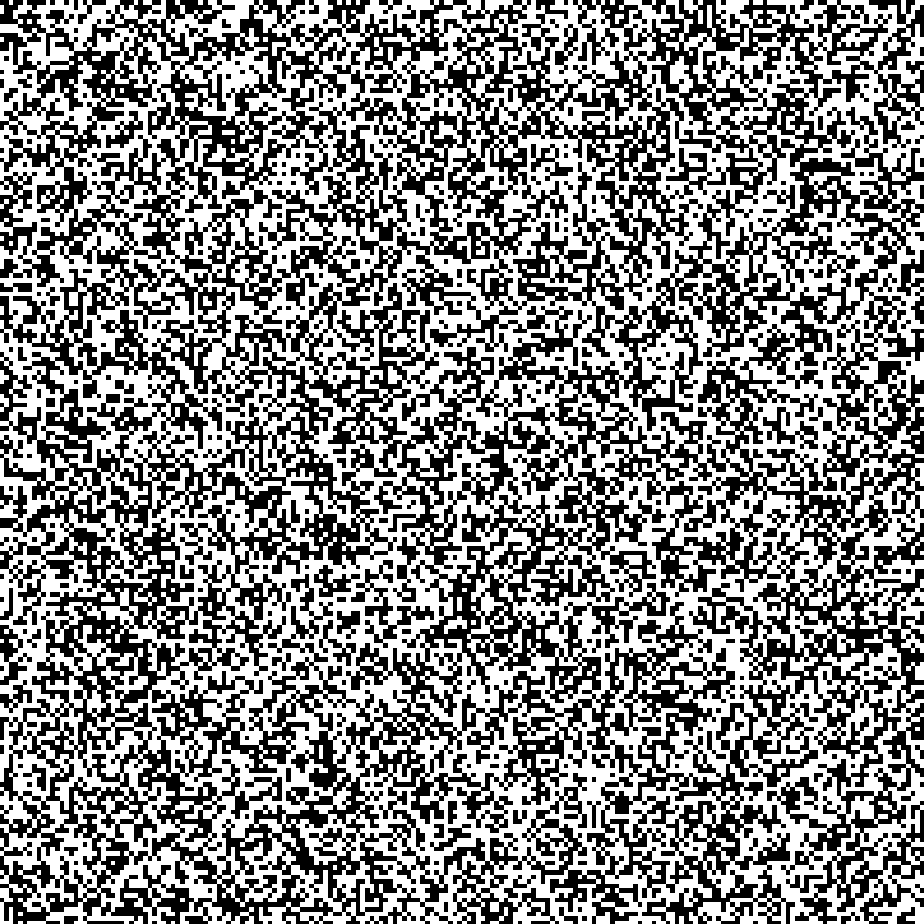}}\\
			{\footnotesize t=1}
		\end{minipage}
		\begin{minipage}{0.188\linewidth}
			\centering
			\fbox{\includegraphics[width=\linewidth]{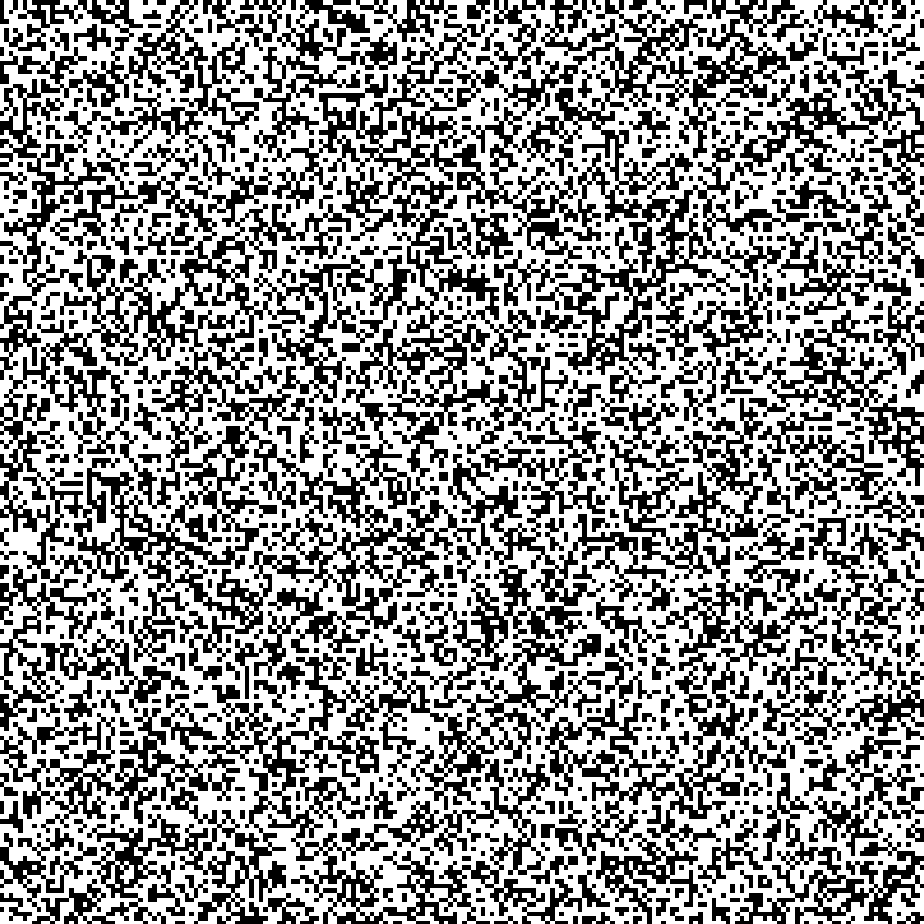}}\\
			{\footnotesize t=10}
		\end{minipage}
		\begin{minipage}{0.188\linewidth}
			\centering
			\fbox{\includegraphics[width=\linewidth]{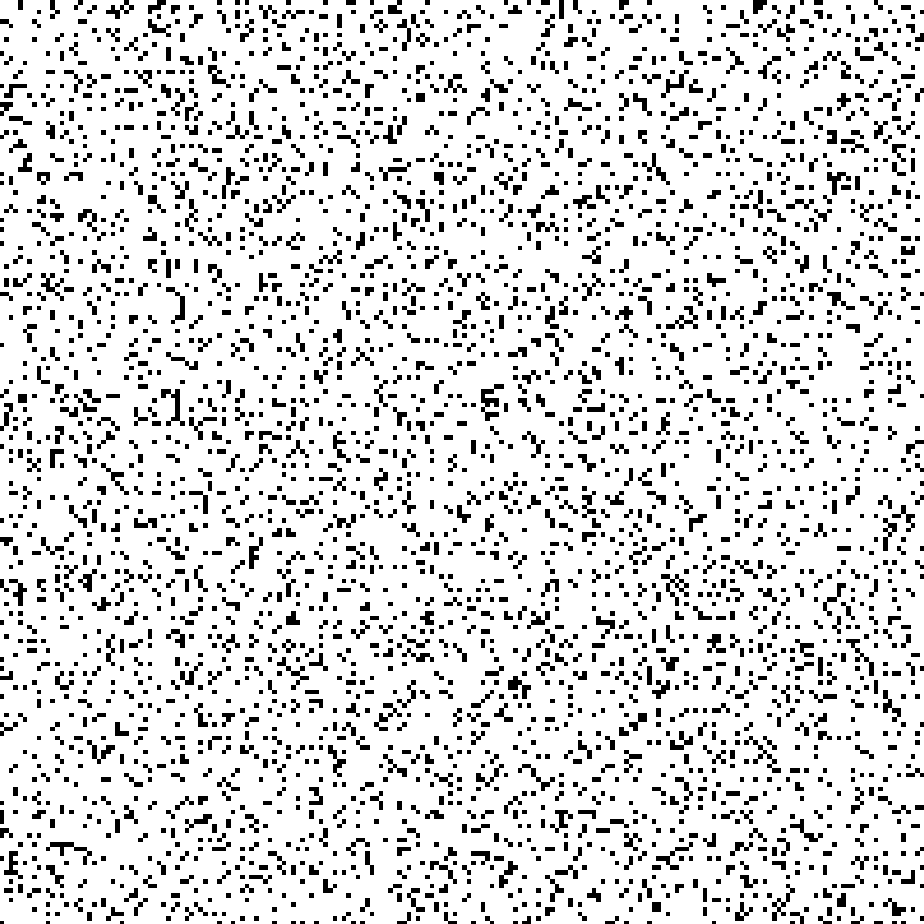}}\\
			{\footnotesize t=100}
		\end{minipage}
		\begin{minipage}{0.188\linewidth}
			\centering
			\fbox{\includegraphics[width=\linewidth]{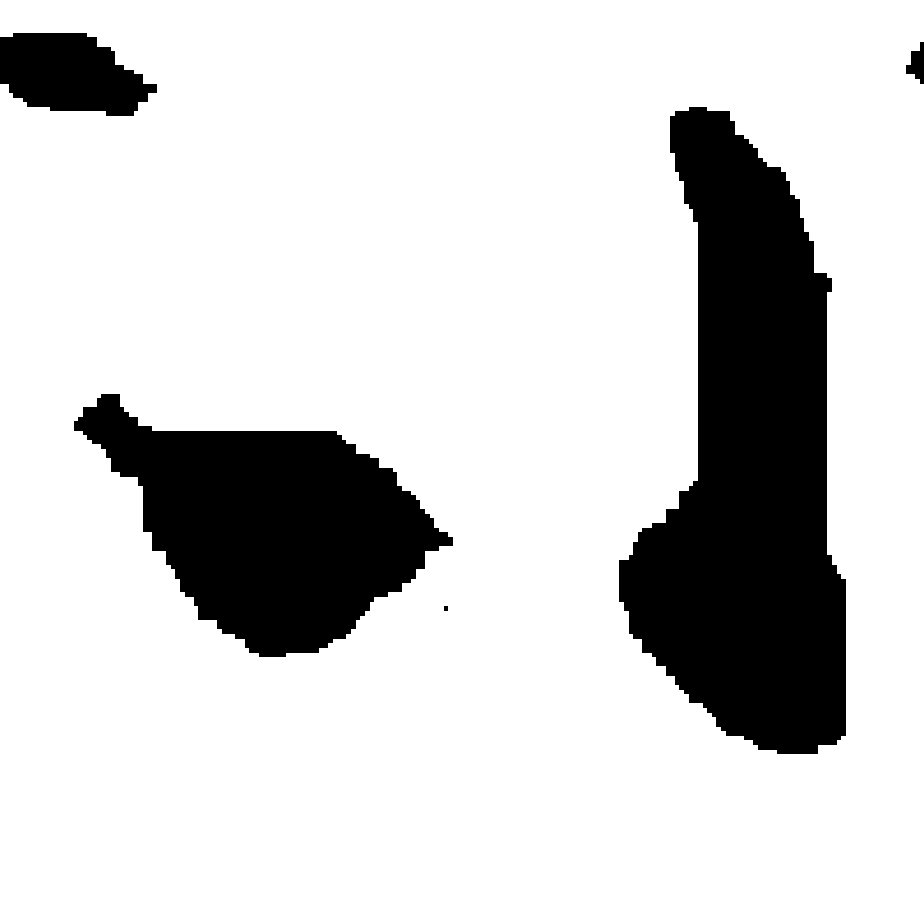}}\\
			{\footnotesize t=1000}
		\end{minipage}
				\\
		[2mm]
		\centering
		{\footnotesize (a) r=3.6}
	\end{minipage}
	\hfill
	\begin{minipage}{0.45\linewidth}
		\begin{minipage}{\linewidth}
			\centering
			\includegraphics[width=\linewidth]{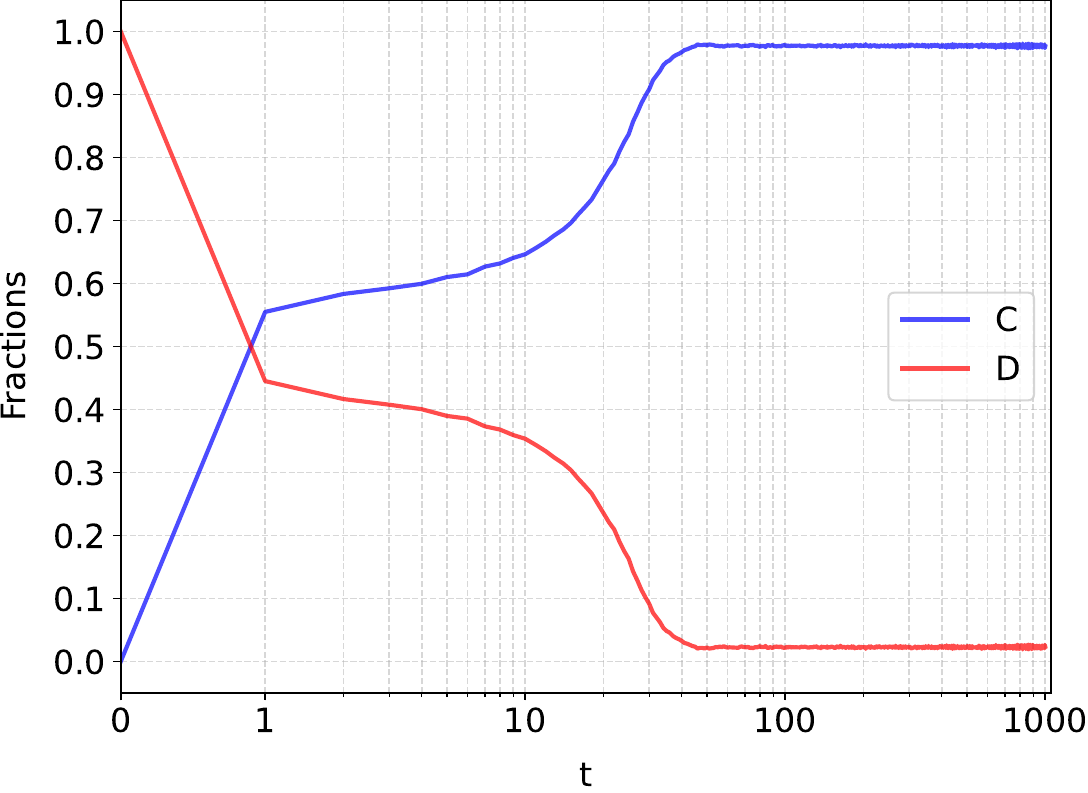}\\
		\end{minipage}
		\vspace{2mm}
		\\
		\begin{minipage}{0.188\linewidth}
			\centering
			\fbox{\includegraphics[width=\linewidth]{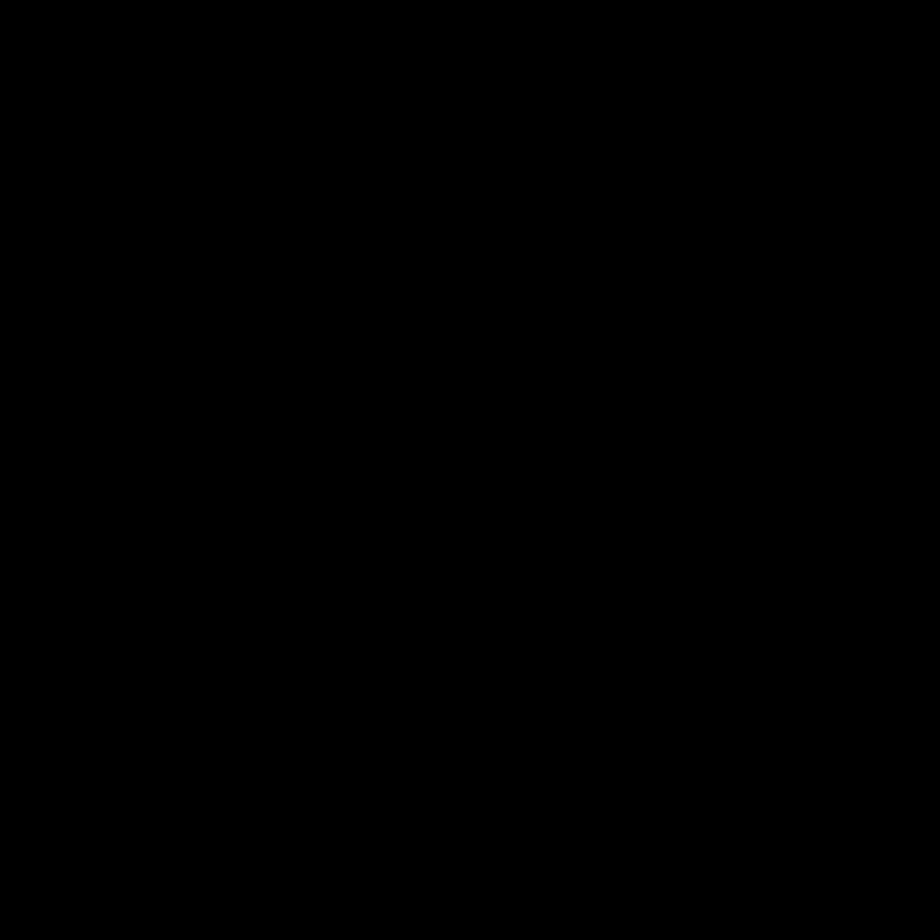}}\\
			{\footnotesize t=0}
		\end{minipage}
		\begin{minipage}{0.188\linewidth}
			\centering
			\fbox{\includegraphics[width=\linewidth]{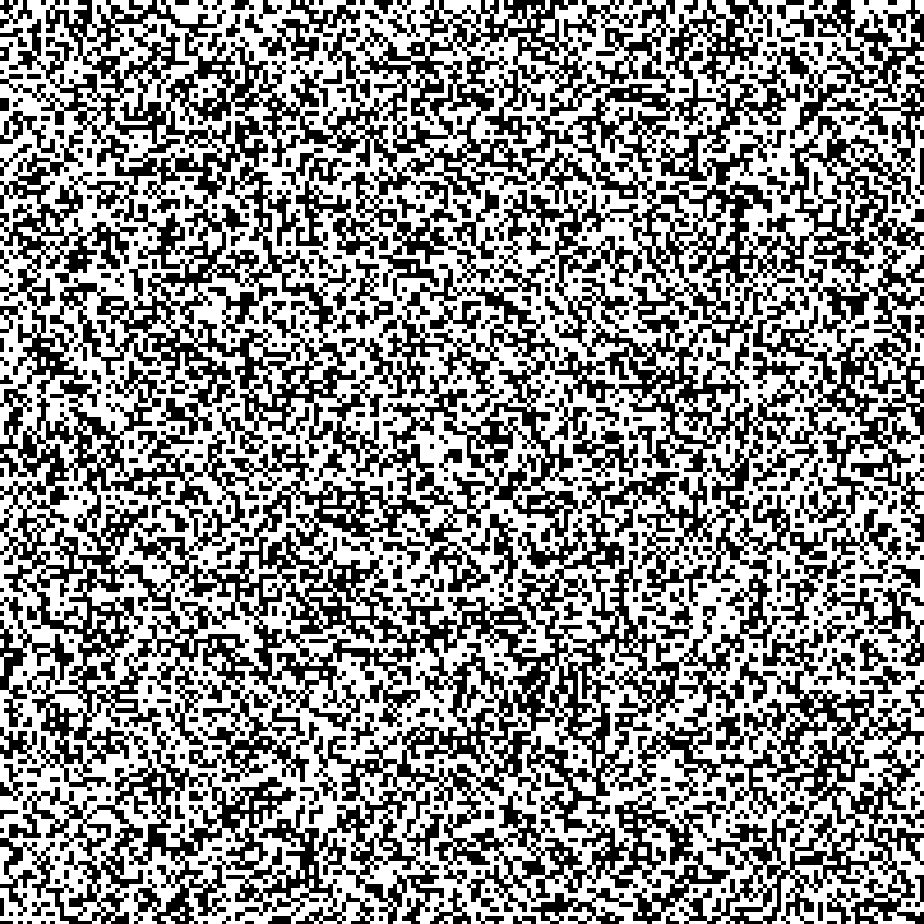}}\\
			{\footnotesize t=1}
		\end{minipage}
		\begin{minipage}{0.188\linewidth}
			\centering
			\fbox{\includegraphics[width=\linewidth]{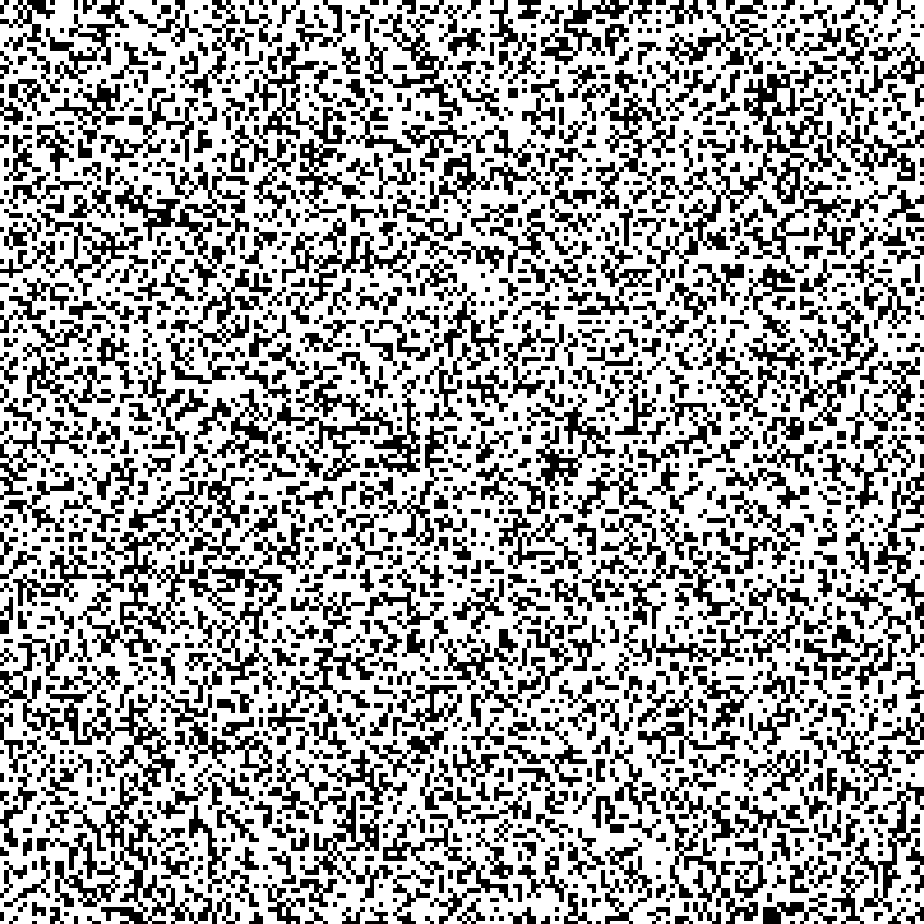}}\\
			{\footnotesize t=10}
		\end{minipage}
		\begin{minipage}{0.188\linewidth}
			\centering
			\fbox{\includegraphics[width=\linewidth]{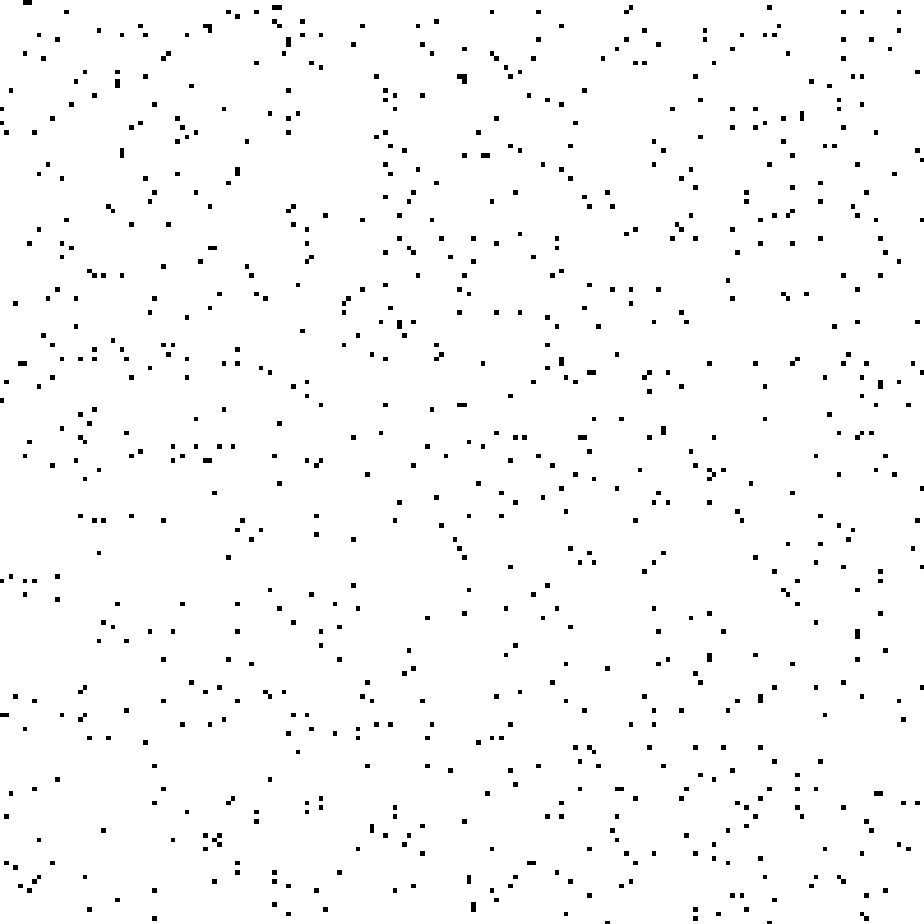}}\\
			{\footnotesize t=100}
		\end{minipage}
		\begin{minipage}{0.188\linewidth}
			\centering
			\fbox{\includegraphics[width=\linewidth]{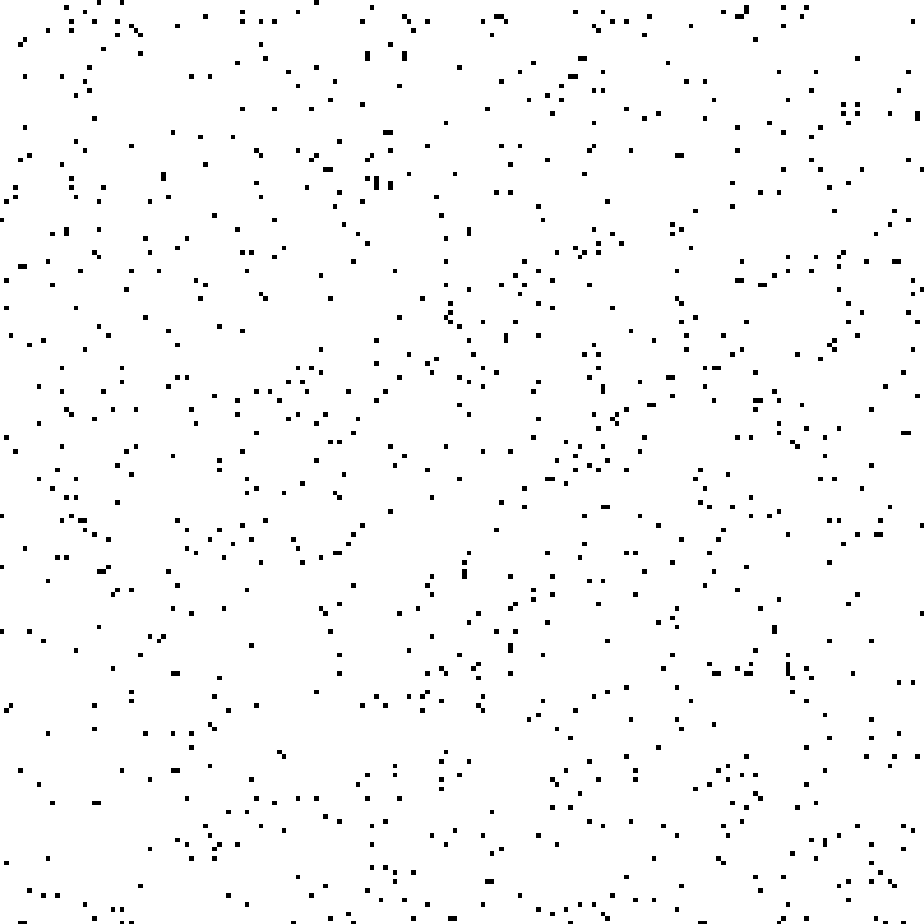}}\\
			{\footnotesize t=1000}
		\end{minipage}
				\\
		[2mm]
		\centering
		{\footnotesize (b) r=4.6}
	\end{minipage}
	\\
	[2mm]
	\begin{minipage}{\linewidth}
		\begin{minipage}{0.188\linewidth}
			\centering
			\includegraphics[width=\linewidth]{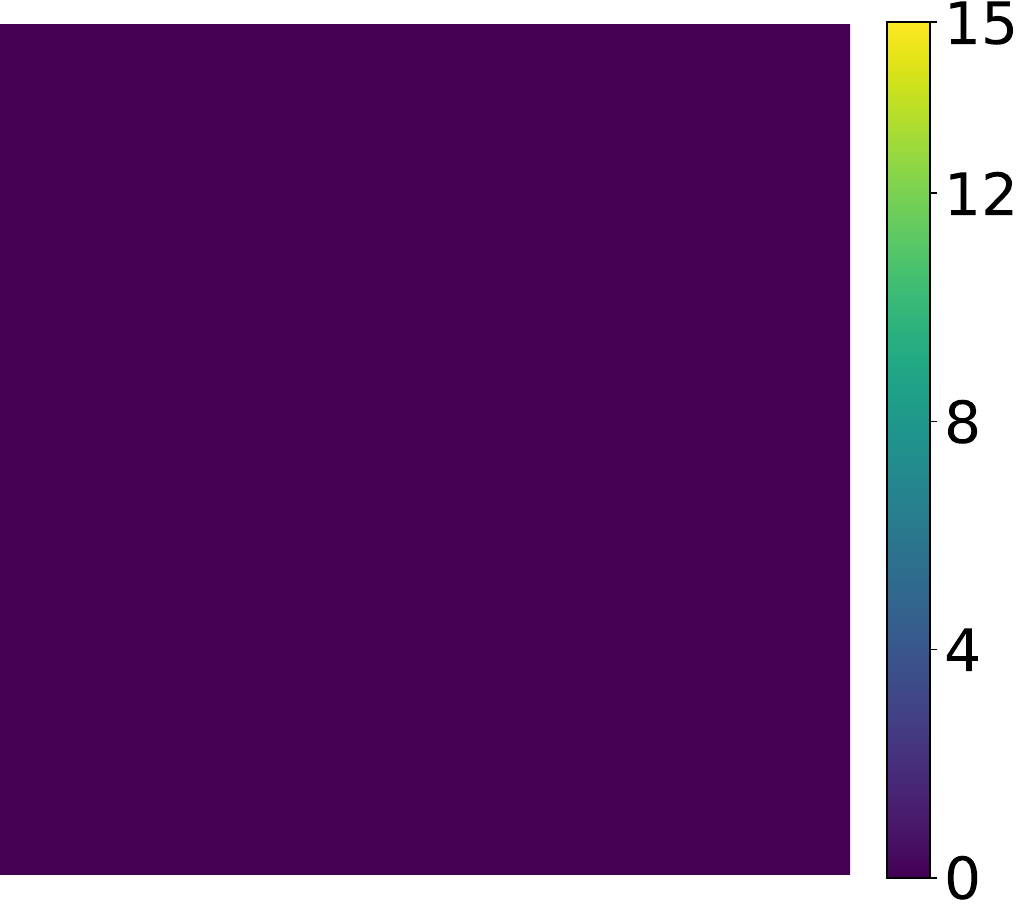}\\
			{\footnotesize t=0}
		\end{minipage}
		\hfill
		\begin{minipage}{0.188\linewidth}
			\centering
			\includegraphics[width=\linewidth]{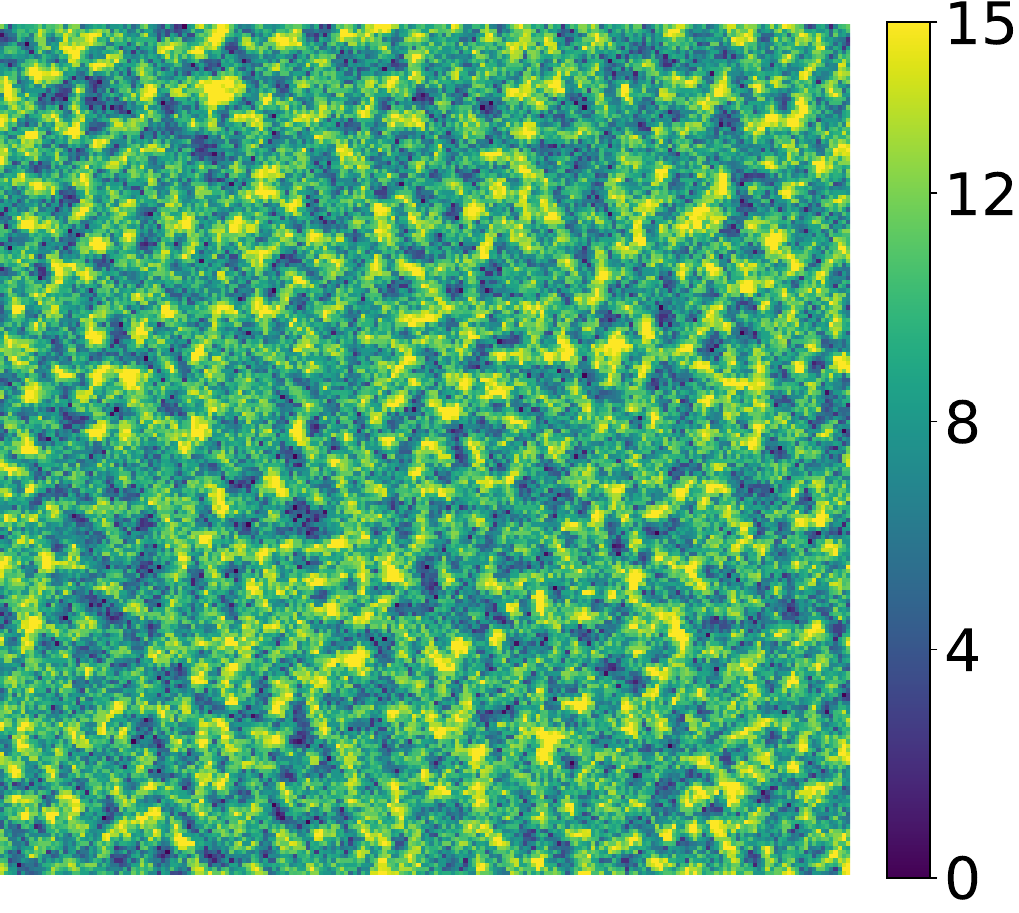}\\
			{\footnotesize t=1}
		\end{minipage}
		\hfill
		\begin{minipage}{0.188\linewidth}
			\centering
			\includegraphics[width=\linewidth]{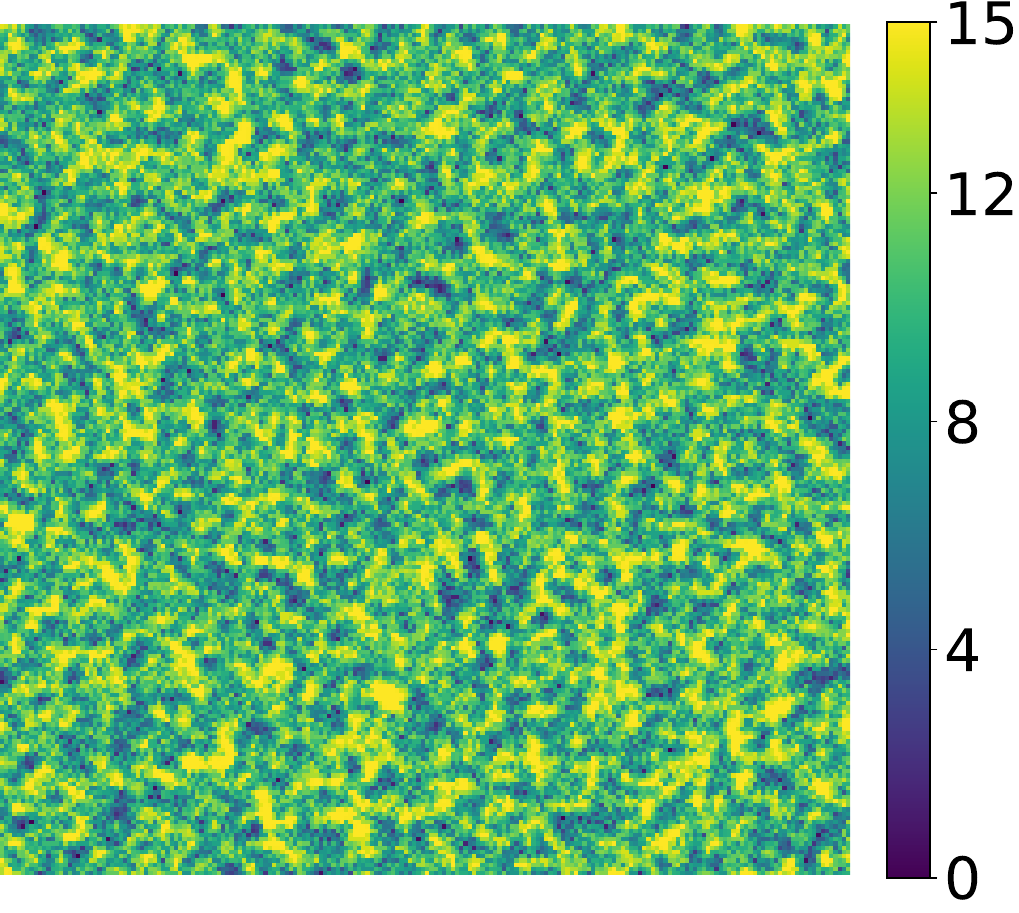}\\
			{\footnotesize t=10}
		\end{minipage}
		\hfill
		\begin{minipage}{0.188\linewidth}
			\centering
			\includegraphics[width=\linewidth]{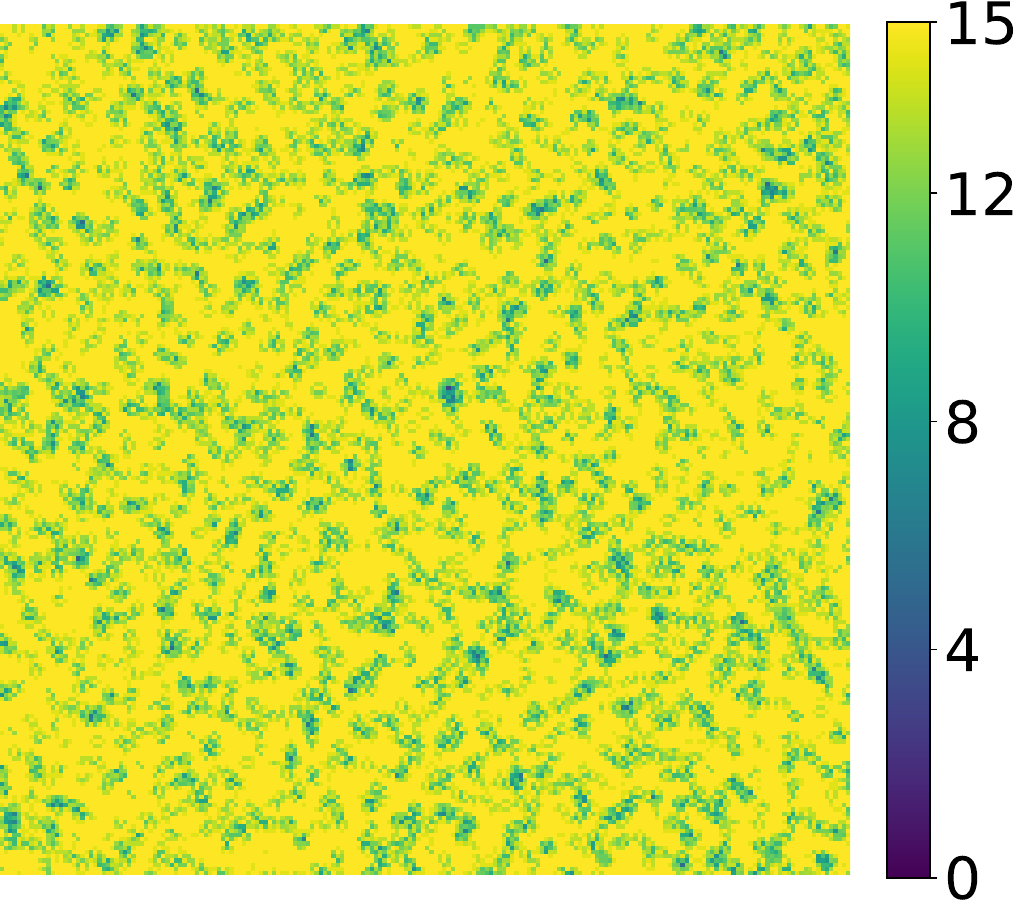}\\
			{\footnotesize t=100}
		\end{minipage}
		\hfill
		\begin{minipage}{0.188\linewidth}
			\centering
			\includegraphics[width=\linewidth]{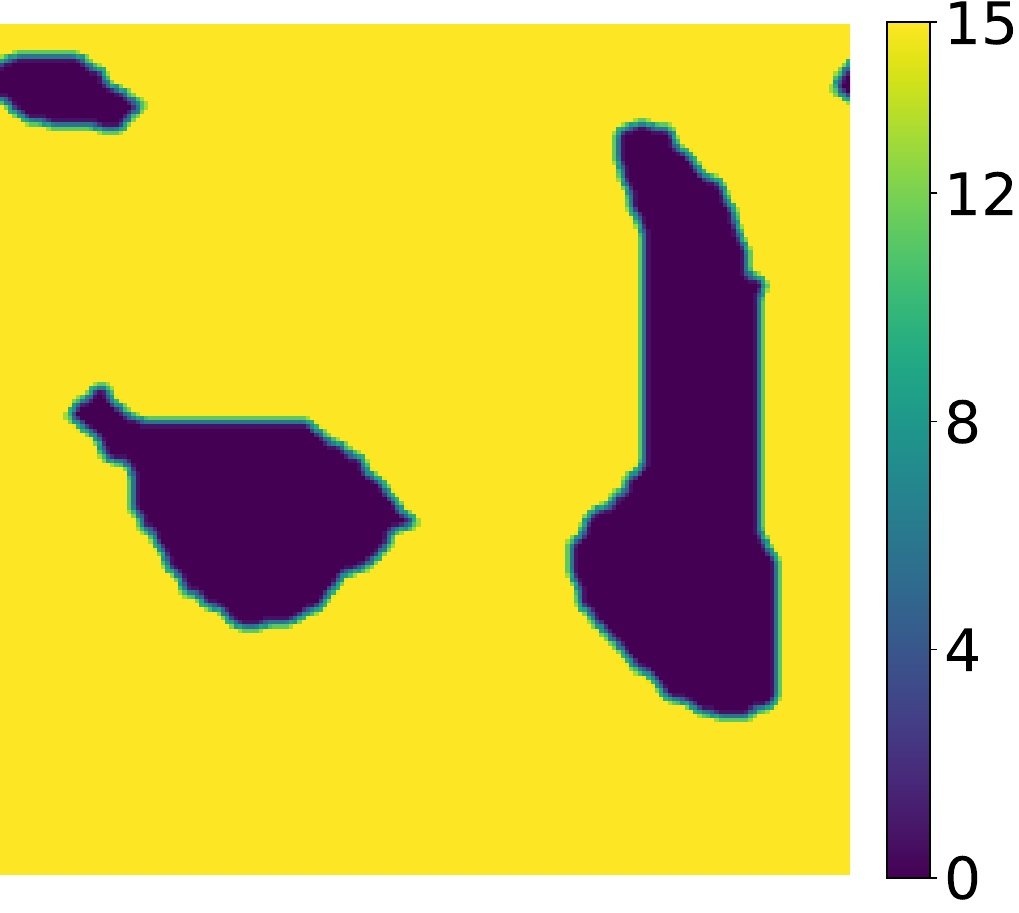}\\
			{\footnotesize t=1000}
		\end{minipage}
				\\
		[2mm]
		\centering
		{\footnotesize (c) r=3.6 (Payoff heatmaps)}
	\end{minipage}
	\\
	[2mm]
	\begin{minipage}{\linewidth}
		\begin{minipage}{0.188\linewidth}
			\centering
			\includegraphics[width=\linewidth]{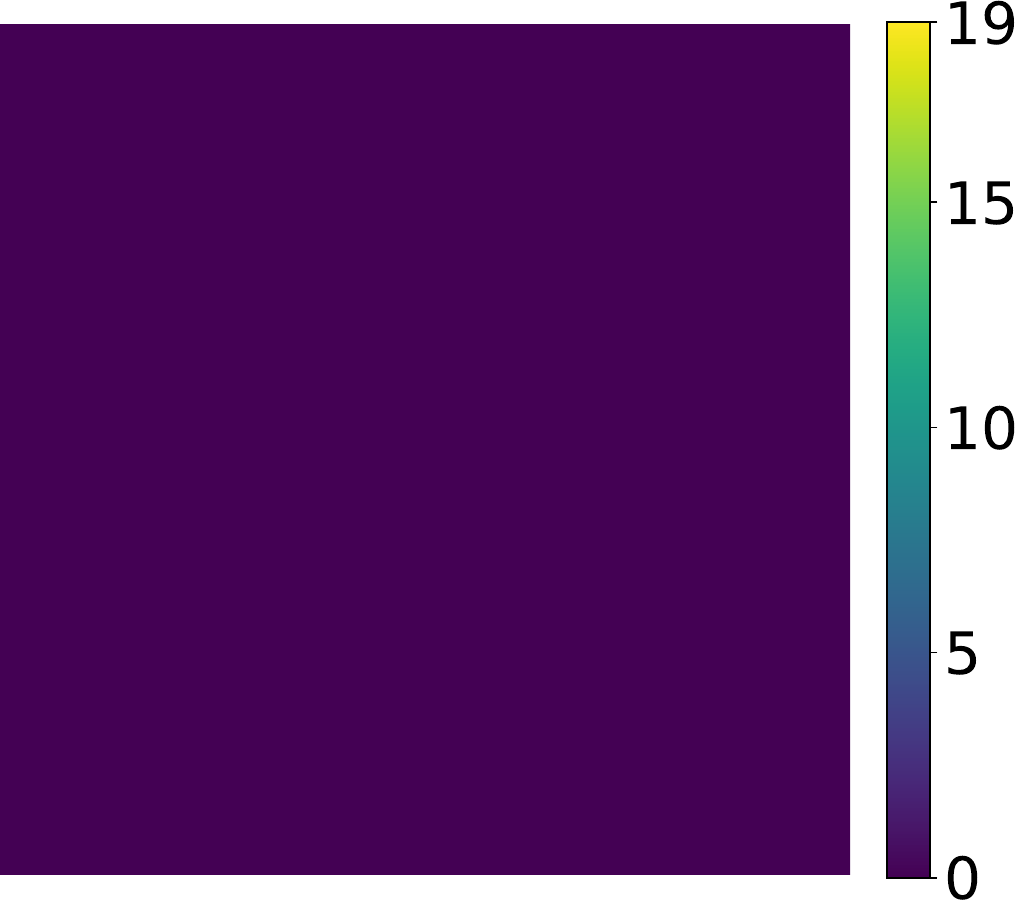}\\
			{\footnotesize t=0}
		\end{minipage}
		\hfill
		\begin{minipage}{0.188\linewidth}
			\centering
			\includegraphics[width=\linewidth]{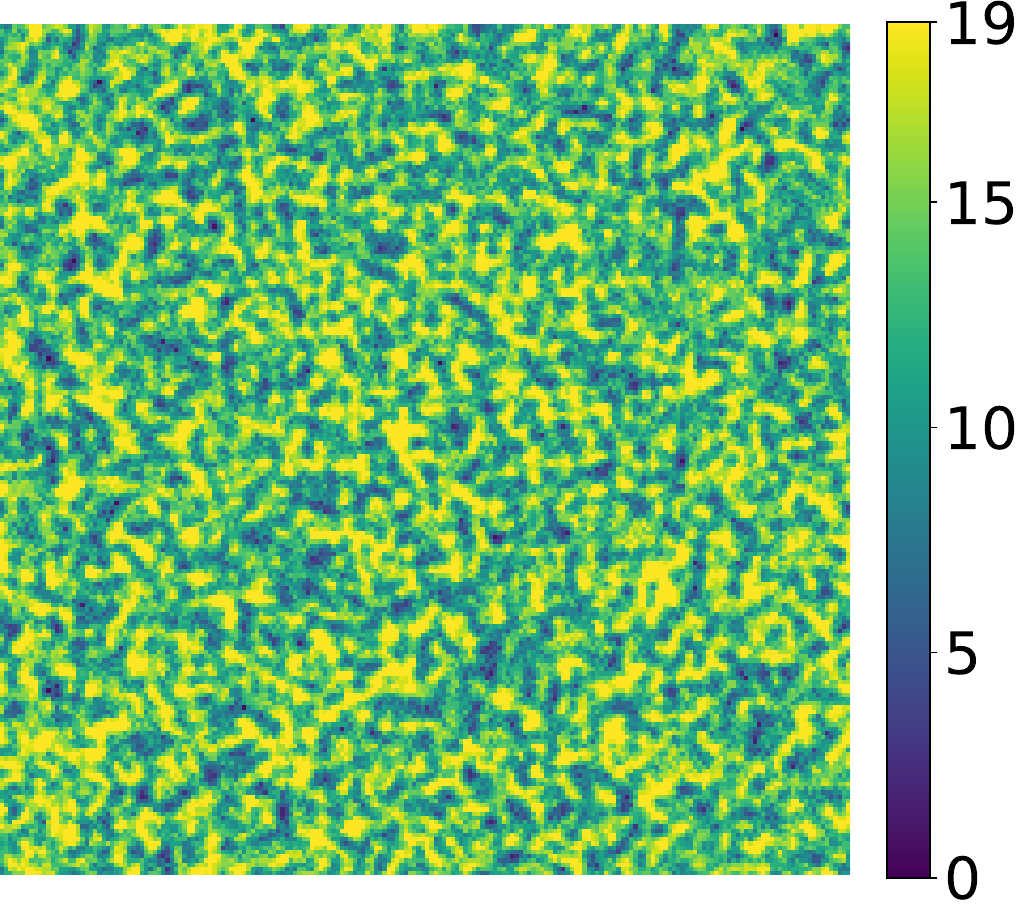}\\
			{\footnotesize t=1}
		\end{minipage}
		\hfill
		\begin{minipage}{0.188\linewidth}
			\centering
			\includegraphics[width=\linewidth]{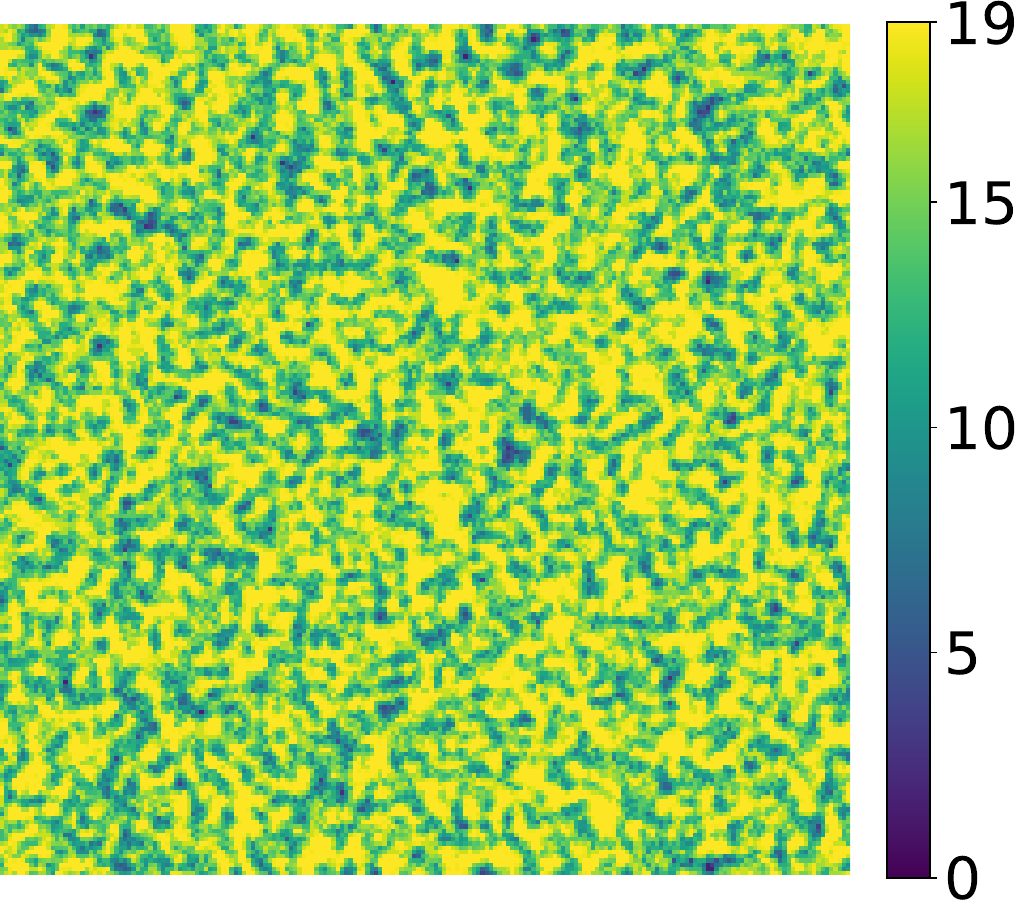}\\
			{\footnotesize t=10}
		\end{minipage}
		\hfill
		\begin{minipage}{0.188\linewidth}
			\centering
			\includegraphics[width=\linewidth]{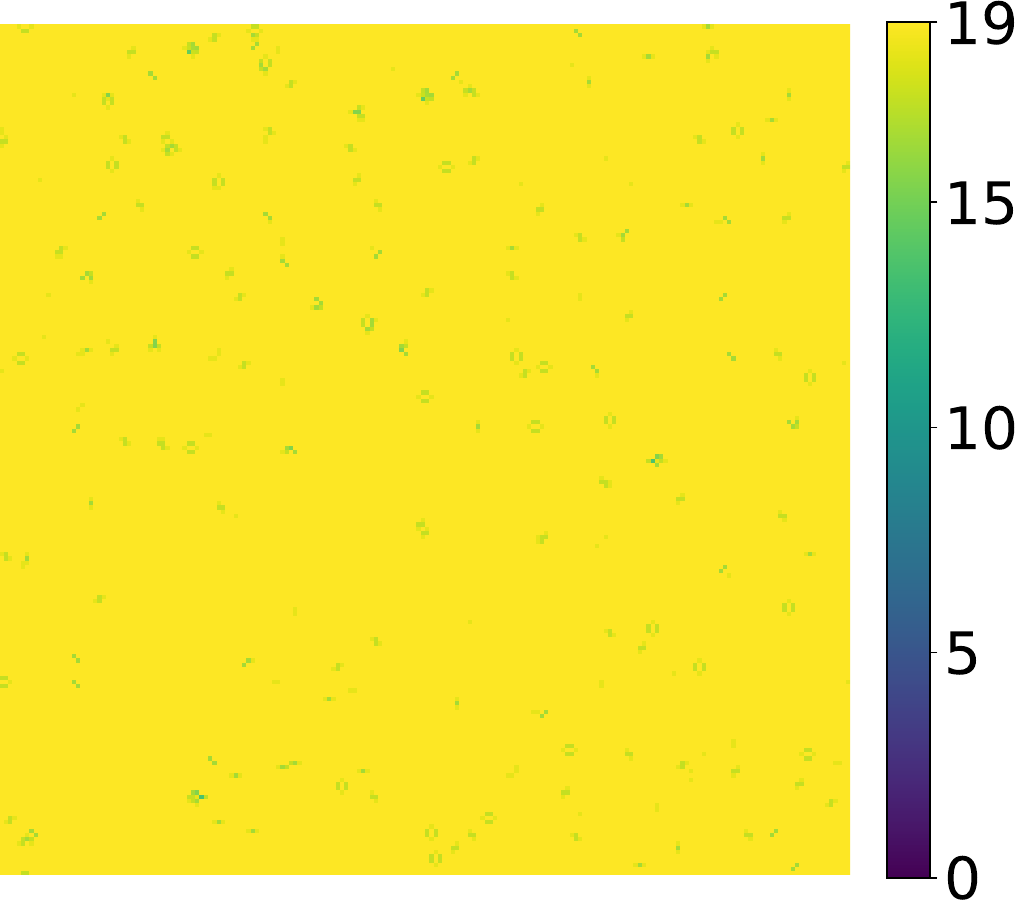}\\
			{\footnotesize t=100}
		\end{minipage}
		\hfill
		\begin{minipage}{0.188\linewidth}
			\centering
			\includegraphics[width=\linewidth]{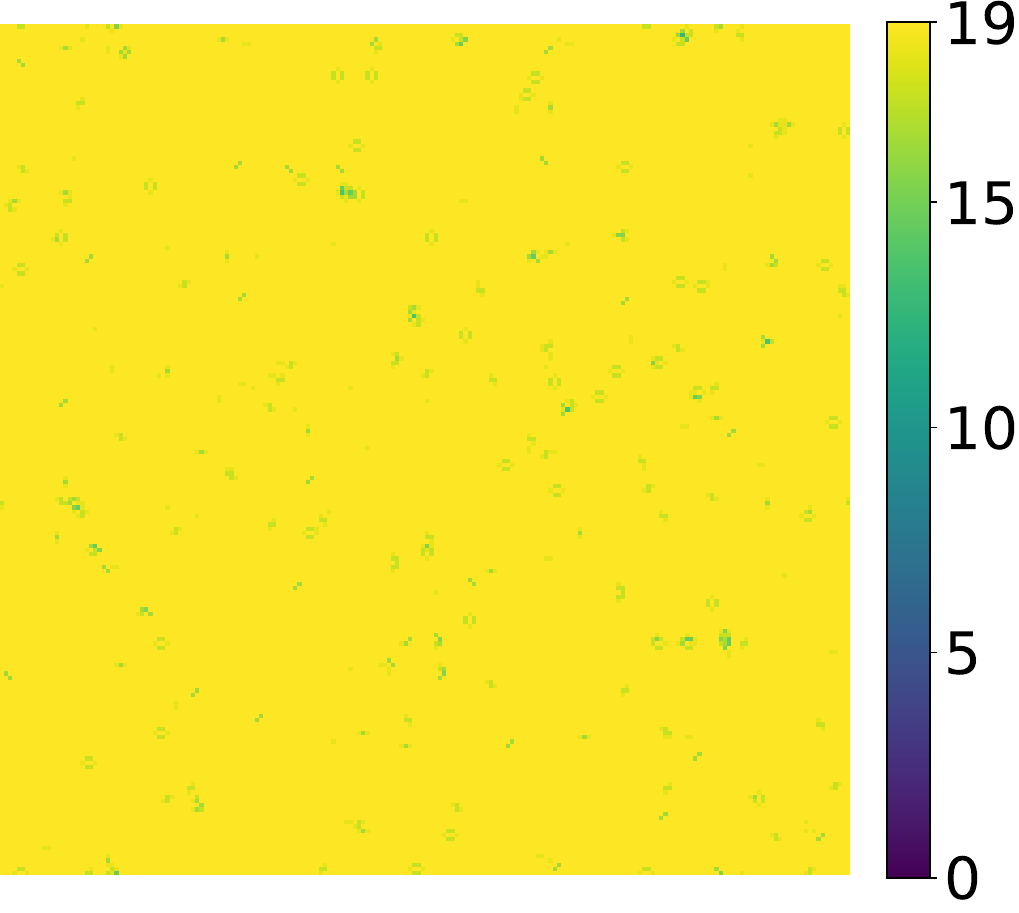}\\
			{\footnotesize t=1000}
		\end{minipage}
				\\
		[2mm]
		\centering
		{\footnotesize (d) r=4.6 (Payoff heatmaps)}
	\end{minipage}
	\caption{GRPO-GCC with all-defectors initialization on a $200 \times 200$ lattice. (a) $r=3.6$ cooperation dynamics and state snapshots. (b) $r=4.6$ cooperation dynamics and state snapshots. (c) $r=3.6$ payoff heatmaps at $t=0,1,10,100,1000$. (d) $r=4.6$ payoff heatmaps at $t=0,1,10,100,1000$. At $r=3.6$, defectors form fewer but larger clusters compared with Bernoulli initialization, while at $r=4.6$ cooperation rapidly dominates with only scattered defectors remaining.}
	\label{fig:GRPO_GCC_unique}
\end{figure*}

The results are summarized in Fig.~\ref{fig:GRPO_GCC_unique}. In subfigure (a) with $r=3.6$, cooperation gradually emerges and stabilizes at a dominant level after several dozen iterations, similar to the case with Bernoulli random initialization. However, the spatial snapshots reveal an important distinction: defectors no longer form many small clusters but instead aggregate into fewer, larger clusters. This indicates that under a fully defective start, GRPO-GCC promotes cooperation broadly across the system. However, the global constraint allows a limited number of sizeable defector communities to persist. The payoff heatmaps in subfigure (c) confirm this observation. Localized regions of low payoff correspond to these larger clusters of defectors, while surrounding cooperative regions achieve higher returns. In subfigure (b) with $r=4.6$, the cooperation fraction rises steeply and reaches about $98\%$ within 50 iterations. The snapshots show that by $t=100$, defectors remain only as scattered individuals with no ability to sustain cluster formation. The payoff heatmap in subfigure (d) corroborates this result, showing widespread high-reward zones dominated by cooperation, while defectors are confined to isolated low-payoff pockets. Taken together, these results demonstrate that GRPO-GCC enables robust cooperative emergence even under all-defectors initialization. The system exhibits dynamics comparable to random initialization but produces fewer and larger defector clusters at moderate $r$.

Taken together, the initialization experiments demonstrate that GRPO-GCC consistently drives the emergence of cooperation across diverse and challenging starting conditions. Under half-and-half initialization, cooperation expands steadily and rapidly dissolves the initial boundary, with higher $r$ producing faster convergence. With Bernoulli random initialization, cooperation remains dominant but exhibits oscillatory dynamics at moderate $r$. In this regime, defectors survive in the form of small clusters due to the self-limiting nature of GCC. Under all-defectors initialization, cooperation re-emerges. At $r=3.6$, defectors aggregate into fewer and larger clusters, while higher $r$ values eliminate clustering and yield near-complete cooperation. These findings highlight the robustness of GRPO-GCC in promoting cooperation. They also demonstrate its distinctive ability to sustain heterogeneous outcomes, such as persistent defector clusters under specific parameter regimes.

\section{Conclusions}
\label{sec:con}

This study introduces the Group Relative Policy Optimization with Global Cooperation Constraint (GRPO-GCC) as a novel DRL framework for SPGG. To our knowledge, this is the first work to extend GRPO into this domain. It establishes a new methodological baseline for studying cooperation in structured populations. The central contribution of GRPO-GCC lies in its integration of group-relative policy optimization with a global cooperation constraint that dynamically reshapes incentives. By amplifying cooperative payoffs at intermediate cooperation levels and attenuating them near extremes, the framework aligns individual decision making with sustainable collective outcomes. This design not only stabilizes policy adaptation but also prevents convergence to fragile equilibria such as universal defection or unconditional cooperation. Beyond performance improvements, GRPO-GCC provides a principled perspective on how simple global signals can coordinate decentralized learning processes. By highlighting the role of constraint-based mechanisms in fostering resilient and interpretable cooperation, GRPO-GCC demonstrates how reinforcement learning can bridge evolutionary principles with modern multi-agent systems. Future work may extend this research toward more generalized forms of global constraints, heterogeneous agent settings, and dynamic network structures. Such extensions would further enrich both the theoretical understanding and practical applications of cooperation in socio-technical systems.







\printcredits

	\section*{CRediT authorship contribution statement}

\textbf{Zhaoqilin Yang}: Writing – original draft, Investigation, Writing – review and editing, Methodology, Conceptualization.
\textbf{Chanchan Li}: Validation, Writing – review and editing, Visualization, Methodology.
\textbf{Tianqi Liu}: Investigation, Supervision, Writing – original draft.
\textbf{Xin Wang}: Conceptualization, Software, Writing – review and editing.
\textbf{Youliang Tian}: Funding acquisition, Resources, Supervision.

\section*{Declaration of competing interest }

The authors declare that they have no known competing financial interests or personal relationships that could have appeared to influence the work reported in this paper.

\section*{Data availability}

No data was used for the research described in the article.

\section*{Acknowledgments}
This work was supported by the Natural Science Special Project (Special Post) Research Foundation of Guizhou University (No.[2024] 39), Guizhou Provincial Basic Research Program (Natural Science) Youth Guidance Project (No. Qiankehe Foundation QN(2025) 054). National Key Research and Development Program of China under Grant 2021YFB3101100; National Natural Science Foundation of China under Grant 62272123; Project of High-level Innovative Talents of Guizhou Province under Grant [2020]6008; Science and Technology Program of Guizhou Province under Grant [2020]5017, [2022]065; Science and Technology Program of Guiyang under Grant [2022]2-4.

\bibliographystyle{cas-model2-names}

\bibliography{cas-refs}



\end{document}